\documentclass[showpacs, amsmath, amsfonts, amssymb, aps, superscriptaddress]{revtex4}

\usepackage{amsmath,amssymb}
\usepackage{graphicx}
\usepackage{mathtools}
\usepackage{physics}
\usepackage[utf8]{inputenc}
\usepackage[T1]{fontenc}
\usepackage{microtype}
\usepackage{caption}
\usepackage{parskip}
\usepackage{xcolor}

\usepackage{amsfonts} 
\usepackage{mathtools}
\usepackage{listings}
\usepackage{booktabs}
\usepackage{float}
\usepackage{natbib}
\begin{document}

    \title{Quasinormal modes of the Kazakov--Solodukhin quantum-corrected black hole: a spectral analysis}
    
    \author{Davide Batic}
    \email{davide.batic@ku.ac.ae}
    \affiliation{Mathematics Department, Khalifa University of Science and Technology, PO Box 127788, Abu Dhabi, United Arab Emirates}

    \author{Denys Dutykh}
    \email{denys.dutykh@ku.ac.ae}
    \affiliation{Mathematics Department, Khalifa University of Science and Technology, PO Box 127788, Abu Dhabi, United Arab Emirates}
    
    \author{Mark Essa Sukaiti}
    \email{100064482@ku.ac.ae}
    \affiliation{Mathematics Department, Khalifa University of Science and Technology, PO Box 127788, Abu Dhabi, United Arab Emirates}
    
    \date{\today}

\begin{abstract}
We compute quasinormal modes of the Kazakov--Solodukhin quantum-corrected black hole using a high-precision Chebyshev spectral method. After factoring out the quasinormal mode asymptotics at the event horizon and at spatial infinity, the radial problem is reduced to a quadratic matrix pencil for the dimensionless frequency $\Omega=M\omega$. We apply this framework to minimally coupled scalar perturbations, electromagnetic perturbations, a non-minimally coupled scalar field, and an axial effective source Regge--Wheeler-type gravitational sector. The latter is treated as an effective axial model, rather than as the full gravitational perturbation problem, because the Kazakov--Solodukhin spacetime is not Ricci flat. Our results reproduce the available WKB, time-domain, Mashhoon, and asymptotic-iteration benchmarks in their common regimes of validity, after accounting for the different normalisations used in the literature. The spectral method also resolves additional overtones and candidate purely imaginary overdamped roots. In several sectors, these roots exhibit a spacing scale close to $M\kappa=1/4$, with occasional multiple gaps in the retained numerical sequence. In the near-extremal, but still subextremal, regime, the detected purely imaginary branches approach an approximately equally spaced surface-gravity-scaled ladder, while the oscillatory spectra remain sector dependent. Within the parameter ranges and resolutions considered here, all retained modes have $\Im{\Omega}<0$, and no stable growing mode is detected.
\end{abstract}

    \maketitle
	
\section{Introduction}\label{sec:intro}

Black holes occupy a central position in general relativity, relativistic astrophysics, and gravitational wave physics. Their response to external perturbations is encoded in a discrete set of complex frequencies known as quasinormal modes (QNMs). These modes are defined by boundary conditions rather than by normalizability in a closed system. More precisely, for an asymptotically flat black hole, the perturbation must be purely ingoing at the event horizon and purely outgoing at spatial infinity. The resulting spectral problem is therefore non-Hermitian, and the corresponding frequencies are complex. With the convention adopted in this work, namely a time dependence of the form $e^{-i\omega t}$, a mode with $\Im{\omega}<0$ is damped, while a positive imaginary part would signal exponential growth. QNMs provide a useful diagnostic of linear stability, characterise the ringdown stage of black-hole perturbations, and play an important role in black-hole spectroscopy and tests of strong-field gravity \cite{Kokkotas1999LR, Berti2009CQG, Konoplya2011RMP}. Beyond classical black hole solutions of general relativity, QNMs also provide a sensitive probe of modified or effective geometries. In particular, quantum-gravity-inspired deformations of the Schwarzschild metric provide a controlled setting in which one can examine how short-distance corrections affect wave propagation, damping rates, and the structure of the complex-frequency spectrum. The spacetime considered in the present work is the Kazakov--Solodukhin quantum-corrected (KSQC) black hole, obtained as a spherically symmetric deformation of the Schwarzschild solution \cite{KazakovNPB1994}. Such a spacetime is not Ricci flat and has a curvature singularity. This non-Ricci-flat character is important when interpreting perturbation sectors, especially the gravitational one. The QNM spectrum of the KS geometry has been studied by several methods. \cite{SalehASS2014} computed scalar field QNMs using a third-order WKB approximation and found that increasing the quantum-correction parameter tends to lower both the oscillation frequency and the damping rate. Gravitational and Dirac perturbations, again using third-order WKB methods, have been considered in \cite{SalehASS2016}. Electromagnetic perturbations were analysed by \cite{WangJAA2017} with the same WKB order. \cite{KonoplyaPLB2020} subsequently revisited the problem using higher-order WKB methods with Padé improvement and time-domain integration, considering minimally coupled scalar, electromagnetic, Dirac, and conformally coupled scalar fields. In that work, it was also emphasised that the KSQC spacetime is not Ricci flat, and that a vacuum gravitational perturbation equation $\delta R_{\mu\nu}=0$ cannot be imposed without further assumptions. \cite{ZhangIJP2023} later studied QNMs in the non-Ricci-flat scalar setting using the Mashhoon, or P\"oschl--Teller, approximation and the asymptotic iteration method (AIM) \cite{Ciftci2003JPA}. More recently, \cite{bolokhov2025overtones} investigated overtones and Hawking evaporation in the same quantum-corrected geometry, showing that overtones and grey-body factors may be more sensitive to the deformation than the fundamental mode. These works provide important benchmarks, but several aspects of the spectrum remain difficult to access with methods designed to target a small number of modes. WKB approximations are especially useful for barrier-controlled modes, but their reliability decreases for low multipoles, high overtones, or regimes where the potential is strongly deformed. Time-domain integration is well suited to identifying dominant ringdown frequencies, but subdominant or highly damped branches may be hidden by the tail or by more slowly damped modes. AIM and P\"oschl--Teller approximations provide complementary checks, but they do not directly expose the global structure of the spectrum in the complex plane. Analytic studies of wave equations in other non-Schwarzschild black-hole backgrounds, including rotating linear-dilaton and Born--Infeld--dilaton geometries, provide complementary examples in which resonant frequencies, boxed modes, greybody factors, and quantisation conditions can be extracted from exact or semi-analytic solutions~\cite{Sakalli2016PRD, SakalliTokgoz2016AnnPhys, SakalliJusufiOvgun2018GRG}. For this reason, a high-precision spectral approach is a natural tool for revisiting the KSQC spectrum. In the present work, we apply a Chebyshev spectral method to the QNM problem of the KS black hole. The method follows the general strategy developed and tested in several recent spectral studies of black holes and wormholes \cite{Batic2024CQG, Batic2024EPJC, Batic2024PRD, Batic2025EPJC, Batic2025CQG, Batic2026EPJC, Batic2025PRSA}. After factoring out the QNM asymptotics at the horizon and at spatial infinity, the remaining radial function is regular on a compactified interval. Expanding this regular part in Chebyshev polynomials and collocating the transformed equation leads to a quadratic matrix pencil for the dimensionless frequency $\Omega=M\omega$. The matrices are assembled with high precision, and the resulting spectra are filtered to ensure stability across three spectral resolutions. We consider minimally coupled scalar perturbations, electromagnetic perturbations, a conformally coupled scalar field, and an axial effective source Regge--Wheeler-type sector. The last sector is not claimed to be the complete gravitational perturbation theory of the KSQC geometry. Rather, the non-Ricci-flat background is interpreted as being supported by an effective anisotropic source, and the axial potential is obtained under the assumption that this effective source carries no independent axial degree of freedom. This provides a consistent model sector that reduces to the ordinary Regge--Wheeler potential in the Schwarzschild limit, while avoiding the direct use of $\delta R_{\mu\nu}=0$ on a non-vacuum background. The goals of the paper are fourfold. First, we validate the spectral implementation by comparing it with the available WKB, time-domain, AIM, and Mashhoon data in their common domains of applicability. Since different papers use different normalisations of the deformation parameter and of the frequency, special attention is paid to converting between the horizon-normalised convention and the mass-normalised convention used here. Second, we extend the numerical spectra to higher overtones and to parameter regimes not previously tabulated. Third, we investigate candidate purely imaginary overdamped roots detected by the spectral method. Such roots are treated cautiously as candidate spectral features. Purely imaginary QNMs are known to occur in black hole spectra, but in an asymptotically flat spacetime, one must also keep in mind the possible role of branch-cut or tail contributions. Fourth, we analyse the near-extremal regime, in which the deformation parameter approaches its maximum at a fixed horizon radius. In this regime, we find a pronounced reorganisation of the detected overdamped branches and an approximate near-extremal degeneracy of the purely imaginary ladders in several sectors, while the oscillatory spectra remain sector dependent.

The paper is organised as follows. In Sec.~\ref{sec:eom} we review the KSQC geometry, define the perturbation sectors, and derive the radial equations and QNM boundary conditions. We also discuss the conformally coupled scalar sector and the axial effective-source Regge--Wheeler model. In Sec.~\ref{subsec:extremal_KS_limit} we analyse the $a_K\to 1^{-}$ limiting geometry and clarify why this limit is singular rather than a regular extremal black-hole limit. In Sec.~\ref{sec:method} we describe the Chebyshev spectral discretisation, the quadratic eigenvalue problem, and the stability criterion used to select physical roots. Section~\ref{sec:NR} contains the numerical results and comparisons with the existing literature for the scalar, electromagnetic, conformally coupled scalar, and axial sectors. We also discuss candidate purely imaginary modes, near-extremal behaviour, and stability. Finally, Sec.~\ref{sec:conclusions} summarises the main findings and outlines directions for further work.

\section{Metric and equations of motion}\label{sec:eom}

To introduce the QNM eigenvalue problem and clarify our conventions, we briefly review the perturbation equations for scalar, electromagnetic, and gravitational disturbances of the KSQC black-hole background, emphasising the boundary conditions and spectral formulation common to the three sectors while keeping their effective potentials distinct. In Planck units, i.e. $c = G_N =\hbar= 1$, the corresponding line element takes the form \cite{KazakovNPB1994}
\begin{equation}\label{LE}
ds^2=-f(r)dt^2+f^{-1}(r)dr^2+r^2\left(d\vartheta^2 + \sin^2{\vartheta} d\varphi^2\right),\quad 
f(r)=\frac{1}{r}\left(\sqrt{r^2-a^2}-2M\right),\quad r>a.
\end{equation}
Here, $M$ denotes the mass parameter of the black hole, while $a^2=4G_R/\pi$ is the deformation parameter that measures the strength of the underlying quantum correction and determines how strongly the short-distance structure of the spacetime differs from its classical counterpart. Under the assumption of a time dependence of the form $e^{-i\omega t}$ and an angular component described by spherical harmonics, the equations governing massless scalar and electromagnetic perturbations are
\begin{equation}\label{ODE01}
    f(r)\frac{d}{dr}\left(f(r)\frac{d\psi}{dr}\right)+\left[\omega^2-U_\epsilon(r)\right]\psi(r)=0,\qquad
    U_\epsilon(r)=f(r)\left[\frac{\epsilon}{r}\frac{df}{dr}+\frac{\ell(\ell+1)}{r^2}\right],\qquad
    \epsilon=1-s^2
\end{equation}
with multipole number $\ell = 0, 1, 2, \ldots$ and $\epsilon = 1$ (massless scalar perturbation $s=0$), and $\epsilon = 0$ (electromagnetic perturbation $s=1$). Since the KSQC spacetime is not Ricci flat, one may also consider a conformally coupled scalar field. In that case, the effective potential is \cite{ChernikovAIHPA1968, KonoplyaPLB2020, ZhangIJP2023, SalehASS2016} 
\begin{equation}\label{Vg}
V_c(r)=f(r)\left[\frac{\ell(\ell+1)}{r^2}+\frac{1}{r}\frac{df}{dr}+\frac{1}{6}R(r)\right],\quad
R(r)=\frac{2}{r^2}+\frac{3a^2-2r^2}{r(r^2-a^2)^{3/2}}.
\end{equation}
Introducing the dimensionless radial coordinate $x=r/(2M)$, the lapse function can be written as
\begin{equation}\label{fx}
    f(x)=\frac{\sqrt{4x^2-\mathfrak{a}^2}-2}{2x},\quad
    \mathfrak{a}=\frac{a}{M},\quad x>\frac{\mathfrak{a}}{2}.
\end{equation}
The event horizon is located at the simple zero of $f$, namely
\begin{equation} \label{eq:horizon}
x_h=\frac{\sqrt{\mathfrak{a}^2+4}}{2}.
\end{equation}
A direct computation gives $f^{'}(x_h)=1$ where the prime denotes differentiation with respect to $x$. Hence, the horizon is nondegenerate for every finite value of the deformation parameter. Finally, rationalizing the numerator in \eqref{fx} yields the equivalent but useful factorized form
\begin{equation}
f(x)=\frac{x+x_h}{x\left(\sqrt{x^2+1-x_h^2}+1\right)}(x-x_h).    
\end{equation}

\subsection{Massless scalar and electromagnetic perturbations in the KSQC spacetime}

\begin{figure}[t]
\centering
\includegraphics[width=0.95\linewidth]{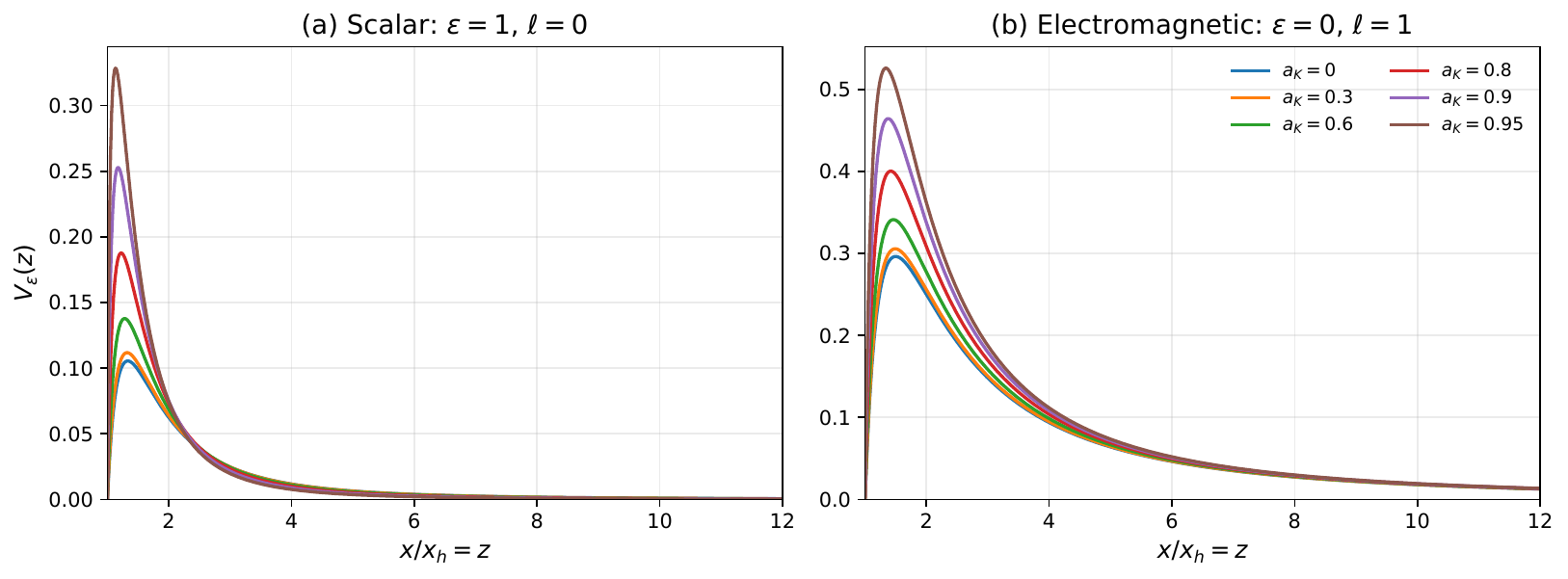}
\caption{
Effective potentials $V_\epsilon(z)$ (defined in \eqref{ourODE}) in the horizon-normalised radial coordinate $z=x/x_h$, with $x_h=(1-a_K^2)^{-1/2}$ and $a_K=a/r_h$ \cite{KonoplyaPLB2020}. The left panel shows the minimally coupled scalar monopole potential, $\epsilon=1$, $\ell=0$, while the right panel shows the electromagnetic dipole potential, $\epsilon=0$, $\ell=1$. Increasing $a_K$ raises the potential barrier and moves its maximum closer to the horizon $z=1$. The extremal value $a_K=1$ is not plotted because $x_h$ diverges in this parametrisation.
}
\label{fig:scalar_em_potentials}
\end{figure}

By means of the substitution $z=x/x_h$ which sends $x=x_h\to 1$, equation \eqref{ODE01} can be expressed in the equivalent form
\begin{equation}\label{ourODE}
    f(z)\frac{d}{dz}\left(f(z)\frac{d\psi}{dz}\right)+\left[4x_h^2\Omega^2-V_\epsilon(z)\right]\psi(z)=0,\qquad
    V_\epsilon(z)=f(z)\left[\frac{\epsilon}{z}\frac{df}{dz}+\frac{\ell(\ell+1)}{z^2}\right],\qquad
    \Omega=M\omega
\end{equation} 
with $f(z)$ given by
\begin{equation}\label{fz}
f(z)=\frac{x_h(z+1)}{z\left(\sqrt{x_h^2 z^2+1-x_h^2}+1\right)}(z-1).
\end{equation}
From a geometrical perspective, the rescalings introduced above have several important consequences. First, they remove the explicit dependence on the mass parameter $M$, thereby rendering the problem manifestly scale-invariant. Second, it allows one to interpret $\mathfrak{a}$ as the sole parameter governing deviations from classical general relativity. Finally, and most importantly for our purposes, it maps the domain of interest to a fixed interval with a distinguished boundary at $x=1$, which greatly facilitates the imposition of quasinormal-mode boundary conditions and the implementation of spectral methods. Before adapting the problem to the Spectral Method, it is useful to display the deformation dependence of the effective potentials. Fig.~\ref{fig:scalar_em_potentials} shows the effective potential $V_\epsilon(z)$ defined in \eqref{ourODE}, for the minimally coupled scalar monopole and the electromagnetic dipole. As $a_k=a/r_h$ increases, both the scalar monopole and electromagnetic dipole barriers become higher and move closer to the horizon in the coordinate $z=x/x_h$. For example, the scalar $\ell=0$ peak moves from approximately $z\sim 1.334$ at $a_K=0$ to $z\sim 1.136$ at $a_K=0.95$, while its height increases from approximately $0.1055$ to $0.3285$. The electromagnetic $\ell=1$ barrier similarly shifts from $z\sim 1.499$ to $z\sim 1.341$, with the peak increasing from approximately $0.2963$ to $0.5260$.

In the following analysis, we focus on computing the QNMs for the spectral problem stated in \eqref{ourODE}. For this purpose, we represent $\Omega$ as $\Omega = \Omega_R + i\Omega_I$, where $\Omega_I < 0$ ensures that the perturbation is damped in time. The boundary conditions are set so that the radial field exhibits inward radiation at the event horizon and outgoing radiation at spatial infinity. This necessitates a detailed examination of the solution asymptotic behaviour in \eqref{ourODE}, both near the event horizon $z \to 1^{+}$ and at spatial infinity ($z \to +\infty$). Moreover, to compute the QNMs using the spectral method, we must recast the differential equation \eqref{ourODE} and the appropriate boundary conditions over the compact interval $[-1,1]$. This adjustment is essential because the method expands the regular part of the eigenfunctions in terms of Chebyshev polynomials. We split our analysis by examining the behaviour of the radial field in two different regions
\begin{enumerate}
\item 
{\underline{Asymptotic behavior as $z\to 1^+$}}: Let us recall that $z=1$ is a simple zero of $f(z)$. Moreover, according to \eqref{fz}, the latter can be represented in the form $f(z) = (z-1) g(z)$ where $g(z)$ is an analytic function at $z=1$ with the property that $f^{'}(1)\neq 0$. If we reformulate \eqref{ourODE} in the form
\begin{equation}\label{ODEZ}
     \frac{d^2\psi}{dz^2}+p(z)\frac{d\psi}{dz}+q(z)\psi(z)=0,\quad
     p(z)=\frac{f^{'}(z)}{f(z)},\quad
     q(z)=\frac{4x_h^2\Omega^2-V_\epsilon(z)}{f^2(z)},   
\end{equation}
since $p(z)$ and $q(z)$ have poles of order one and two at $z = 1$, respectively, this point is classified as a regular singular point of \eqref{ODEZ}. Hence, by means of Frobenius theory \cite{ince1956ordinary}, we can construct solutions of the form
\begin{equation}
\psi(z) = (z-1)^\rho\sum_{\kappa=0}^\infty a_\kappa(z-1)^\kappa.
\end{equation}
The leading behavior at $z = 1$ is represented by the term $(z-1)^\rho$ where $\rho$ is determined by the indicial equation
\begin{equation}\label{indicial0}
        \rho(\rho-1) + p_0\rho + q_0 = 0
\end{equation}
with
\begin{equation}
        p_0 = \lim_{z \to 1}(z-1)p(z) = 1, \qquad
        q_0=\lim_{z \to 1}(z-1)^2 q(z) = 4\Omega^2.
\end{equation}
The roots of \eqref{indicial0} are $\rho_\pm = \pm 2i\Omega$ and the correct QNM boundary condition at $z = 1$ reads
\begin{equation}\label{QNMBCz1}
    \psi\underset{{z\to 1^+}}{\longrightarrow} (z-1)^{-2i\Omega}.
\end{equation}
\item 
{\underline{Asymptotic behaviour as $z\to +\infty$}}: Following the method outlined in \cite{Batic2024EPJC}, it is straightforward to verify that the point at infinity is an irregular singular point of rank one. Hence, the asymptotic behaviour of the solutions to equation \eqref{ODEZ} can be deduced using the method outlined in \cite{Olver1994MAA}. For this purpose, we start by observing that
\begin{equation}
    p(z) = \sum_{\kappa=0}^\infty\frac{\mathfrak{f}_\kappa}{z^\kappa} = \mathcal{O}\left(\frac{1}{z^2}\right), \qquad
    q(z) = \sum_{\kappa=0}^\infty\frac{\mathfrak{g}_\kappa}{z^\kappa}=4x_h^2\Omega^2+\frac{8x_h\Omega^2}{z}+\mathcal{O}\left(\frac{1}{z^2}\right).
\end{equation}
Given that at least one of the coefficients $\mathfrak{f}_0$, $\mathfrak{g}_0$, $\mathfrak{g}_1$ is nonzero, a formal solution to \eqref{ODEZ} is represented by \cite{Olver1994MAA}
\begin{equation}\label{olvers}
    \psi^{(j)}(z) = z^{\mu_j}e^{\lambda_j z}\sum_{\kappa=0}^\infty\frac{a_{\kappa,j}}{z^\kappa}, \qquad j \in \{1,2\},
\end{equation}
where $\lambda_1$, $\lambda_2$, $\mu_1$ and $\mu_2$ are the roots of the characteristic equations
\begin{equation}\label{chareqns}
   \lambda^2+\mathfrak{f}_0\lambda+\mathfrak{g}_0=0,\quad
   \mu_j=-\frac{\mathfrak{f}_1\lambda_j+\mathfrak{g}_1}{\mathfrak{f}_0+2\lambda_j}.
\end{equation}
A straightforward computation shows that $\lambda_\pm = \pm 2ix_h\Omega$ and $\mu_\pm = \pm 2i\Omega$. As a result, the QNM boundary condition at spatial infinity can be expressed as
\begin{equation}\label{QNMBCzinf}
    \psi\underset{{z\to +\infty}}{\longrightarrow} z^{2i\Omega}e^{2i x_h\Omega z}.
\end{equation}
It is gratifying to observe that it reproduces correctly the corresponding condition for the classical Schwarzschild case.
\end{enumerate}
Having established the asymptotic behaviour of the radial solution both near the event horizon and at spatial infinity, we now proceed to reformulate the problem in a manner that is particularly well-suited for spectral analysis. The primary objective of this transformation is to factor out the singular and oscillatory behaviour associated with the QNM boundary conditions, thereby reducing the problem to the study of a new radial function that is manifestly regular across the entire physical domain. More precisely, we seek to introduce a new radial function $\Phi(z)$ such that the original field $\psi(z)$ can be written as a product of known asymptotic factors and a residual function $\Phi(z)$ which is smooth and finite at both boundaries, namely at the event horizon $z=1$ and at spatial infinity $z \to +\infty$. This strategy is standard in the spectral treatment of QNM problems, as it allows one to isolate the essential singular behaviour and ensures that the remaining function is amenable to polynomial approximation. Therefore, we introduce the transformation
\begin{equation}\label{ansatz}
\psi(z)=z^{4i\Omega}(z-1)^{-2i \Omega}e^{2i x_h\Omega(z-1)}\Phi(z).
\end{equation}
Substituting the ansatz \eqref{ansatz} into the original radial equation \eqref{ODEZ} yields a differential equation of the form
\begin{equation} \label{transformedEq1}
P_2(z)\Phi^{''}(z)+P_1(z)\Phi^{'}(z)+P_0(z)\Phi(z)=0,
\end{equation}
where the coefficient functions are given by
\begin{eqnarray}
P_2(z)&=&z^2(z-1)^2 f^2(z),\\
P_1(z)&=&z^2(z-1)^2 f(z)f^{'}(z)+4i\Omega z(z-1)f^2(z)\left[x_h z^2-(x_h-1)z-2\right],\\
P_0(z)&=&\Omega^2 Q_{2}(z)+ i\Omega Q_1(z)-z^2(z-1)^2 V_\epsilon(z)
\end{eqnarray}
with
\begin{eqnarray}
Q_2 (z)&=&-4\left\{f^2(z)\left[x_h z^2-(x_h-1)z-2\right]^2-x_h^2 z^2(z-1)^2\right\},\\
Q_1(z)&=&2z(z-1)\left[x_h z^2-(x_h-1)z-2\right]f(z)f^{'}(z)-2(z^2-4z+2)f^2(z).
\end{eqnarray}
In order to implement a spectral method, it is convenient to map the semi-infinite domain $z \in (1,+\infty)$ onto a finite interval. To this end, we introduce the transformation
\begin{equation}\label{compactification}
z=\frac{2}{1-y},
\end{equation}
which maps the event horizon $z=1$ to $y=-1$ and spatial infinity to $y=1$. Furthermore, a dot denotes differentiation with respect to the new variable $y$. Then, equation \eqref{transformedEq1} becomes
\begin{equation}\label{ODEynone}
S_2(y)\ddot{\Phi}(y)+S_1(y)\dot{\Phi}(y)+S_0(y)\Phi(y)=0,
\end{equation}
with
\begin{eqnarray}
S_2(y)&=&(1+y)^2 f^2(y),\label{S2onone}\\
S_1(y)&=&-\frac{2(1+y)^2}{1-y}f^2(y)+(1+y)^2 f(y)\dot{f}(y)+8i\Omega(1+y)f^2(y)\left[\frac{2x_h}{(1-y)^2}-\frac{x_h-1}{1-y}-1\right],\label{S1onone}\\
S_0(y)&=&\Omega^2\Sigma_2(y)+i\Omega\Sigma_1(y)+\Sigma_0(y),\label{S0onone}
\end{eqnarray}
where
\begin{eqnarray}
\Sigma_2(y)&=&\frac{16 x_h^2(1+y)^2}{(1-y)^4}-16f^2(y)\left[\frac{2x_h}{(1-y)^2}-\frac{x_h-1}{1-y}-1\right]^2,\\
\Sigma_1(y)&=&4(1+y)\left[\frac{2x_h}{(1-y)^2}-\frac{x_h-1}{1-y}-1\right]f(y)\dot{f}(y)-\frac{4(y^2+2y-1)}{(1-y)^2}f^2(y),\\
\Sigma_0(y)&=&-\frac{4(1+y)^2}{(1-y)^4}V_\epsilon(y).
\end{eqnarray}
As a result of the transformation introduced above, we have 
\begin{eqnarray}
f(y)&=&g(y)(1+y),\quad
g(y)=\frac{x_h(3-y)}{2\left[\sqrt{(1-x_h^2)(1-y)^2+4x_h^2}+(1-y)\right]},\label{fv}\\ 
V_\epsilon(y)&=&\frac{(1-y)^2(1+y)}{4}g(y)\left\{\epsilon\left[(1-y^2)\dot{g}(y)+(1-y)g(y)\right]+\ell(\ell+1)\right\}.
\end{eqnarray}
\begin{table}
\caption{Classification of the points $y=\pm 1$ for the relevant functions defined by \eqref{S2onone}-\eqref{S0onone}, and \eqref{fv}. The abbreviations $z$ ord $n$ and $p$ ord $m$ stand for zero of order $n$ and pole of order $m$, respectively.}
\begin{center}
\begin{tabular}{ | c | c | c | c | c | c | c | c }
\hline
$y$  & $f(y)$  & $V_\epsilon(y)$ & $S_2(y)$ & $S_{1}(y)$ & $S_{0}(y)$\\ \hline
$-1$ & z \mbox{ord} 1 & z \mbox{ord} 1 & z \mbox{ord} 4& z \mbox{ord} 3 & z \mbox{ord} 3 \\ \hline
$+1$ & $+1$  & z \mbox{ord} 2 & $+4$ & p \mbox{ord} 2 & p \mbox{ord} 2\\ \hline
\end{tabular}
\label{tableEinsnone}
\end{center}
\end{table}
Table~\ref{tableEinsnone} reveals that the coefficients \eqref{S2onone}-\eqref{S0onone} share a common zero of order three at $y=-1$, while $y=1$ corresponds to a pole of order two for $S_1(y)$ and $S_0(y)$. In order to regularize the equation \eqref{ODEynone} and ensure compatibility with the spectral method, we multiply \eqref{ODEynone} by the factor $(1-y)^2/(1+y)^3$. This leads to a differential equation of the form
\begin{equation}\label{ODEM}
M_2(y)\ddot\Phi(y)+M_1(y)\dot\Phi(y)+M_0(y)\Phi(y)=0,
\end{equation}
where the coefficients are given by
\begin{eqnarray}
M_2(y)&=&(1+y)(1-y)^2 g^2(y),\label{M2}\\
M_1(y)&=&i\Omega N_1(y)+N_0(y),\quad
M_0(y)=\Omega^2 C_2(y)+i\Omega C_1(y)+C_0(y).
\end{eqnarray}
The functions $N_1(y)$, $N_0(y)$, $C_2(y)$, $C_1(y)$, and $C_0(y)$ are defined explicitly by
\begin{eqnarray}
N_1(y)&=&-8g^2(y)\left[y^2-(x_h+1)y-x_h\right],\quad
N_0(y)=-2g^2(y)(1-y^2)+g(y)(1-y)^2\left[(1+y)\dot{g}(y)+g(y)\right],\\
C_2(y)&=&\frac{16}{(1+y)(1-y)^2}\left\{x_h^2-g^2(y)\left[2x_h-(x_h-1)(1-y)-(1-y)^2\right]^2\right\},\\
C_1(y)&=&\frac{4}{1+y}\left\{g(y)\left[2x_h-(x_h-1)(1-y)-(1-y)^2\right]\left[(1+y)\dot{g}(y)+g(y)\right]-g^2(y)(y^2+2y-1)\right\},\\
C_0(y)&=&-g(y)\left\{\epsilon\left[(1-y^2)\dot{g}(y)+(1-y)g(y)\right]+\ell(\ell+1)\right\}.\label{C0}
\end{eqnarray}
It can be easily verified with Maple that
\begin{eqnarray}
&&\lim_{y\to 1^{-}}M_2(y)=0=\lim_{y\to -1^{+}}M_2(y),\label{lim1}\\
&&\lim_{y\to 1^{-}}M_1(y)=4ix_h\Omega,\quad
\lim_{y\to -1^{+}}M_1(y)=i\Omega\Lambda_1+\Lambda_0,\label{lim2}\\
&&\lim_{y\to 1^{-}}M_0(y)=A_2\Omega^2 +A_0,\quad
\lim_{y\to -1^{+}}M_0(y)=B_2\Omega^2+i\Omega B_1+B_0,\label{lim3}
\end{eqnarray}
where
\begin{eqnarray}
\Lambda_0&=&x_h^2,\quad \Lambda_1=-4\Lambda_0,\quad
A_0=-\frac{\ell(\ell+1)}{2},\quad
A_2=2(1+x_h)^2,\label{Acoefnone}\\
B_2&=&2x_h^2(x_h^2+2x_h+5),\quad
B_1=\frac{x_h^2}{2}(x_h^2+2x_h+5),\quad
B_0=-\frac{x_h}{2}\left[\ell(\ell+1)+\epsilon x_h\right].\label{Dcoefnone}
\end{eqnarray}
In the final step leading to the application of the spectral method, we recast the differential equation \eqref{ODEM} into the following form
\begin{equation}\label{TSCH}
L_0\left[\Phi, \dot{\Phi}, \ddot{\Phi}\right]+  iL_1\left[\Phi, \dot{\Phi}, \ddot{\Phi}\right]\Omega+  
L_2\left[\Phi, \dot{\Phi}, \ddot{\Phi}\right]\Omega^2 = 0
\end{equation}
with
\begin{eqnarray}
L_0\left[\Phi, \dot{\Phi}, \ddot{\Phi}\right]&=&L_{00}(y)\Phi+L_{01}(y)\dot{\Phi}+L_{02}(y)\ddot{\Phi},\label{L0none}\\
L_1\left[\Phi, \dot{\Phi}, \ddot{\Phi}\right]&=&L_{10}(y)\Phi+L_{11}(y)\dot{\Phi}+L_{12}(y)\ddot{\Phi}, \label{L1none}\\
L_2\left[\Phi, \dot{\Phi}, \ddot{\Phi}\right]&=&L_{20}(y)\Phi+L_{21}(y)\dot{\Phi}+ L_{22}(y)\ddot{\Phi}.\label{L2none}
\end{eqnarray}
Moreover, in Table~\ref{tableZweinone}, we have summarized the $L_{ij}$ appearing in (\ref{L0none})--(\ref{L2none}) and their limiting values at $y = \pm 1$.

\begin{table}
\caption{Definitions of the coefficients $L_{ij}$ and their corresponding behaviours at the endpoints of the interval $-1 \leq y \leq 1$. The symbols appearing in this table have been defined in \eqref{M2}-\eqref{C0} and \eqref{Acoefnone}-\eqref{Dcoefnone}.}
\begin{center}
\begin{tabular}{ | c | c | c | c | c | c | c | c }
\hline
$(i,j)$  & $\displaystyle{\lim_{y\to -1^+}}L_{ij}$  & $L_{ij}$ & $\displaystyle{\lim_{y\to 1^-}}L_{ij}$  \\ \hline
$(0,0)$ &  $B_0$          & $C_0$                  & $A_0$\\ \hline
$(0,1)$ &  $\Lambda_0$    & $N_0$                  & $0$\\ \hline
$(0,2)$ &  $0$            & $M_2$                  & $0$\\ \hline 
$(1,0)$ &  $B_1$          & $C_1$                  & $0$\\ \hline 
$(1,1)$ &  $\Lambda_1$    & $N_1$                  & $4x_h$\\ \hline 
$(1,2)$ &  $0$            & $0$                    & $0$\\ \hline 
$(2,0)$ &  $B_2$          & $C_2$                  & $A_2$\\ \hline
$(2,1)$ &  $0$            & $0$                    & $0$\\ \hline
$(2,2)$ &  $0$            & $0$                    & $0$\\ \hline
\end{tabular}
\label{tableZweinone}
\end{center}
\end{table} 

\subsection{Non-minimally coupled scalar field perturbations}

\begin{figure}[t]
\centering
\includegraphics[width=0.95\linewidth]{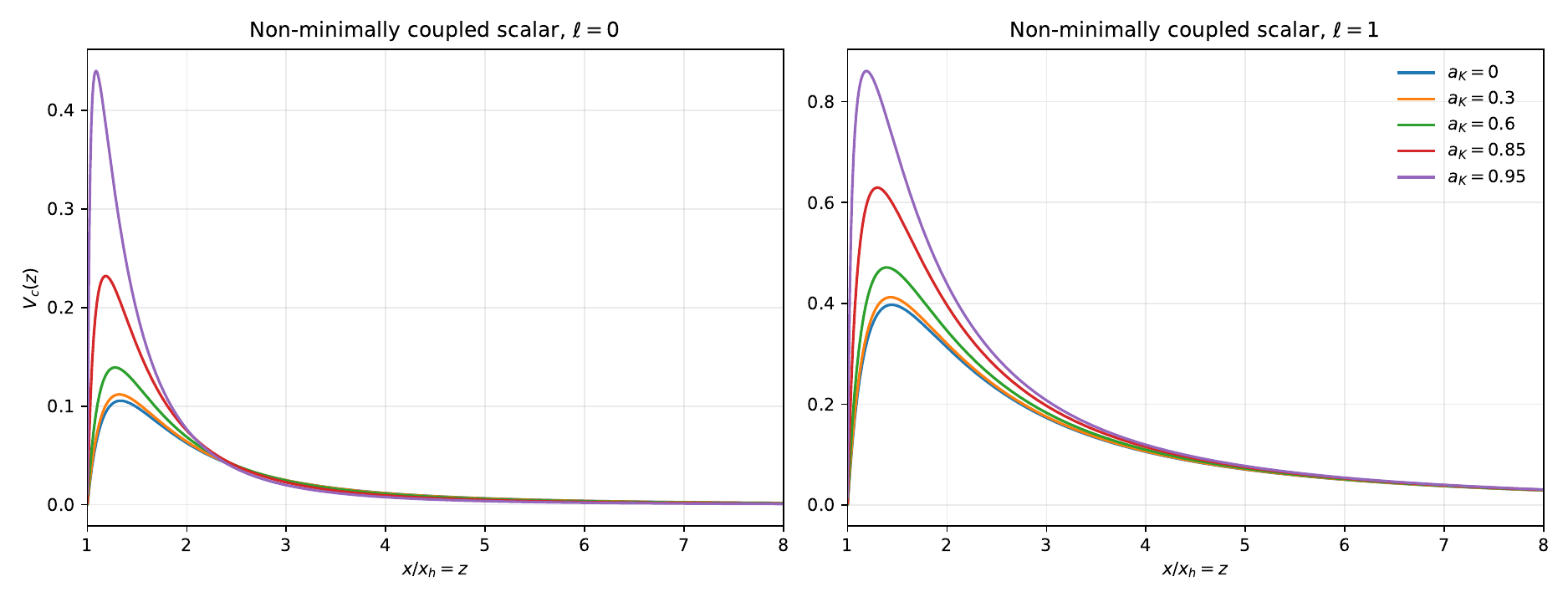}
\caption{
Effective potential $V_c(z)$ for the non-minimally coupled scalar field, plotted in the horizon-normalised coordinate $z=x/x_h$, with $x_h=(1-a_K^2)^{-1/2}$ and $a_K=a/r_h$. The left and right panels correspond to $\ell=0$ and $\ell=1$, respectively. As in the minimally coupled scalar and electromagnetic sectors, increasing $a_K$ raises the barrier and shifts its maximum towards the horizon $z=1$. The extremal value $a_K=1$ is not shown because $x_h$ diverges in this parametrisation.}
\label{fig:conformal_scalar_potential}
\end{figure}

For the non-minimally coupled scalar sector, we employ the same dimensionless radial variable introduced above, namely $r=2Mx_h z$, so that the event horizon is again located at $z=1$. With this rescaling, the radial equation retains the same structure as \eqref{ourODE}. The only modification is that the scalar/electromagnetic potential $V_\epsilon(z)$ must be replaced by the dimensionless Regge-Wheeler potential
\begin{equation}\label{Voddz}
  V_{c}(z)=f(z)\left[\frac{3\ell(\ell+1)+1}{3z^2}+\frac{f^{'}(z)}{z}-\frac{4x_h\left[2x_h^2 z^2+3(1-x_h^2)\right]}{3z\left[4x_h^2 z^2+4(1-x_h^2)\right]^{3/2}}\right],
\qquad \ell\geq 0.
\end{equation}
Before adapting the problem to the Spectral Method, it is useful to display the deformation dependence of the effective potential \eqref{Voddz}. Fig.~\ref{fig:conformal_scalar_potential} shows the effective potential $V_c(z)$ for the non-minimally coupled scalar field. As $a_k=a/r_h$ increases for fixed $\ell$, the barrier becomes higher and moves closer to the horizon in the coordinate $z=x/x_h$. For example, the scalar $\ell=0$ peak moves from about $z\sim 1.333$ at $a_K=0$ to $z\sim 1.086$ at $a_
K=0.95$, while the peak height increases from about $0.1055$ to $0.4399$.

The Frobenius analysis at the two endpoints proceeds exactly as in the scalar and electromagnetic cases. Indeed, since $V_{c}(z)$ is proportional to $f(z)$, it vanishes at both the event horizon and at spatial infinity. Therefore, it does not modify the leading indicial behaviour of the radial equation. Hence, the QNM boundary condition at the event horizon $z=1$ is still given by \eqref{QNMBCz1}, while the corresponding outgoing condition at spatial infinity, namely \eqref{QNMBCzinf}, is unchanged as well. We may therefore use the same ansatz as in \eqref{ansatz}, which factors out the horizon asymptotics and leaves a radial function regular at both endpoints. Substitution into the corresponding radial equation yields an equation of the same form as \eqref{transformedEq1}. The only change is the replacement of $V_\epsilon(z)$ by $V_{c}(z)$ in the coefficient $P_0(z)$. The compactification to the Chebyshev interval can also be performed using \eqref{compactification}. Consequently, one obtains the same transformed equation as \eqref{ODEynone}, with the replacement $V_\epsilon(y)\to V_{c}(y)$ in $S_0(y)$. More precisely, we have
\begin{equation}
  V_{c}(y)=(1-y)^2 f(y)\left[\frac{3\ell(\ell+1)+1}{12}+\frac{1}{4}(1-y)\dot{f}(y)-\frac{2x_h\left[8x_h^2+3(1-x_h^2)(1-y)^2\right]}{3\left[16x_h^2+4(1-x_h^2)(1-y)^2\right]^{3/2}}\right],\qquad \ell\geq 0.
\end{equation}
The classification of the endpoints is unchanged with respect to the scalar and electromagnetic cases, and is therefore still summarised by Table~\ref{tableEinsnone}. In particular, the coefficient functions have the same vanishing behaviour at $y = \pm 1$. Hence, as before, we can divide the ODE in the compact coordinate $y$ by the common factor $(1-y)^2/(1+y)^3$. This yields an equation formally identical to \eqref{ODEM}. The only modification occurs in the coefficient $C_0(y)$, which now reads
\begin{equation}
C_0(y)=-g(y)\left\{
\frac{3\ell(\ell+1)+1}{3}+(1-y)\left[(1+y)\dot{g}(y)+g(y)\right]-
\frac{8x_h\left[8x_h^2+3(1-x_h^2)(1-y)^2\right]}{3\left[16x_h^2+4(1-x_h^2)(1-y)^2\right]^{3/2}}
\right\}.
\end{equation}
The only endpoint limit in \eqref{lim1}--\eqref{lim3} that needs to be changed is
\begin{equation}
\lim_{y\to -1^{+}}M_0(y)=B_2\Omega^2+i\Omega B_1+\widetilde{B}_0,
\end{equation}
where $B_2$ and $B_1$ are given in \eqref{Dcoefnone} and
\begin{equation}
\widetilde{B}_0=-\frac{x_h}{2}\left[\frac{x_h}{6}(x_h^2+3)+\ell(\ell+1)+\frac{1}{3}\right].
\end{equation}
Finally, the resulting equation can be written in the same quadratic spectral form as Eq.~\eqref{TSCH}. The corresponding coefficient functions are defined exactly as in \eqref{L0none}-\eqref{L2none}, with the replacement $V_\epsilon(y)\to V_{c}(y)$, and their endpoint behaviour is summarised in Table~\ref{tableZweinone}.

\subsection{Axial gravitational perturbations as an effective source model}
\label{subsec:axial_grav_potential}

\begin{figure}[t]
\centering
\includegraphics[width=0.95\linewidth]{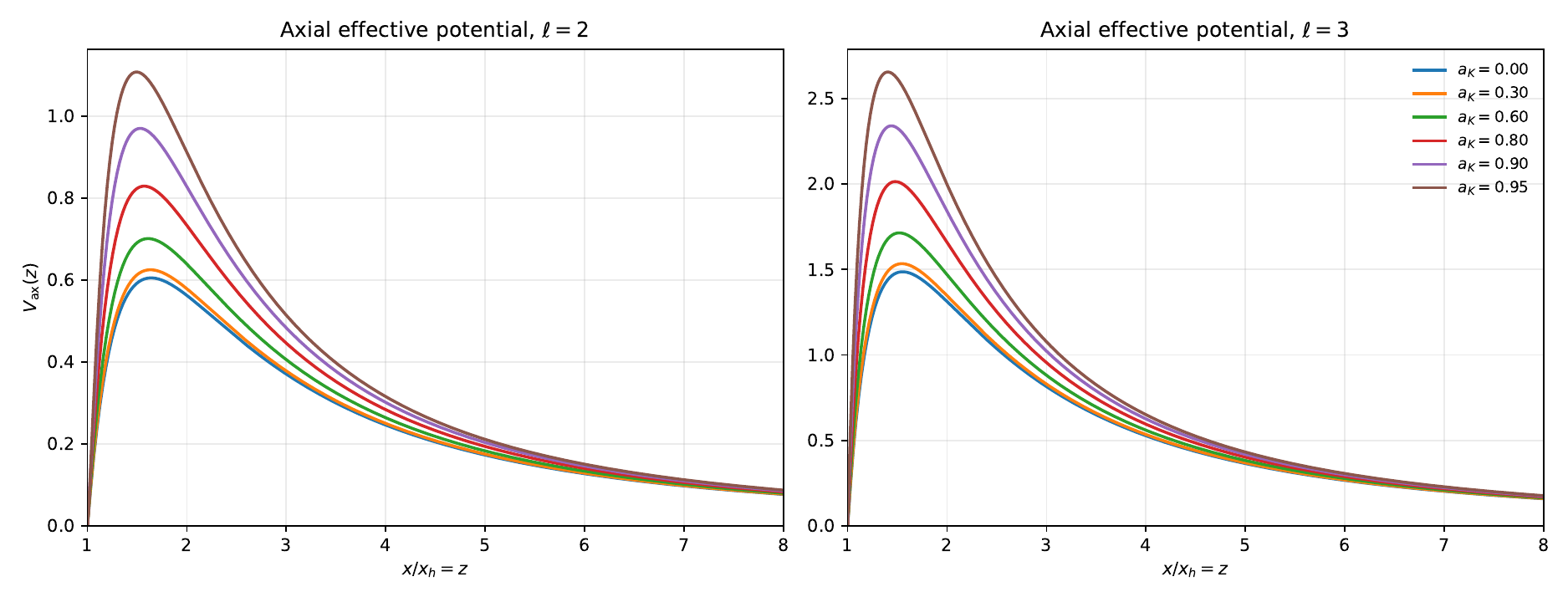}
\caption{
Axial effective-source potential $V_{\rm ax}(z)$, plotted in the horizon-normalised radial coordinate $z=x/x_h$, with $x_h=(1-a_K^2)^{-1/2}$ and $a_K=a/r_h$. The left and right panels show the first two allowed axial multipoles, $\ell=2$ and $\ell=3$, respectively. Increasing $a_K$ raises the potential barrier and shifts its maximum towards the horizon $z=1$, in agreement with the qualitative behaviour found in the scalar, electromagnetic, and non-minimally coupled scalar sectors. The extremal value $a_K=1$ is not plotted because $x_h$ diverges in this parametrisation.}
\label{fig:axial_potential}
\end{figure}

The gravitational sector requires more care than the scalar and electromagnetic test field sectors. The KS spacetime is not Ricci flat. Therefore, the vacuum perturbation equation $\delta R_{\mu\nu}=0$ cannot be imposed on this background without additional assumptions. This point was emphasised in \cite{KonoplyaPLB2020}, where it was noted that the earlier gravitational treatment of \cite{SalehASS2016} is not consistent as a vacuum perturbation problem. In the present work, we therefore do not interpret the gravitational sector as a perturbation of a Ricci-flat vacuum spacetime. Instead, we formulate an axial Regge--Wheeler-type effective source model. Let the KS line element be given by \eqref{LE}. Since the spacetime is not Ricci flat, we formally rewrite the background Einstein equations as
\begin{equation}\label{eq:eff_source_background}
  G_{\mu\nu}=8\pi T^{\mathrm{eff}}_{\mu\nu},
\end{equation}
where $T^{\mathrm{eff}}_{\mu\nu}$ denotes the effective stress-energy tensor supporting the KS geometry. The purpose of this construction is not to claim a unique microscopic matter model, but rather to keep track of the non-vacuum character of the background when deriving the axial perturbation equation. It is useful to introduce the Misner--Sharp mass function $m(r)$ by
\begin{equation}\label{eq:mass_function_def}
f(r)=1-\frac{2m(r)}{r}.
\end{equation}
For the KS lapse \eqref{LE}, one obtains
\begin{equation}\label{eq:mass_function_KS}
m(r)=M+\frac{1}{2}\left(r-\sqrt{r^2-a^2}\right),
\end{equation}
and therefore
\begin{equation}\label{eq:mass_function_derivative_KS}
m'(r)=\frac{1}{2}\left(1-\frac{r}{\sqrt{r^2-a^2}}\right).
\end{equation}
Writing the effective source in the form
\begin{equation}\label{eq:anisotropic_source}
T^\mu{}_\nu=\mbox{diag}\left(-\rho,p_r,p_t,p_t\right),
\end{equation}
the $tt$- and $rr$-components of the background Einstein equations give
\begin{equation}\label{eq:rho_pr}
8\pi\rho=\frac{2m'(r)}{r^2},\qquad
8\pi p_r=-\frac{2m'(r)}{r^2}.
\end{equation}
Thus
\begin{equation}\label{eq:rho_minus_pr}
p_r=-\rho,\qquad
4\pi(\rho-p_r)=\frac{2m'(r)}{r^2}.
\end{equation}
We now consider axial metric perturbations. In the Regge--Wheeler gauge, the odd-parity perturbation can be written as \cite{Regge1957PR}
\begin{equation}\label{eq:axial_perturbation}
h_{\mu\nu}^{ax}dx^\mu dx^\nu=2\sum_{\ell,m}e^{-i\omega t}\left[h_0^{\ell m}(r)dt+
h_1^{\ell m}(r)dr\right]S_A^{\ell m}dx^A,
\end{equation}
where $A\in\{\theta,\phi\}$, and
\begin{equation}
S_A^{\ell m}=\epsilon_A{}^{B}D_B Y_{\ell m}
\end{equation}
are the axial vector spherical harmonics on the unit two-sphere. The admissible gravitational multipoles satisfy $\ell\geq 2$. The perturbation equations must be obtained from
\begin{equation}\label{eq:perturbed_effective_Einstein}
    \delta G_{\mu\nu}=8\pi\delta T^{\mathrm{eff}}_{\mu\nu},
\end{equation}
not from $\delta R_{\mu\nu}=0$. At this point, a physical closure assumption is required. The background relation $p_r=-\rho$ follows from the static, spherically symmetric KS geometry, but it does not by itself exclude odd-parity perturbations of the effective source. In particular, a generic anisotropic source could carry axial, or toroidal, perturbations of the form
\begin{equation}
\delta T^{eff}_{tA}=\tau_0(r)e^{-i\omega t}S_A^{\ell m},\qquad
\delta T^{eff}_{rA}=\tau_1(r)e^{-i\omega t}S_A^{\ell m},\qquad
\delta T^{eff}_{AB}=\tau_2(r)e^{-i\omega t}S_{AB}^{\ell m},
\end{equation}
where $S_A^{\ell m}$ and $S_{AB}^{\ell m}$ are axial vector and tensor harmonics. The functions $\tau_0$, $\tau_1$, and $\tau_2$ would describe independent transverse momentum and anisotropic-stress perturbations of the effective source. Since the KS metric is not derived here from a specified four-dimensional matter action, there is no unique evolution equation, equation of state, or constitutive relation fixing these functions. In the present work, we therefore define a restricted axial effective-source model by setting the independent axial source perturbation to zero,
\begin{equation}
\delta T^{ind}_{\rm ax}=0.
\end{equation}
This does not mean that the background source is ignored: the non-vacuum character of the geometry still enters the axial potential through the background quantities \(\rho\) and \(p_r\). The assumption only removes additional propagating odd-parity source degrees of freedom, thereby closing the metric perturbation problem. Under this closure, the axial perturbation is a Regge--Wheeler-type metric perturbation on a fixed effective source, and the axial master variable \(\Psi_{ax}\) satisfies
\begin{equation}\label{eq:axial_master_equation}
\frac{d^2\Psi_{ax}}{dr_*^2}+\left[\omega^2-V_{ax}(r)\right]\Psi_{ax}=0,\qquad
\frac{dr_*}{dr}=\frac{1}{f(r)},
\end{equation}
with the effective source Regge--Wheeler potential
\begin{equation}\label{eq:axial_potential_general}
V_{ax}(r)=f(r)\left[\frac{\ell(\ell+1)}{r^2}-\frac{6m(r)}{r^3}+4\pi(\rho-p_r)\right].
\end{equation}
Using \eqref{eq:rho_minus_pr}, this becomes
\begin{equation}\label{eq:axial_potential_mass_function}
V_{ax}(r)=f(r)\left[\frac{\ell(\ell+1)}{r^2}-\frac{6m(r)}{r^3}+\frac{2m'(r)}{r^2}\right].
\end{equation}
Notice that this expression reduces to the standard Regge--Wheeler potential in vacuum, where $m(r)=M$ and $m'(r)=0$. Substituting \eqref{eq:mass_function_KS} and
\eqref{eq:mass_function_derivative_KS} into \eqref{eq:axial_potential_mass_function}, we obtain as in \cite{SalehASS2016}
\begin{equation}\label{eq:axial_potential_KS}
V_{ax}(r)=f(r)\left[\frac{(\ell-1)(\ell+2)}{r^2}-\frac{6M}{r^3}+\frac{3\sqrt{r^2-a^2}}{r^3}
-\frac{1}{r\sqrt{r^2-a^2}}\right].
\end{equation}
It is instructive to rewrite \eqref{eq:axial_potential_KS} directly in terms of $f(r)$. From $m(r)=r[1-f(r)]/2$, we have $m'(r)=\left[1-f(r)-r f'(r)\right]/2$. Substitution into \eqref{eq:axial_potential_mass_function} gives the equivalent compact expression
\begin{equation}\label{eq:axial_potential_compact}
V_{ax}(r)=f(r)\frac{\ell(\ell+1)-2+2f(r)-r f'(r)}{r^2}.
\end{equation}
If the independent axial source perturbation were retained, the right-hand side of the master equation would no longer vanish. Schematically, one would obtain
\begin{equation}
\left[\frac{d^2}{dr_*^2}+\omega^2-V_{ax}(r)\right]\psi_{ax}=
{\cal S}_{ax}[\tau_0,\tau_1,\tau_2],
\end{equation}
together with additional equations specifying the dynamics of the source amplitudes $\tau_0$, $\tau_1$, and $\tau_2$. If these amplitudes were prescribed externally, the problem would describe a forced response rather than a free QNM spectrum. If they were dynamical and coupled to the metric, the QNM frequencies would be the zeros of the coupled metric-source system and would, in general, be shifted relative to the frequencies reported below. Additional source-led axial modes could also appear. Therefore, the axial frequencies quoted in this work are not universal gravitational frequencies of the KS geometry. They are the frequencies of the closed metric-led axial effective-source model defined by $\delta T^{\rm ind}_{ax}=0$. This form is particularly useful for the spectral implementation. Moreover, it is straightforward to check that \eqref{eq:axial_potential_compact} correctly reproduces the usual Schwarzschild Regge--Wheeler potential for $a=0$. We emphasize that \eqref{eq:axial_potential_KS} coincides algebraically with Eq.~(21) of \cite{SalehASS2016}. The interpretation, however, is different. In \cite{SalehASS2016}, the potential was obtained after imposing the vacuum equation $\delta R_{\mu\nu}=0$. Here we instead obtain the same algebraic expression from the source-corrected axial equation $\delta G_{\mu\nu}=8\pi\delta T^{\mathrm{eff}}_{\mu\nu}$, under the assumption that the effective source carries no independent axial perturbation. In this sense, the spectrum computed from \eqref{eq:axial_potential_KS} should be interpreted as the axial effective source Regge--Wheeler spectrum of the KS geometry, not as the full gravitational spectrum of a specified four-dimensional quantum gravity action. For the axial gravitational sector, we use the same dimensionless radial variable introduced above, namely $r=2Mx_h z$, so that the event horizon is again located at $z=1$. With this rescaling, the radial equation retains the same structure as \eqref{ourODE}. The only modification is that the scalar/electromagnetic potential $V_\epsilon(z)$ must be replaced by the dimensionless potential
\begin{equation}\label{Vaxz}
V_{ax}(z)=f(z)\frac{\ell(\ell+1)-2+2f(z)-zf^{'}(z)}{z^2},
\qquad \ell\geq 2.
\end{equation}
Before adapting the problem to the Spectral Method, it is useful to display the deformation dependence of the effective potential \eqref{Vaxz}. Fig.~\ref{fig:axial_potential} shows the effective potential $V_{ax}(z)$ for the non-minimally coupled scalar field. As $a_k=a/r_h$ increases for fixed $\ell$, the barrier becomes higher and moves closer to the horizon in the coordinate $z=x/x_h$. For example, for $\ell=2$, the peak moves from approximately $z_{max}\sim 1.640$ with $V_{max}\sim 0.605$ at $a_K=0$, to $z_{max}\sim 1.495$ with $V_{max}\sim 1.108$ at $a_K=0.95$. For $\ell=3$, the peak moves from approximately $z_{max}\sim 1.553$ to $z_{max}\sim 1.405$, while the height increases from $V_{max}\sim 1.487$ to $V_{max}\sim 2.655$.

The Frobenius analysis at the two endpoints proceeds exactly as in the scalar and electromagnetic cases. Indeed, since $V_{ax}(z)$ is proportional to $f(z)$, it vanishes at both the event horizon and at spatial infinity. Therefore, it does not modify the leading indicial behaviour of the radial equation. Hence, the QNM boundary condition at the event horizon $z=1$ is still given by \eqref{QNMBCz1}, while the corresponding outgoing condition at spatial infinity, namely \eqref{QNMBCzinf}, is unchanged as well. Therefore, we may use the same ansatz as in \eqref{ansatz}, which factors out the horizon asymptotics and leaves a radial function regular at both endpoints. Substitution into the corresponding radial equation yields an equation of the same form as \eqref{transformedEq1}. The only change is the replacement of $V_\epsilon(z)$ by $V_{ax}(z)$ in the coefficient $P_0(z)$. The compactification to the Chebyshev interval can also be achieved using \eqref{compactification}. Consequently, one obtains the same transformed equation as \eqref{ODEynone}, with the replacement $V_\epsilon(y)\to V_{ax}(y)$ in $S_0(y)$. More precisely, we have
\begin{equation}
  V_{ax}(y)=\frac{1}{4}(1-y)^2 f(y)\left[\ell(\ell+1)-2+2f(y)-(1-y)\dot{f}(y)\right],\qquad \ell\geq 2.
\end{equation}
The classification of the endpoints is unchanged with respect to the scalar and electromagnetic cases, and is therefore still summarised by Table~\ref{tableEinsnone}. In particular, the coefficient functions have the same vanishing behaviour at $y=\pm 1$. Hence, as before, we can divide the ODE in the compact coordinate $y$ by the common factor $(1-y)^2/(1+y)^3$. This yields an equation formally identical to \eqref{ODEM}. The only modification occurs in the coefficient $C_0(y)$, which now reads
\begin{equation}
C_0(y)=-g(y)\left[\ell(\ell+1)-2+(3y+1)g(y)-(1-y^2)\dot{g}(y)\right].
\end{equation}
The only endpoint limit in \eqref{lim1}--\eqref{lim3} that needs to be changed is
\begin{equation}
\lim_{y\to -1^{+}}M_0(y)=B_2\Omega^2+i\Omega B_1+\widehat{B}_0,
\end{equation}
where $B_2$ and $B_1$ are given in \eqref{Dcoefnone} and
\begin{equation}
\widehat{B}_0=-\frac{x_h}{2}\left[\ell(\ell+1)-x_h-2\right].
\end{equation}
Finally, the resulting equation can be written in the same quadratic spectral form as Eq.~\eqref{TSCH}. The corresponding coefficient functions are defined exactly as in \eqref{L0none}-\eqref{L2none}, with the replacement $V_\epsilon(y)\to V_{ax}(y)$, and their endpoint behaviour is summarised in Table~\ref{tableZweinone}.

\section{The $a_K\to 1^{-}$ limiting geometry}\label{subsec:extremal_KS_limit}

We now discuss the limiting geometry obtained when the horizon-normalised deformation parameter in \cite{KonoplyaPLB2020} approaches its maximal value. As before, let $a$ denote the dimensional Kazakov--Solodukhin deformation parameter. The KS metric is given by \eqref{LE} and the horizon is located according to 
\begin{equation}\label{eq:KS_horizon_extremal_section}
    r_h=\sqrt{4M^2+a^2},
\end{equation}
while the curvature singularity is situated at $r=a$ \cite{KonoplyaPLB2020,bolokhov2025overtones}. For a genuine black hole configuration, one has $a<r_h$. It is useful to introduce the dimensionless parameter
\begin{equation}    \label{eq:p_definition_extremal}
p:=a_K=\frac{a}{r_h},\qquad 0\leq p<1,
\end{equation}
which is the deformation parameter used in the tables of \cite{KonoplyaPLB2020}, where the horizon radius is fixed to $r_h=1$. Solving \eqref{eq:KS_horizon_extremal_section} for the mass parameter gives
\begin{equation}\label{eq:M_in_terms_of_p}
    2M=r_h\sqrt{1-p^2}.
\end{equation}
Thus, if $r_h$ is kept fixed and $p\to1^{-}$, then $M\to 0$ as noted by \cite{KonoplyaPLB2020}. Let $\rho=r/r_h$. Substituting $a=pr_h$ and \eqref{eq:M_in_terms_of_p} into the second equation in \eqref{LE}, the lapse function becomes
\begin{equation}\label{eq:fp_R}
f_p(\rho)=\frac{\sqrt{\rho^2-p^2}-\sqrt{1-p^2}}{\rho}.
\end{equation}
For every $0\leq p<1$, the event horizon is at $\rho=1$, since $f_p(1)=0$, whereas the curvature singularity is at $R=p<1$, strictly inside the horizon. Therefore, for $p<1$, the geometry describes a black hole. The formal limiting geometry is obtained by taking $p\to 1^{-}$ at fixed $r_h$. For every $\rho>1$, the limit of \eqref{eq:fp_R} exists and gives
\begin{equation}\label{eq:fext_R}
f_{e}(\rho)=\lim_{p\to 1^-} f_p(\rho)=\frac{\sqrt{\rho^2-1}}{\rho}.
\end{equation}
Equivalently, in terms of the areal radius $r$, with $r_0:=r_h$, one obtains
\begin{equation}\label{eq:fext_r}
f_{e}(r)=\sqrt{1-\frac{r_0^2}{r^2}},\qquad r>r_0.
\end{equation}
The corresponding limiting line element is therefore
\begin{equation}\label{eq:extremal_KS_metric}
ds_{e}^2=-f_e(r)dt^2+\frac{dr^2}{f_e(r)}+r^2\left(d\vartheta^2 + \sin^2{\vartheta} d\varphi^2\right),\qquad r>r_0.
\end{equation}
We stress that the subscript ``e'' refers only to the limiting procedure $p\to 1^{-}$. The metric \eqref{eq:extremal_KS_metric} is not an extremal black hole in the Reissner--Nordstr\"om sense. Indeed, for $p<1$, the derivative of the lapse at the horizon is
\begin{equation}
\left.\frac{df_p}{dr}\right|_{r=r_h}=\frac{1}{r_h\sqrt{1-p^2}},
\end{equation}
so the surface gravity is
\begin{equation}\label{eq:surface_gravity_near_extremal_KS}
\kappa_p=\frac{1}{2r_h\sqrt{1-p^2}}=\frac{1}{4M}.
\end{equation}
Thus, as $p\to 1^{-}$ at fixed $r_h$, the surface gravity diverges rather than vanishing. This behaviour is opposite to what happens in ordinary extremal limits characterised by a regular degenerate horizon. The limiting surface $r=r_0$ is singular. This follows directly from the Ricci scalar of the KS geometry given by the second equation in \eqref{Vg}, which diverges as $r\to a$ \cite{KonoplyaPLB2020, bolokhov2025overtones}. In the limiting geometry $a\to r_0$, the singularity coincides with the surface $r=r_0$. Explicitly,
\begin{equation}\label{eq:Ricci_scalar_ext}
R_{e}(r)=\frac{2}{r^2}\left[1-\left(1-\frac{r_0^2}{r^2}\right)^{-1/2}\right]+
\frac{r_0^2}{r^4}\left(1-\frac{r_0^2}{r^2}\right)^{-3/2},
\end{equation}
which diverges for $r\to r_0^{+}$. Therefore, the limiting surface is not a regular horizon. In fact, near $r=r_0$,
\begin{equation}
f_{e}(r)=\sqrt{1-\frac{r_0^2}{r^2}}\sim\sqrt{\frac{2(r-r_0)}{r_0}},\qquad r\to r_0^{+}.
\end{equation}
Consequently, the tortoise coordinate of the limiting metric is finite at $r=r_0$, that is
\begin{equation}\label{eq:tortoise_ext_finite}
r_*^{e}=\int \frac{dr}{f_{e}(r)}=\sqrt{r^2-r_0^2}+const.
\end{equation}
This is qualitatively different from the logarithmic divergence of the tortoise coordinate at a regular nonextremal black-hole horizon. The limit $p\to 1^{-}$ should therefore be interpreted as the coalescence of the event horizon with the curvature singularity, rather than as the formation of a regular extremal horizon. Finally, we verify that the limiting metric is asymptotically flat. From \eqref{eq:fext_r}, as $r\to\infty$,
\begin{equation}\label{eq:fext_asymptotic}
f_{e}(r)=1-\frac{r_0^2}{2r^2}-\frac{r_0^4}{8r^4}+\mathcal{O}(r^{-6}).
\end{equation}
Hence,
\begin{equation}
g_{tt}=-1+\frac{r_0^2}{2r^2}+\mathcal{O}(r^{-4}),\qquad
g_{rr}=1+\frac{r_0^2}{2r^2}+\mathcal{O}(r^{-4}),
\end{equation}
and the angular part is the standard spherical area term. Thus, the metric approaches the Minkowski metric in spherical coordinates, with corrections of order $r^{-2}$. In particular, the $1/r$ term is absent, so the ADM mass of the limiting geometry is zero, consistently with $M\to 0$ in \eqref{eq:M_in_terms_of_p}. The Ricci scalar also vanishes at infinity, since \eqref{eq:Ricci_scalar_ext} gives
\begin{equation}
R_{e}(r)=\frac{3r_0^4}{4r^6}+\mathcal{O}(r^{-8}),\qquad r\to\infty .
\end{equation}
Therefore, \eqref{eq:extremal_KS_metric} is asymptotically flat, but it contains a singular null boundary at $r=r_0$. In this sense, the $a_K\to 1^{-}$ limit is more appropriately described as a singular horizon, or null naked singularity, rather than as a regular extremal KS black hole. This observation is important for the interpretation of the QNM frequencies. For every $p=a_K<1$, the singularity lies strictly inside the event horizon, and the usual black-hole QNM boundary conditions are imposed at a regular, nondegenerate horizon. The limiting procedure $p\to 1^{-}$, however, is not a limit to a regular zero-temperature extremal black hole. Instead, at fixed $r_h$
\begin{equation}
M=\frac{r_h}{2}\sqrt{1-p^2}\to 0,\qquad
\kappa_p=\frac{1}{2r_h\sqrt{1-p^2}}=\frac{1}{4M}\to\infty .
\end{equation}
Therefore, a finite value of the mass-normalised frequency $\Omega=M\omega$ corresponds to
\begin{equation}
r_h\omega=\frac{2\Omega}{\sqrt{1-p^2}}\to\infty ,
\end{equation}
or equivalently to a physical frequency that diverges in proportion to the surface gravity. The finite quantity in this scaling limit is instead $\omega/\kappa_p=4\Omega$. Consequently, a near-extremal ladder $\Omega_n\simeq-i(n+1)/4$ should be read as $\omega_n\simeq -i(n+1)\kappa_p$, namely as a surface-gravity-scaled overdamped structure of the black-hole family $p<1$. It is not a finite-frequency QNM spectrum of the singular limiting metric \eqref{eq:extremal_KS_metric}.

\section{Spectral method \label{sec:method}}

We now describe the numerical implementation used to solve the quadratic eigenvalue problem derived in the previous section. After the QNM asymptotics have been factored out by the ansatz \eqref{ansatz}, the remaining function $\Phi(y)$ is regular on the compact interval $[-1,1]$. Therefore, no additional boundary conditions have to be imposed at the endpoints. The boundary conditions are already encoded in the prefactor of \eqref{ansatz}. The only requirement left is the regularity of $\Phi(y)$ at $y=\pm1$. The transformed equation for the minimally coupled scalar and electromagnetic sectors has the form
\begin{equation}
    L_0\left[\Phi,\dot{\Phi},\ddot{\Phi}\right]
    +i\Omega
    L_1\left[\Phi,\dot{\Phi},\ddot{\Phi}\right]
    +\Omega^2
    L_2\left[\Phi,\dot{\Phi},\ddot{\Phi}\right]
    =0,
\end{equation}
as in \eqref{TSCH}. The operators $L_j$ with $j\in\{0,1,2\}$ are defined in \eqref{L0none}--\eqref{L2none}, and their coefficient functions are summarized in Table~\ref{tableZweinone}. The non-minimally coupled scalar sector and the axial effective-source gravitational sector lead to the same quadratic structure. In those cases, the only change is the replacement of the potential-dependent coefficient $C_0(y)$ by the corresponding expression obtained from $V_c(y)$ or from the axial potential $V_{ax}(y)$. Thus, all sectors considered in this work are reduced to the same type of quadratic differential eigenvalue problem for the frequency $\Omega=M\omega$. Since the problem is posed on the finite interval $[-1,1]$, it is natural to approximate the regular part of the radial function by a Chebyshev expansion \cite{Trefethen2000, Boyd2000}. We denote by $N$ the number of retained Chebyshev polynomials, reserving $n$ for the overtone number, and write
\begin{equation}\label{eq:cheb_expansion_KS}
    \Phi(y)=\sum_{k=0}^{N-1} a_k T_k(y),
\end{equation}
where $T_k(y)=\cos\!\left(k\arccos y\right)$ with $-1\leq y\leq1$ are the Chebyshev polynomials of the first kind. The unknowns of the discrete problem are the coefficients
\begin{equation}
    \mathbf a=(a_0,a_1,\ldots,a_{N-1})^T .
\end{equation}
Substituting \eqref{eq:cheb_expansion_KS} into the quadratic differential equation gives a residual which is required to vanish at $N$ collocation points. In the computations reported in this work, we use the Chebyshev--Gauss roots grid
\begin{equation}\label{eq:cheb_roots_grid_KS}
    y_j=\cos\left(\frac{(2j-1)\pi}{2N}\right),
    \qquad
    j=1,\ldots,N.
\end{equation}
This choice avoids the endpoints $y=\pm1$, where the QNM asymptotics have already been removed analytically. The theoretical convergence properties of Chebyshev roots and extrema grids for smooth spectral problems are comparable \cite{Fox1968, Boyd2000}. Throughout the present work, however, all matrices were assembled using the roots grid
\eqref{eq:cheb_roots_grid_KS}. The collocation procedure leads to a quadratic matrix pencil
\begin{equation}\label{eq:quadratic_pencil_KS}
    \left(M_0+i\Omega M_1+\Omega^2 M_2\right)\mathbf a=0,
\end{equation}
where $M_0$, $M_1$, and $M_2$ are real $N\times N$ matrices. These matrices are obtained by evaluating the three differential operators $L_0$, $L_1$, $L_2$ at the collocation points. The matrix assembly is performed in \textsc{Maple}. For each chosen sector, angular number, and deformation parameter, the Chebyshev expansion \eqref{eq:cheb_expansion_KS}, together with its first and second derivatives, is inserted into the differential equation symbolically. The coefficients of the unknowns $a_k$ are then extracted at each collocation point, producing the entries of $M_0$, $M_1$, and $M_2$. The matrices are exported in ASCII format and then read by the \textsc{Julia} code. Separate \textsc{Maple} assemblers are used for the minimally coupled scalar/electromagnetic sectors, for the non-minimally coupled scalar sector, and for the axial effective-source gravitational sector, but all of them produce the same quadratic pencil \eqref{eq:quadratic_pencil_KS}. In \textsc{Julia}, the quadratic eigenvalue problem \eqref{eq:quadratic_pencil_KS} is linearised as a generalised eigenvalue problem of twice the dimension. Introducing
\begin{equation}
\mathbf u=
    \begin{pmatrix}
        \mathbf a\\
        \Omega\mathbf a
    \end{pmatrix},
\end{equation}
we solve
\begin{equation}\label{eq:linearized_pencil_KS}
    \begin{pmatrix}
        0 & I\\
        -M_0 & -iM_1
    \end{pmatrix}
    \mathbf u
    =
    \Omega
    \begin{pmatrix}
        I & 0\\
        0 & M_2
    \end{pmatrix}
    \mathbf u.
\end{equation}
The $2N$ generalized eigenvalues obtained from \eqref{eq:linearized_pencil_KS} form the raw numerical spectrum. With the convention $e^{-i\omega t}$, damped QNMs have $\Im{\Omega}<0$. Thus, for the QNM tables, we retain only roots satisfying $\Im{\Omega}<-10^{-18}$. We do not remove one of the two $\pm\Re{\Omega}$ branches. Both are kept at the level of computation, and the tables display the representative roots chosen for the discussion. The matrices used for the tables were generated in \textsc{Maple} with $200$ decimal digits. The generalised eigenvalue problem and all post-processing steps were performed in \textsc{Julia} using $220$ decimal digits. The physical frequencies are identified by a convergence test across three different spectral resolutions. More precisely, we use $N=190$, $195$, and $200$. The notation $N=200$ in the captions of the numerical tables refers to the largest resolution in this triplet. A candidate frequency is retained only if it can be matched consistently across the three resolutions, i.e., for each root $\Omega_{190}$ in the spectrum at $N=190$, we find the nearest root $\Omega_{195}$ in the spectrum at $N=195$, and then the nearest root $\Omega_{200}$ in the spectrum at $N=200$. The triplet is accepted only if all three pairwise distances satisfy
\begin{equation}\label{eq:stability_criterion_KS}
    \max\left\{
    |\Omega_{190}-\Omega_{195}|,
    |\Omega_{195}-\Omega_{200}|,
    |\Omega_{190}-\Omega_{200}|
    \right\}
    \leq
    \max\left[
    \varepsilon_{abs},
    \varepsilon_{rel}
    \max\{1,|\Omega_{190}|,|\Omega_{195}|,|\Omega_{200}|\}
    \right],
\end{equation}
with $\varepsilon_{\rm abs}=\varepsilon_{\rm rel}=10^{-4}$. For every accepted triplet, the value reported in the tables is the average
\begin{equation}\label{eq:Omega_average_KS}
    \Omega_{av}=\frac{\Omega_{190}+\Omega_{195}+\Omega_{200}}{3}.
\end{equation}
The pairwise differences $|\Omega_{190}-\Omega_{195}|$, $|\Omega_{195}-\Omega_{200}|$, and $|\Omega_{190}-\Omega_{200}|$ are stored as convergence diagnostics. The frequencies reported in the tables are selected exclusively by the matching procedure described above. After the stable damped roots have been identified, we split them into oscillatory and purely imaginary subsets. A root is classified as purely imaginary when $|\Re{\Omega}|<10^{-10}$. This classification is used only for organising the output tables.  Finally, we also perform an instability diagnostic. Since the time dependence is $e^{-i\omega t}$, a root with $\Im{\Omega}>0$ would correspond to exponential growth. The \textsc{Julia} code therefore inspects the unfiltered finite spectra at each resolution and writes all roots satisfying $\Im{\Omega}>10^{-18}$ to separate candidates with positive imaginary part. These roots are then subjected to the same three-resolution matching criterion \eqref{eq:stability_criterion_KS}. This provides a direct check that the damped spectra reported below do not hide a stable growing mode that the QNM damping filter removed.

\section{Numerical Results}\label{sec:NR}

In order to compare our results with those of \cite{KonoplyaPLB2020}, one has to take into account the different normalisations used for both the deformation parameter and the frequency. Let $a$ denote the dimensional Kazakov--Solodukhin parameter entering the metric function. In Ref.~\cite{KonoplyaPLB2020}, the horizon radius is fixed to $r_h=1$. Therefore, the deformation parameter appearing in the numerical tables is $a_{K}=a/r_h$. In the present work, following the convention adopted in \cite{SalehASS2014,SalehASS2016,WangJAA2017,ZhangIJP2023}, we use instead the mass-normalized parameter
\begin{equation}
\mathfrak{a}=\frac{a}{M},\qquad x_h=\frac{r_h}{2M}=\frac{\sqrt{\mathfrak{a}^2+4}}{2}.
\end{equation}
The two deformation parameters are therefore related by
\begin{equation}
a_{K}=\frac{a}{r_h}=\frac{\mathfrak{a}}{2x_h}=\frac{\mathfrak{a}}{\sqrt{\mathfrak{a}^2+4}},
\end{equation}
or, equivalently,
\begin{equation}
    \mathfrak{a}=\frac{2a_{K}}{\sqrt{1-a_{K}^2}}.
\end{equation}
The frequencies must also be converted. Our dimensionless frequency is $\Omega=M\omega$, whereas the frequencies tabulated in \cite{KonoplyaPLB2020} are expressed in units of $r_h^{-1}$. Denoting \cite{KonoplyaPLB2020}'s dimensionless frequency by $\omega_{K}=r_h\omega$, we have
\begin{equation}
\omega_{K}=r_h\omega=2x_h\Omega=\sqrt{\mathfrak{a}^2+4}\Omega,
\end{equation}
or equivalently,
\begin{equation}\label{conversion}
\Omega=\frac{\sqrt{1-a_{K}^2}}{2}\omega_{K}.
\end{equation}
All comparisons with \cite{KonoplyaPLB2020} are made after applying these two conversion rules. We also recall that the phrase ``near-extremal'' in \cite{KonoplyaPLB2020} is used in a parametric sense. Since the horizon radius is fixed there, the admissible range of the deformation parameter is $0\leq a_{K}<1$, and the near-extremal regime corresponds to
\begin{equation}
    a_{K}=\frac{a}{r_h}\to 1^{-}.
\end{equation}
In this scenario, the curvature singularity located at $r=a$ approaches the event horizon $r=r_h$. This should not be confused with the extremal limit of the Reissner--Nordstr\"om black hole, where two horizons merge, and the surface gravity vanishes. In the Kazakov--Solodukhin geometry, the terminology instead refers to the deformation parameter approaching its maximum value consistent with a fixed horizon radius. In our normalisation, this same limit is
\begin{equation}
\mathfrak{a}\to \infty,\qquad a_{K}=\frac{\mathfrak{a}}{\sqrt{\mathfrak{a}^2+4}} \to 1^{-}.
\end{equation}

\begin{figure}[t]
\centering
\includegraphics[width=0.98\linewidth]{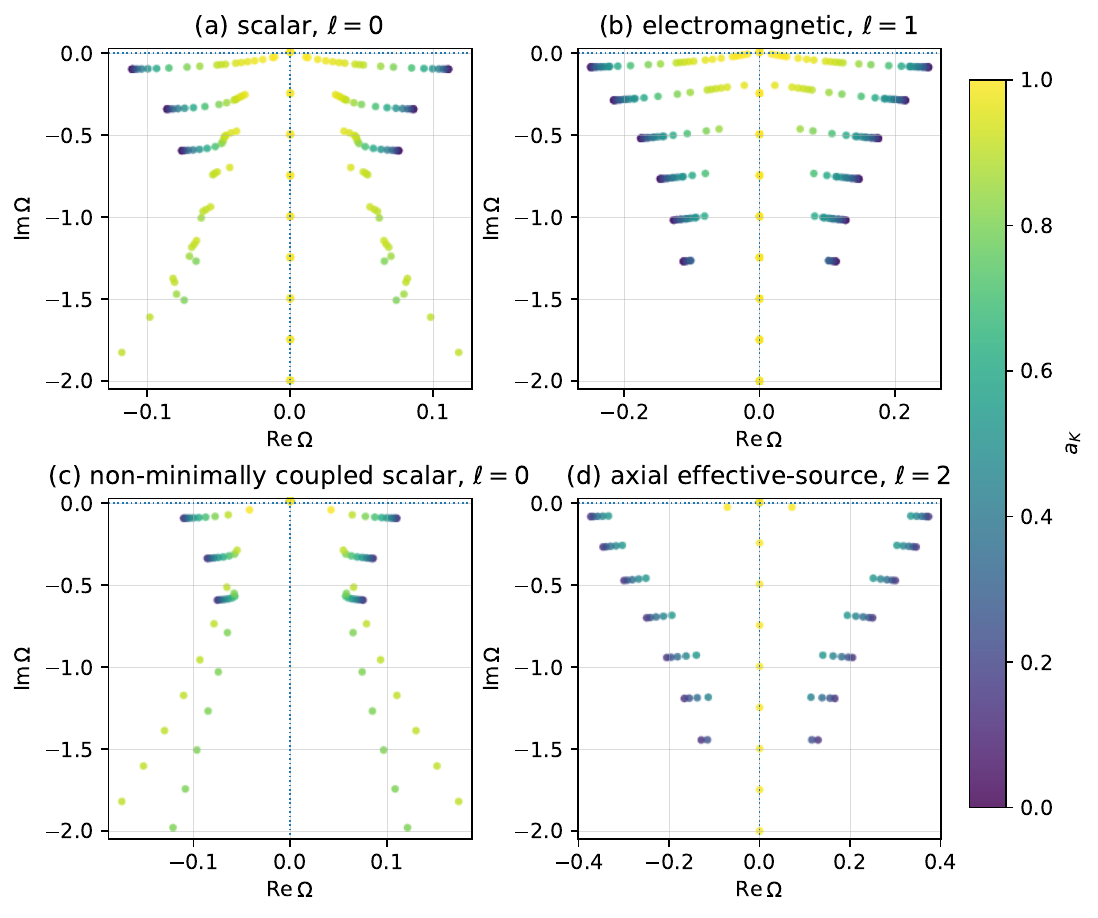}
\caption{
Complex-frequency scatter plot of the retained roots in the moderate damping window. The axes show $\mathrm{Re}\Omega$ and $\mathrm{Im}\,\Omega$, and the colour encodes the horizon-normalised deformation parameter $a_K$. The four panels show the lowest multipoles in the sectors analysed in the paper, i.e.  minimally coupled scalar $(\ell=0)$, electromagnetic $(\ell=1)$, non-minimally coupled scalar $(\ell=0)$, and axial effective source $(\ell=2)$ perturbations. The oscillatory branches appear in symmetric pairs under $\mathrm{Re}\,\Omega\to-\mathrm{Re}\,\Omega$, while the purely imaginary candidates lie on the negative imaginary axis.}
\label{fig:complex_scatter_representative}
\end{figure}

\begin{figure}[t]
\centering
\includegraphics[width=0.98\linewidth]{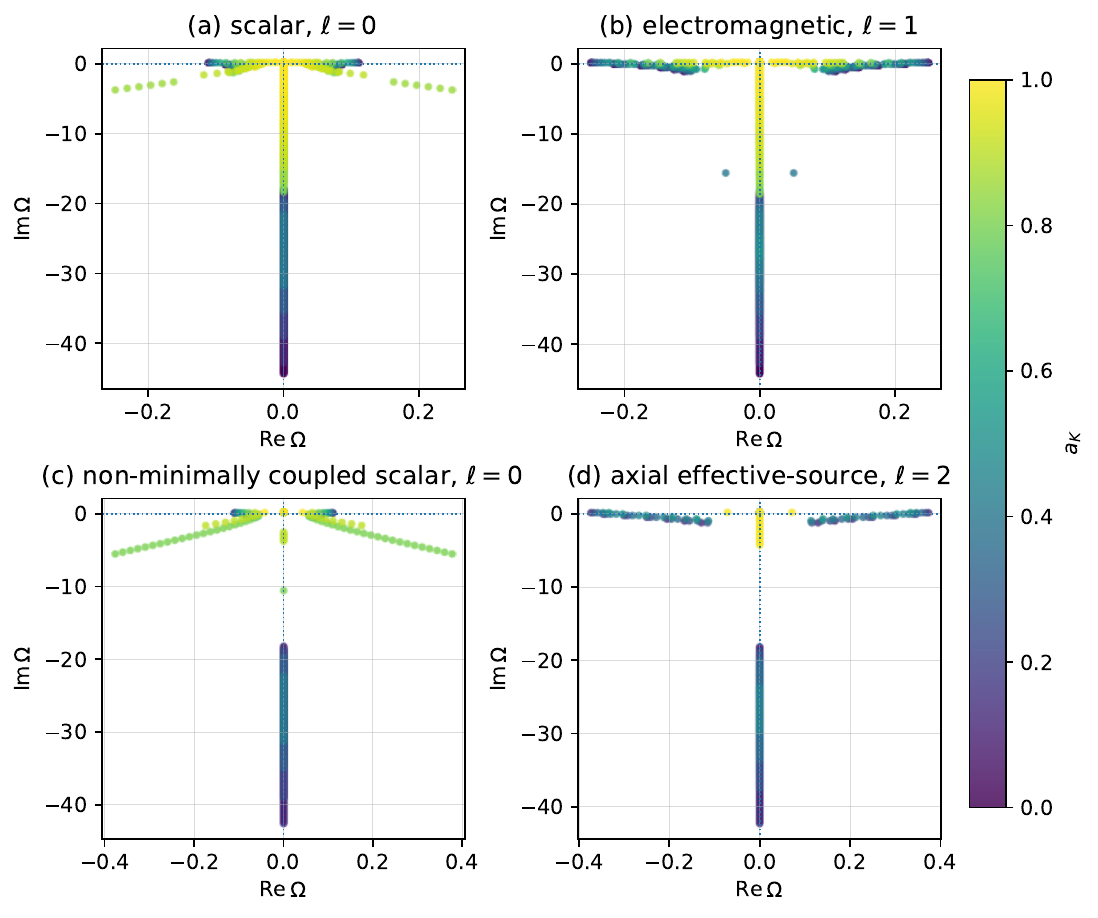}
\caption{
Same representative sectors as in Fig.~\ref{fig:complex_scatter_representative}, displayed over the full damping range present in the corresponding reports. This view makes the overdamped purely imaginary branches along $\mathrm{Re}\,\Omega=0$ explicit.}
\label{fig:complex_scatter_representative_full}
\end{figure}

\begin{figure}[t]
\centering
\includegraphics[width=0.98\linewidth]{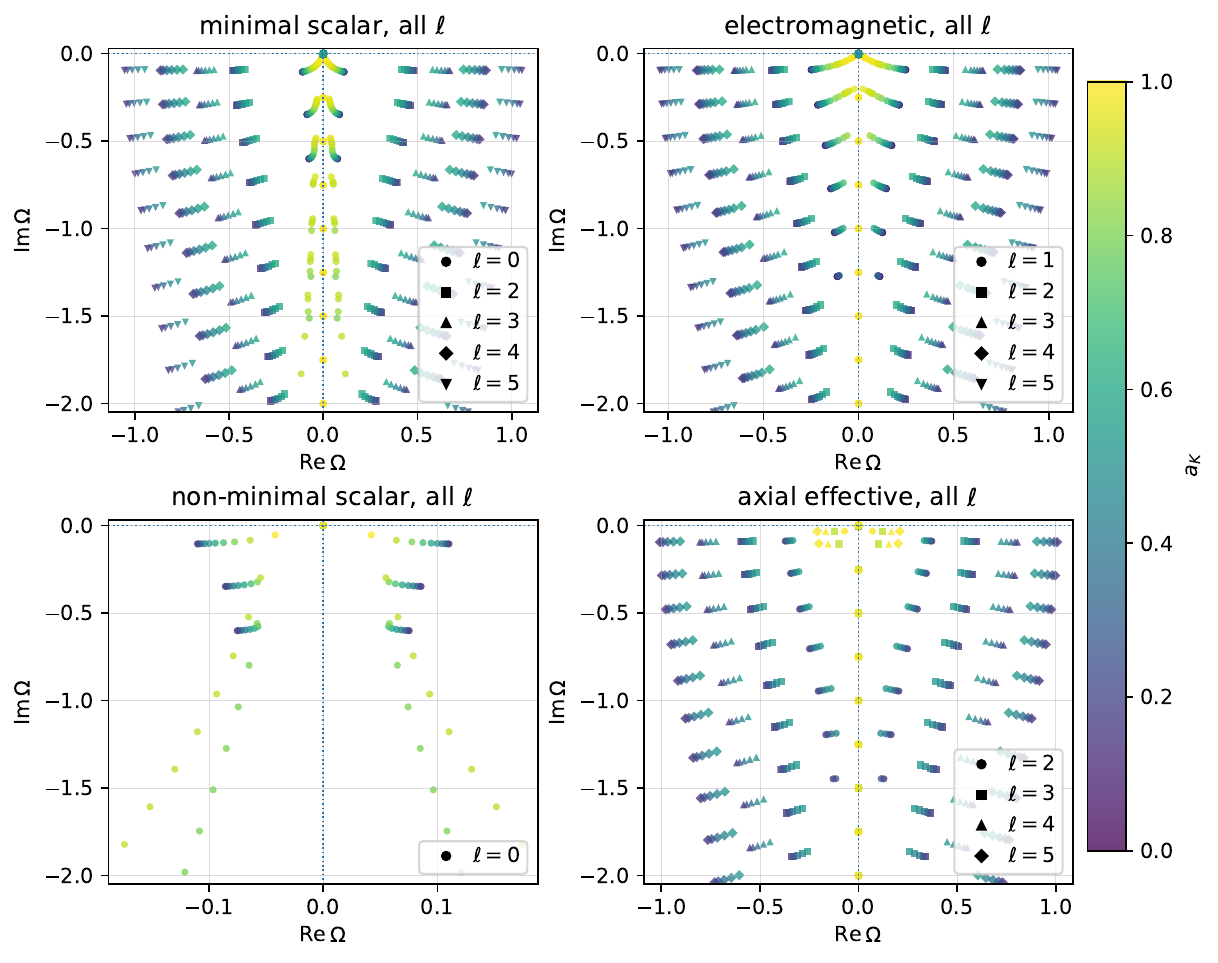}
\caption{Complex-frequency scatter plot of all available multipoles in the moderate damping window $-2\leq \operatorname{Im}\Omega\leq 0$. The four panels show the minimally coupled scalar, electromagnetic, non-minimally coupled scalar, and axial effective-source sectors. The marker shape labels the multipole number $\ell$, while the colour encodes the horizon-normalised deformation parameter $a_K$. The vertical dotted line marks $\operatorname{Re}\Omega=0$, and the horizontal dotted line marks the stability boundary $\operatorname{Im}\Omega=0$. The oscillatory roots appear in pairs related by $\operatorname{Re}\Omega\to-\operatorname{Re}\Omega$, whereas the candidate overdamped roots lie on the negative imaginary axis. The non-minimally coupled scalar panel contains the multipoles available in the present data set.}
\label{fig:complex_scatter_all_multipoles}
\end{figure}

\begin{figure}[t]
\centering
\includegraphics[width=0.98\linewidth]{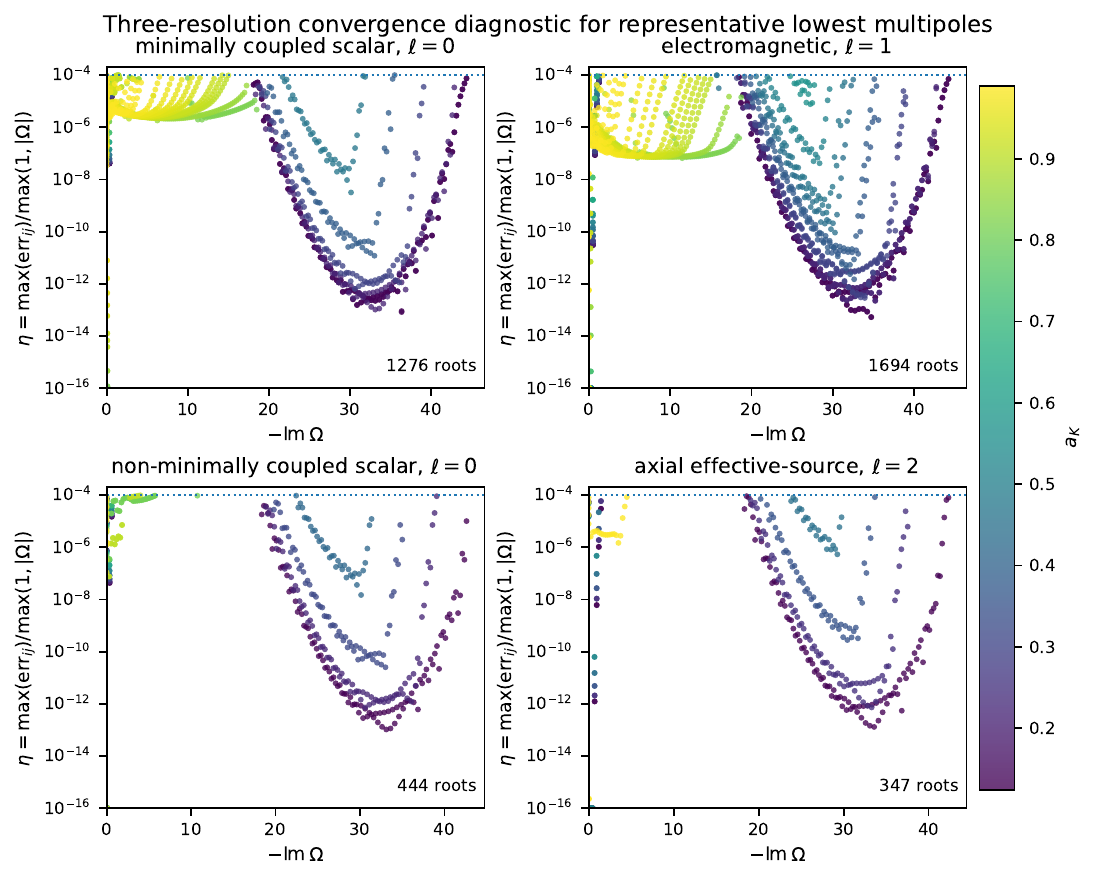}
\caption{
Three-resolution convergence diagnostic for representative lowest multipoles. For each retained triplet obtained at $N\in\{190,195,200\}$, we plot the normalised error \eqref{ne} as a function of the damping rate $-\operatorname{Im}\Omega$. The colour encodes the horizon-normalised deformation parameter $a_K$. The horizontal dotted line marks the matching threshold $10^{-4}$ used in the stability filter. All retained roots lie below this line; the expected loss of accuracy is confined to the most highly damped roots.}
\label{fig:convergence_eta_damping}
\end{figure}

\begin{figure}[t]
\centering
\includegraphics[width=0.98\linewidth]{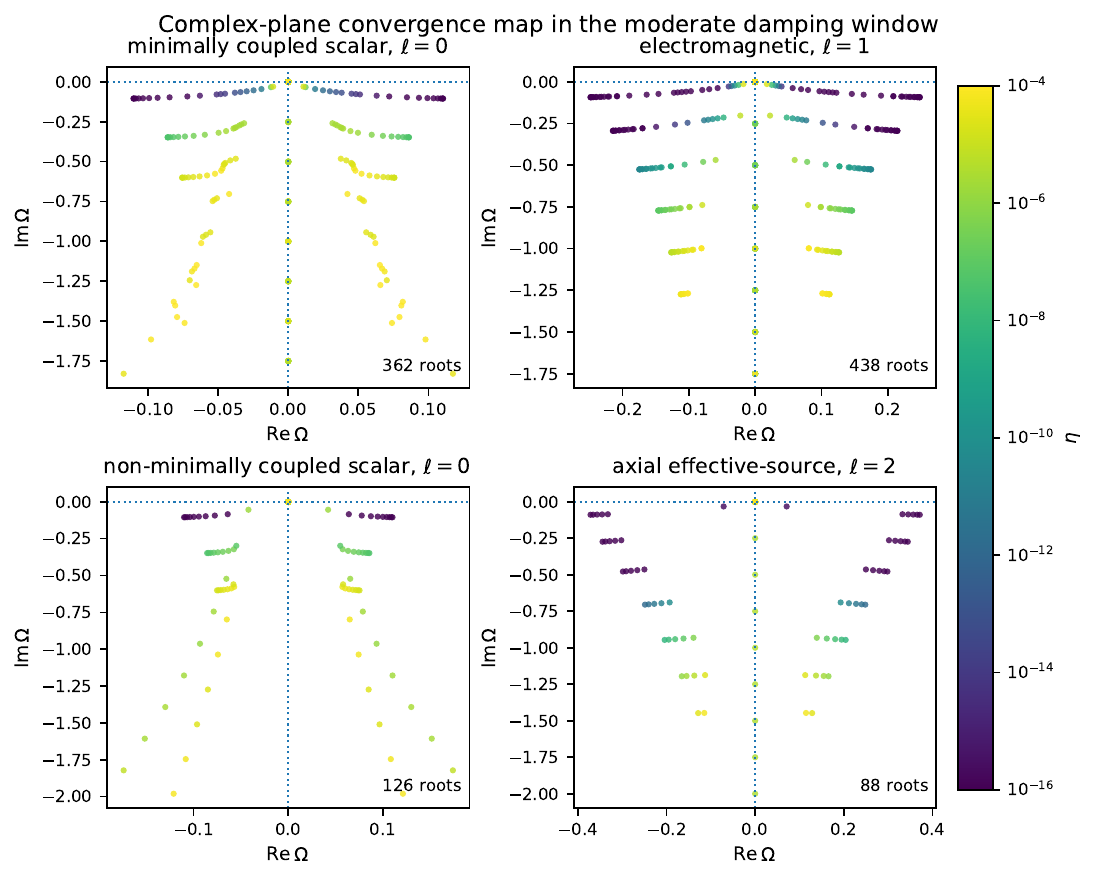}
\caption{
Complex-plane convergence map for the same representative sectors in the moderate damping window $-2\leq\operatorname{Im}\Omega\leq 0$. The axes show the retained roots in the $(\operatorname{Re}\Omega, \operatorname{Im}\Omega)$ plane, while the colour shows the normalised triplet error $\eta$ computed from \eqref{ne}. The vertical and horizontal dotted lines mark $\operatorname{Re}\Omega=0$ and $\operatorname{Im}\Omega=0$, respectively. The plot shows that the oscillatory branches and the purely imaginary candidates displayed in the spectral scatter plots are stable under changes of spectral resolution.}
\label{fig:complex_convergence_map}
\end{figure}

\begin{figure}[t]
\centering
\includegraphics[width=0.98\linewidth]{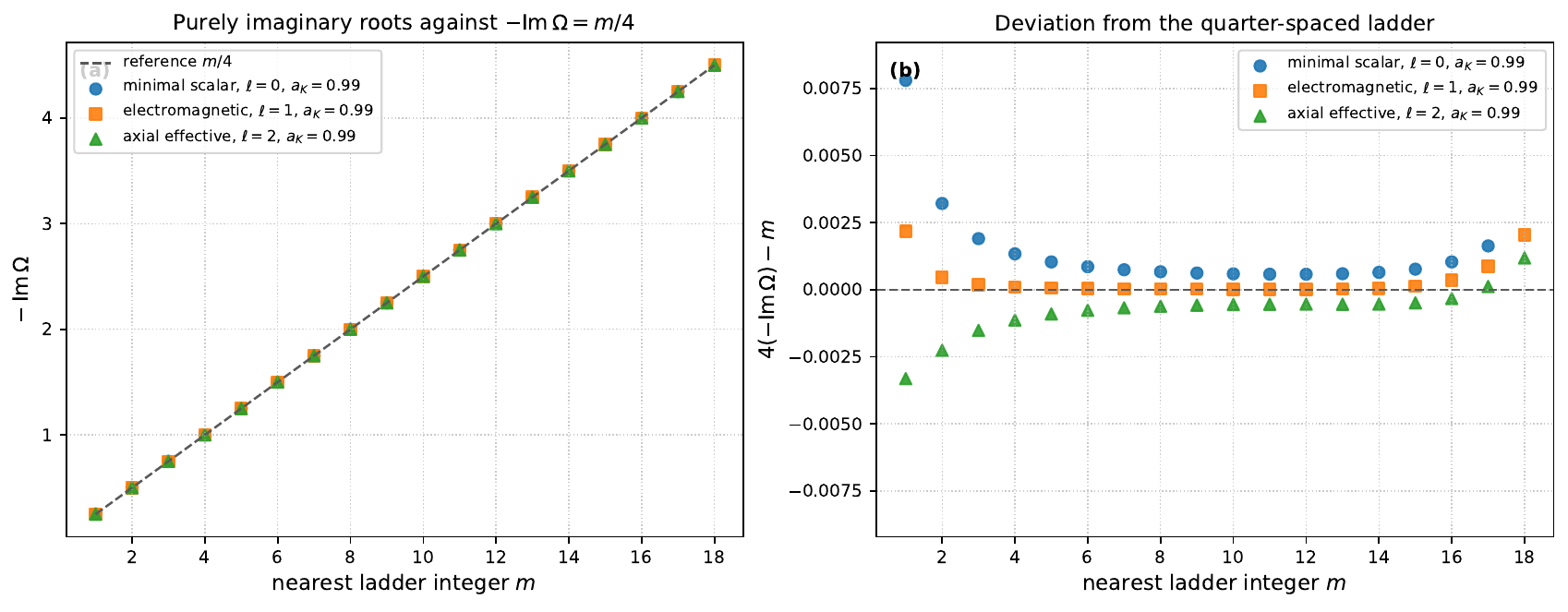}
\caption{
Near-extremal purely imaginary branches extracted from the retained spectral roots. The left panel compares the damping rates with the reference ladder $-\operatorname{Im}\Omega=m/4$, where $m=\operatorname{round}[4(-\operatorname{Im}\Omega)]$. Equivalently, $m=n+1$ in the notation $\Omega_n\simeq-i(n+1)/4$. The right panel shows the residual $4(-\operatorname{Im}\Omega)-m$. The small roots whose nearest ladder integer is $m=0$ are omitted from this diagnostic. The minimally coupled scalar monopole, electromagnetic dipole, and axial effective-source quadrupole branches shown here are all close to the quarter-spaced ladder in the near-extremal data set.}
\label{fig:near_extremal_ladder}
\end{figure}

\begin{figure}[t]
\centering
\includegraphics[width=0.98\linewidth]{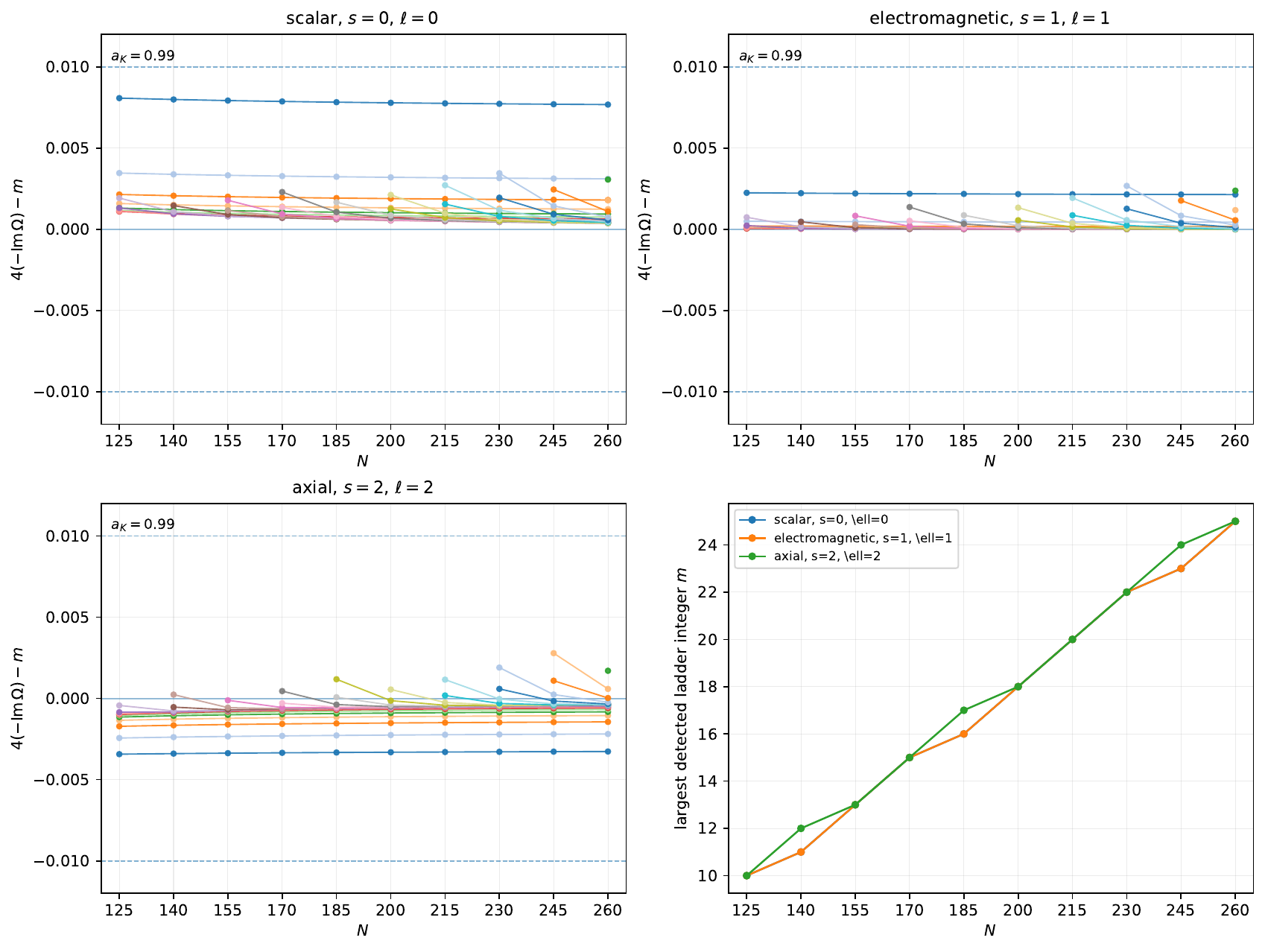}
\caption{Wide-resolution drift test for the imaginary-axis roots at $a_K=0.99$. For each sector, the spectral problem was recomputed in resolution triplets $(N-5, N, N+5)$, with $N\in\{125,140,\ldots,260\}$. For every retained root with $|\operatorname{Re}\Omega|$ numerically zero and $\operatorname{Im}\Omega<0$, we set $m=\operatorname{round}[4(-\operatorname{Im}\Omega)]$ and plot the residual $4(-\operatorname{Im}\Omega)-m$. The first three panels show the minimally coupled scalar monopole, electromagnetic dipole, and axial effective-source quadrupole sectors. The dashed horizontal lines indicate a one-per-cent band in the scaled variable $4(-\operatorname{Im}\Omega)$. The fourth panel shows the largest detected ladder integer as the resolution is increased. Fixed ladder levels remain nearly horizontal as $N$ changes, while increasing the resolution mainly extends the branch to larger damping.}
\label{fig:wide_resolution_drift_test}
\end{figure}

Before discussing the individual perturbation sectors in detail, we first present a graphical overview of the spectra. Figures~\ref{fig:complex_scatter_representative}, \ref{fig:complex_scatter_representative_full}, and \ref{fig:complex_scatter_all_multipoles} summarise the retained roots obtained from the three-resolution stability filter. The representative moderate-damping plot displays the low-lying structure of the four sectors and makes explicit the symmetry of the oscillatory modes under $\operatorname{Re}\Omega \to -\operatorname{Re}\Omega$. It also shows the presence of roots on the negative imaginary axis, which we interpret as candidate purely imaginary overdamped modes. The full-damping plot extends the vertical range and shows that these purely imaginary candidates form long branches deep in the lower half-plane, while no retained root crosses the stability boundary $\operatorname{Im}\Omega=0$. Finally, Fig.~\ref{fig:complex_scatter_all_multipoles} shows that the same qualitative organisation persists when all available multipoles are displayed together. Larger multipoles populate oscillatory branches with larger $|\operatorname{Re}\Omega|$, whereas the purely imaginary candidates remain concentrated on $\operatorname{Re}\Omega=0$. These figures, therefore, complement the tables by providing a direct visual map of the QNM spectrum in the complex $\Omega$-plane. The convergence of the retained roots under changes of spectral resolution is shown in Figs.~\ref{fig:convergence_eta_damping} and \ref{fig:complex_convergence_map}. For each candidate mode, we compare the roots obtained at $N\in\{190, 195, 200\}$ and retain only triplets whose three pairwise distances satisfy the absolute/relative matching criterion. To display this criterion uniformly for both low-lying and highly damped roots, we use the normalised error
\begin{equation}\label{ne}
  \eta=\frac{\max(\mathrm{err}_{12},\mathrm{err}_{23},\mathrm{err}_{13})}{\max(1,|\Omega|)}.
\end{equation}
All roots shown in the spectral plots satisfy $\eta<10^{-4}$. The convergence is strongest for the low-lying oscillatory modes, whereas the largest normalised errors occur, as expected, for the most highly damped roots. The complex-plane convergence map shows that both the oscillatory branches and the purely imaginary overdamped candidates remain stable under the three-resolution matching procedure. The near-extremal organisation of the purely imaginary roots is displayed in Fig.~\ref{fig:near_extremal_ladder}. To make the spacing visible, each retained purely imaginary root is assigned the nearest ladder integer $m=\operatorname{round}\!\left[4(- \operatorname{Im}\Omega)\right]$. The reference near-extremal formula $\Omega_n\simeq -i(n+1)/4$ then corresponds to $-\operatorname{Im}\Omega=m/4$, with $m=n+1$. The left panel of Fig.~\ref{fig:near_extremal_ladder} shows that the minimally coupled scalar monopole, electromagnetic dipole, and axial effective-source quadrupole roots lie close to this quarter-spaced ladder. The right panel shows the residual $4(-\operatorname{Im}\Omega)-m$, which remains small compared with the unit spacing in $4(-\operatorname{Im}\Omega)$. For the displayed $a_K=0.99$ data, the largest residuals are approximately $7.8\times10^{-3}$ in the minimally coupled scalar sector, $2.2\times10^{-3}$ in the electromagnetic sector, and $3.3\times10^{-3}$ in the axial effective-source sector. The largest deviations occur near the beginning or the end of the retained sequence, where the spectral convergence is least favourable. To address the possibility that the imaginary-axis roots are merely resolution-dependent samplers of the negative-imaginary-axis cut, we performed an additional wide-resolution drift test. Instead of using only the triplet $N\in\{190,195,200\}$, we repeated the computation for resolution triplets $(N,N+5,N+10)$, with central resolution $N\in\{125,140,\ldots,260\}$, at $a_K=0.99$. For each retained imaginary-axis root, we defined
\begin{equation}
  \gamma=-\operatorname{Im}\Omega,\qquad m=\operatorname{round}(4\gamma),
\end{equation}
and monitored the residual $4\gamma-m$ as $N$ was varied. The result is shown in Fig.~\ref{fig:wide_resolution_drift_test}. For fixed ladder index $m$, the scalar, electromagnetic, and axial branches remain close to horizontal plateaus over the whole resolution window. In contrast, increasing $N$ mainly increases the largest resolved value of $m$, i.e. it extends the branch to more highly damped roots. This behaviour is not what one would expect from a simple drift of spurious roots with the discretisation. Rather, within the present Chebyshev formulation, the imaginary-axis roots displayed in the near-extremal ladder behave as stable spectral features. This test does not by itself constitute a rigorous proof that every member of the imaginary-axis sequence is an isolated pole of the analytically continued Green function. A direct Wronskian, continued-fraction, or resolvent analysis would still be valuable. It does, however, rule out the simplest numerical interpretation in which the displayed ladder is produced by roots whose locations drift appreciably with the spectral resolution.

\subsection{Scalar sector}

We first consider the scalar monopole, $s=0$, $\ell=0$, summarised in Table~\ref{table_s0_L0}. In the Schwarzschild limit $a_K=0$, our first three modes provide a benchmark of the normalisation and of the implementation of the QNM boundary conditions. For $0.01\le a_K\le 0.6$, \cite{KonoplyaPLB2020} provides both seventh-order WKB and time-domain data for the scalar fundamental mode. Since \cite{KonoplyaPLB2020} fixes $r_h=1$, while our frequency is $\Omega=M\omega$, the comparison is made after applying \eqref{conversion}. After this conversion, the agreement with the time-domain values is at the sub-per cent level. The relative differences in the real and imaginary parts remain below about $0.6\%$ throughout the interval $0.01\le a_K\le 0.6$. This agreement validates the frequency normalisation, the horizon and infinity boundary conditions, and the spectral implementation for the low-lying scalar modes. For larger values of $a_K$, the comparison with \cite{KonoplyaPLB2020} must be interpreted more cautiously. The time-domain data for the scalar $\ell=0$ mode are not available there for $a_K\in\{0.7,0.8,0.9\}$, and the comparison is only with the seventh-order WKB values. In this regime, the discrepancy becomes substantial. At $a_K=0.8$, for instance, the converted WKB value is $\Omega_{ WKB}=0.106513-0.123592i$, whereas the spectral method gives $\Omega_{SM}=0.0721-0.0917i$. The relative differences are approximately $30\%$ in the real part and $25\%$ in the magnitude of the imaginary part. At $a_K=0.9$, the converted WKB value $\Omega_{WKB}=0.107303-0.022386i$ differs even more strongly from the spectral value $\Omega_{SM}=0.0512-0.0792i$. Since this is a low-multipole mode in a strongly deformed regime, and since no time-domain value was extracted in \cite{KonoplyaPLB2020}, we do not regard the WKB-only entries at large $a_K$ as reliable benchmarks. We next compare with the results of \cite{ZhangIJP2023}, reported in Table~\ref{table_s0_L0_Zhang}, who computed the fundamental $\ell=n=0$ mode with the Mashhoon, or P\"oschl--Teller, approximation and with the asymptotic iteration method (AIM). A comment is necessary before making this comparison. \cite{ZhangIJP2023} introduces the non-Ricci-flat scalar equation with an additional $R(r)/6$ term and state that their QNM calculation is performed with the corresponding potential. However, their tabulated WKB benchmark values coincide with the minimally coupled scalar data of \cite{KonoplyaPLB2020}, rather than with the conformally coupled scalar values reported separately in that reference. Moreover, we reproduce the frequencies in \cite{ZhangIJP2023} with the minimally coupled scalar potential. For this reason, in what follows, we use the numerical values of \cite{ZhangIJP2023} as an additional benchmark for the minimally coupled scalar sector, while noting the apparent mismatch between the potential stated in that paper and the numerical data reported there. With this interpretation, the comparison shows that the spectral results are much closer to AIM than to the Mashhoon approximation. For example, at $a_K=0.01$, the relative difference between SM and AIM is approximately $0.14\%$ in the real part and $0.25\%$ in the magnitude of the imaginary part, while the difference between SM and the Mashhoon method value is about $3.8\%$ and $8.6\%$, respectively. The same pattern persists throughout the range displayed in Table~\ref{table_s0_L0_Zhang}. At $a_K=0.6$, for instance, the relative difference between the AIM and the SM value is about $0.89\%$ in the real part and $0.07\%$ in the damping rate. At $a_K=0.7$, the corresponding relative differences are about $2.1\%$ in the real part and $1.3\%$ in the magnitude of the imaginary part. This comparison is consistent with the conclusion of \cite{ZhangIJP2023} that AIM is more accurate than the Mashhoon approximation for this problem. Further validation is provided by comparison with the third-order WKB results in \cite{SalehASS2014}. In that work, the numerical tables are computed with $M=1$, so their frequencies can be compared directly with our $\Omega=M\omega$. Tables~\ref{table_s0_a0_25_and_a0_5}, \ref{table_s0_a0_75_and_a1}, and \ref{table_s0_a1_25_and_a1_5} show the comparison for $\mathfrak{a}\in\{0.25,0.5,0.75,1,1.25,1.5\}$. and $\ell\in\{2,3,4,5\}$. For fundamental modes, the agreement is very close. For higher overtones, the discrepancy increases, as expected for a low-order WKB approximation. For instance, at $\mathfrak{a}=1.5$, $\ell=2$, and $n=4$, \cite{SalehASS2014} gives $\Omega_{WKB}=0.2562-0.8897i$, whereas the spectral method yields $\Omega_{SM}=0.2796-0.9522i$. Thus, the comparison confirms the WKB values for the lowest modes and shows the expected deterioration of the third-order WKB approximation for higher overtones. The spectral method provides access to parts of the scalar spectrum not covered by earlier WKB, Mashhoon, AIM, and time-domain studies. First, in Tables~\ref{table_s0_a0_25_and_a0_5}, \ref{table_s0_a0_75_and_a1}, and \ref{table_s0_a1_25_and_a1_5}, we report additional overtones beyond those tabulated in \cite{SalehASS2014}. For many entries, the WKB column is labelled N/A while the spectral method still returns stable modes. This extends the available scalar data for $\ell\in\{2,3,4,5\}$ and for the deformation parameters $\mathfrak{a}=0.25,\ldots,1.5$. Second, Tables~\ref{scalargoverdamped}, \ref{table_s0_a0_25_and_a0_5_overdamped}, and \ref{table_s0_a0_75_and_a1_overdamped} display purely imaginary scalar modes. These modes are not reported in the comparison tables of \cite{KonoplyaPLB2020, SalehASS2014, ZhangIJP2023}. For the scalar monopole, the first such modes occur at comparatively large damping. The spacings are close to $0.25i$ in the quantity $\Delta\Omega=\Omega_n-\Omega_{n+1}$. Third, Tables~\ref{table_s0_nearly_extremal}, \ref{scalargoverdampednearextremal}, and \ref{scalargoverdampednearextremalrefined} explore the scalar monopole in the near-extremal regime $a_K\to 1^{-}$. This regime was not covered by time-domain scalar data in \cite{KonoplyaPLB2020}. Our spectral data show that the low-lying oscillatory modes become increasingly small in both real part and damping rate as $a_K$ approaches unity. For $a_K=0.96$, the stable oscillatory roots retained by our criterion are $0.029705-0.059421i$, and $0.031539-0.259637i$, whereas for $a_K\in\{0.97,0.98,0.99,0.992\}$, the present stability criterion retains only one low-lying oscillatory mode in each case, with frequencies $0.0248-0.0533i$, $0.0192-0.0453i$, $0.0124-0.0335i$, and $0.0108-0.0303i$, respectively. We interpret this as a statement about the roots resolved and retained by the present stability criterion, not as a proof that no other QNMs exist in this part of the spectrum. In the mass-normalised frequency $\Omega=M\omega$, the scalar fundamental mode decreases as the deformation is increased. In Table~\ref{table_s0_L0}, the fundamental frequency changes from $0.1105-0.1049i$ at $a_K=0$ to $0.0631-0.0869i$ at $a_K=0.85$. Thus, in this normalisation, both the oscillation frequency and the damping rate decrease over this range. This behaviour should not be confused with the trend in the horizon-normalised frequency $r_h\omega$, because $r_h/M$ itself depends on the deformation parameter. The higher scalar overtones exhibit a richer pattern. In particular, the real parts of the stable roots do not always vary monotonically with $a_K$, especially for $a_K\gtrsim 0.7$. This can be seen in Table~\ref{table_s0_L0} and in the near-extremal data of Table~\ref{table_s0_nearly_extremal}. We regard this behaviour as evidence for a nontrivial rearrangement of the scalar spectrum in the strongly deformed regime. Since the number of stable roots retained by the convergence filter also changes near $a_K\simeq1$, the identification of individual overtone branches should be made with some care. The purely imaginary modes display a more regular pattern. Whenever such modes are detected, their spacing is nearly constant, i.e. $\Delta\Omega=\simeq 0.25i$. This value is naturally compared with the surface gravity scale $M\kappa=1/4$, which is independent of the deformation parameter for the KS geometry. For $0.4<a_K<0.8$, purely imaginary modes are not detected by the present stability criterion, whereas they reappear for $a_K\ge 0.8$. In the near-extremal tables, the first purely imaginary mode moves toward $-i/4$. Moreover, the refined near-extremal data suggest the limiting pattern
\begin{equation}
    \Omega_{n,(s=0)}\simeq-\frac{i}{4}(n+1),
    \qquad n=0,1,2,\ldots,
\end{equation}
with the overtone indexing $n$ used in the tables. This observation is numerical and should be checked independently. In particular, the strict limit $a_K\to1^{-}$ is singular because, at fixed $r_h$, one has $M\to 0$ and the curvature singularity approaches the horizon. Therefore, a detailed analysis of this limiting regime is left for future work. With the convention $e^{-i\omega t}$, stability requires $\Im\Omega<0$. All scalar modes reported in the present tables satisfy this condition.  The results support three main conclusions. First, the spectral method reproduces the available time-domain, WKB, AIM, and Mashhoon benchmarks within their common range of reliability, with the closest agreement with time-domain and AIM data. Second, the method extends the scalar QNM data to additional overtones and to purely imaginary modes not displayed in the earlier comparison tables. Third, the near-extremal scalar spectrum shows a distinctive reorganisation, i.e. the low-lying oscillatory modes become less oscillatory and less damped in the $\Omega=M\omega$ normalisation, while the purely imaginary modes approach an approximately equally spaced ladder with spacing close to $M\kappa=1/4$. The electromagnetic and gravitational sectors will be discussed separately in the next sections.

\subsection{Non-minimally coupled scalar sector}\label{subsec:nonminimal_scalar_results}

We now consider the non-minimally coupled scalar field with conformal coupling. Since the KSQC spacetime is not Ricci flat, this sector differs from the minimally coupled scalar field by the additional curvature term $R(r)/6$ in the effective potential. The numerical results for $\ell=0$ are summarised in Table~\ref{table_s0_L0_minimally coupled}, where we compare our spectral method with the seventh-order WKB and time-domain data of \cite{KonoplyaPLB2020}. As before, the frequencies in \cite{KonoplyaPLB2020} are given in units of $r_h^{-1}$, while our dimensionless frequency is $\Omega=M\omega$. Therefore, the comparison is made after applying \eqref{conversion}. For the fundamental mode, the agreement with the time-domain data is very good over the entire range for which such data are available. For example, at $a_K=0.1$, the time-domain value converted to our normalisation is $\Omega_{TD}=0.10988-0.10529i$, whereas the spectral method gives $\Omega_{SM}=0.1101-0.1048i$. The relative differences are approximately $0.17\%$ in the real part and $0.46\%$ in the magnitude of the imaginary part. At $a_K=0.6$, the comparison gives $\Omega_{TD}=0.09480-0.10068i$, and  $\Omega_{\rm SM}=0.0944-0.1005i$, with relative differences of about $0.39\%$ and $0.19\%$, respectively. Even at $a_K=0.9$, where the deformation is strong, the time-domain and spectral values remain close with $\Omega_{TD}=0.06400-0.08330i$, and $\Omega_{SM}=0.0640-0.0841i$. The relative difference is about $0.08\%$ in the real part and about $0.90\%$ in the damping rate. This comparison provides an additional validation of the spectral implementation for the conformally coupled scalar sector. The comparison with the seventh-order WKB values is also reasonable for small and intermediate deformation, but it deteriorates in the strongly deformed regime. At $a_K=0.1$, the WKB value is $\Omega_{WKB}=0.110754-0.104204i$, while the spectral method gives $\Omega_{SM}=0.1101-0.1048i$. The corresponding relative differences are about $0.62\%$ in the real part and $0.58\%$ in the magnitude of the imaginary part. At $a_K=0.8$, the WKB value is $\Omega_{WKB}=0.079792-0.092359i$, whereas the spectral result is $\Omega_{SM}=0.0776-0.0933i$. Here, the relative differences are approximately $2.76\%$ and $0.98\%$. At $a_K=0.9$, the discrepancy becomes more pronounced with $\Omega_{WKB}=0.076856-0.088481i$, and $\Omega_{SM}=0.0640-0.0842i$. The relative difference is then about $16.8\%$ in the real part and $5.0\%$ in the damping rate. Thus, as in the minimally coupled scalar sector, the WKB approximation should be used with caution for low multipoles in the strongly deformed regime. It is also useful to compare the non-minimally coupled scalar sector with the minimally coupled scalar sector discussed above. The effect of the curvature coupling is very small for weak deformation. For instance, at $a_K=0.1$, the minimally coupled scalar fundamental mode is $0.1101-0.1048i$ and it practically coincides with the non-minimally coupled value. At $a_K=0.4$, the corresponding values are $0.1037-0.1033i$ and $0.1039-0.1033i$. The difference grows as $a_K$ increases. At $a_K=0.8$, the minimally coupled scalar value is $0.0721-0.0917i$, while the non-minimally coupled value is $0.0776-0.0933i$. At $a_K=0.9$, using the minimally coupled scalar value from the near-extremal table, $\Omega_{min}=0.0512-0.0792i$, whereas the non-minimally coupled scalar gives $\Omega_{c}=0.0640-0.0841i$. Thus, the conformal curvature term produces only a small correction at weak deformation, but it becomes increasingly visible in the strongly deformed regime. The higher overtones in Table~\ref{table_s0_L0_minimally coupled} display a non-monotonic pattern for $a_K=0.8$ and $a_K=0.9$. At $a_K=0.8$, the real part decreases from $n=0$ to $n=1$, but then begins to increase. At $a_K=0.9$, the effect is even clearer. This behaviour indicates a nontrivial rearrangement of the overtone spectrum in the strongly deformed regime. Since the modes are identified through stability under spectral resolution, we interpret this cautiously as a property of the retained sequence of roots, rather than as a proof of a particular branch-crossing mechanism. Finally, the purely imaginary modes of the non-minimally coupled scalar sector are reported in Table~\ref{scalargoverdampedmincoup}. These modes are not listed in the comparison tables of \cite{KonoplyaPLB2020}. For $a_K=0.1$, the first six purely imaginary modes exhibit spacings close to $0.25i$. Similarly, for $a_K=0.4$, the first six overdamped modes have spacings between approximately $0.2475i$ and $0.2483i$. At $a_K=0.9$, the detected purely imaginary sequence is much less damped, but again with spacing close to $0.25i$. This spacing is naturally compared with the surface-gravity scale $M\kappa=1/4$, but, as was the case in the minimally coupled scalar sector. It is worth noting that the detection pattern is not uniform in $a_K$. No purely imaginary roots are reported in Table~\ref{scalargoverdampedmincoup} for $a_K\in\{0.5,0.6,0.7\}$, or $a_K=0.99$, while at $a_K=0.8$ only one such root is retained. This should not be read as proof of the absence of purely imaginary modes in the exact spectral problem. Rather, it states what is resolved and retained by the present spectral stability criterion.

\subsection{Electromagnetic sector}

We now turn to electromagnetic perturbations, $s = 1$. In this case, the admissible multipoles satisfy $\ell\geq 1$. We first compare the electromagnetic $\ell = 1$ fundamental mode with \cite{KonoplyaPLB2020}. As in the scalar case, \cite{KonoplyaPLB2020} uses $r_h = 1$, and therefore, the tabulated frequency therein must be converted to our normalisation by means of the formula \eqref{conversion}. The comparison is displayed in Table~\ref{table_s1_L1}. The agreement with both the seventh-order WKB values and the time-domain values is very close over the range where the data overlap. For example, at $a_K=0.01$, the converted time-domain value is $\Omega_{TD}=0.24827-0.09248i$, whereas the spectral method gives $\Omega_{SM}=0.2483-0.0925i$. Similarly, at $a_K=0.8$, the converted time-domain value is $\Omega_{TD}=0.16362-0.07709i$, while the spectral result is $\Omega_{SM}=0.1636-0.0771i$. Thus, in contrast with the scalar monopole at large deformation where the seventh-order WKB becomes unreliable, the electromagnetic fundamental mode remains very well reproduced by the spectral method and the WKB approximation throughout the non-near-extremal comparison range shown in Table~\ref{table_s1_L1}. The agreement persists in the near-extremal regime. Tables~\ref{table_s1_nearly_extremal1} and \ref{table_s1_nearly_extremal2} compare our spectral results with the time domain data of \cite{KonoplyaPLB2020}. At $a_K=0.9$, \cite{KonoplyaPLB2020} gives, after conversion to our convention, $\Omega_{WKB}=0.118392-0.068621i$, $\Omega_{TD}=0.12252-0.06553i$. Our spectral value is $\Omega_{SM}=0.1225-0.0655i$, in direct agreement with the time domain result. The same level of agreement is observed at larger values of $a_K$. This comparison provides a stringent validation of the spectral implementation and the QNM boundary conditions in the electromagnetic sector. A second benchmark is provided by the third-order WKB results of \cite{WangJAA2017}. In that work, the mass is fixed to $M=1$, so the frequencies are directly comparable with our $\Omega=M\omega$. Tables~\ref{table_s1_a0_2_and_a0_4}, \ref{table_s1_a0_6_and_a0_8}, \ref{table_s1_a1_0_and_a1_2}, and \ref{table_s1_a1_5_QNM_and_overdamped} give the comparison for $\mathfrak{a}\in\{0.2,0.4,0.6,0.8,1,1.2,1.5\}$, and $\ell\in\{1,2,3,4,5\}$. For fundamental modes, the agreement is generally very good. The agreement is also good for the lower overtones when $\ell$ is not too small. As expected, the discrepancy with the third-order WKB approximation increases for low multipoles and higher overtones. A representative example is $\mathfrak{a}=1.5$, $\ell=1$, $n=3$, for which \cite{WangJAA2017} gives $\Omega_{WKB}=0.04029-0.69570i$, whereas the spectral method yields $\Omega_{SM}=0.0990-0.7507i$. This large difference is consistent with the known limitation of low-order WKB approximations for modes with relatively high overtone number $n$ compared with the multipole number. By contrast, for larger $\ell$ the agreement remains good even for $n=3$. For example, at $\mathfrak{a}=1.5$, $\ell=5$, $n=3$, the values are $\Omega_{WKB}=0.83841-0.64875i$, $\Omega_{SM}=0.8352-0.6524i$. Thus, the comparison with \cite{WangJAA2017} validates the spectral method for the electromagnetic sector and, at the same time, quantifies the deterioration of the third-order WKB approximation in the expected low $\ell$, high $n$ regime. The spectral method also gives access to electromagnetic modes not listed in the previous WKB tables. In Tables~\ref{table_s1_a0_2_and_a0_4}, \ref{table_s1_a0_6_and_a0_8}, \ref{table_s1_a1_0_and_a1_2}, and \ref{table_s1_a1_5_QNM_and_overdamped}, several entries with $n\geq 4$ have no WKB counterpart in \cite{WangJAA2017}, whereas our spectral computation still returns stable roots. These entries extend the available electromagnetic QNM data beyond the third-order WKB results. We next discuss the purely imaginary electromagnetic modes. These are listed in Tables~\ref{emoverdamped}, \ref{table_s1_a0_2_and_a0_4_overdamped},  \ref{table_s1_a0_6_and_a0_8_overdamped}, \ref{table_s1_a1_0_and_a1_2_overdamped}, \ref{scalargoverdampednearextremals1}, and \ref{scalargoverdampednearextremalrefineds1}. Such modes are not reported in the comparison tables of \cite{KonoplyaPLB2020, WangJAA2017}. For moderate values of the deformation parameter, the detected purely imaginary modes are typically highly damped. Moreover, the elementary spacing scale is again close to $i/4$, but the spacing between two consecutive roots retained by the stability criterion is not always exactly one elementary step. For instance, Table~\ref{emoverdamped} shows $\Delta\Omega=0.7480i$ for $a_K=0.3$ between the first two retained purely imaginary modes, while Table~\ref{table_s1_a0_6_and_a0_8_overdamped} shows $\Delta\Omega=4.5954i$ for $\mathfrak{a}=0.8$, $\ell=1$, again between the first two retained purely imaginary modes. We therefore interpret the data as evidence for an underlying spacing scale close to $i/4$, with possible missing or unresolved intermediate members in the numerically retained sequence, rather than as a strictly uniform ladder at all parameter values. The detection of purely imaginary modes also depends sensitively on the deformation parameter and on $\ell$. In Table~\ref{emoverdamped}, no purely imaginary $\ell=1$ modes are detected by our stability criterion for $a_K=0.6$ and $a_K=0.7$, whereas they reappear for $a_K=0.8$ and $a_K=0.85$. In other words, purely imaginary modes are detected for several values of $\mathfrak{a}$ and $\ell$, but not uniformly throughout the parameter space. In particular, Table~\ref{table_s1_a1_0_and_a1_2_overdamped} shows no detected purely imaginary modes for $\mathfrak{a}=1.2$ with $ell\in\{3,4,5\}$,  and Table~\ref{table_s1_a1_5_QNM_and_overdamped} shows no detected purely imaginary modes for $\mathfrak{a}=1.5$ with $\ell\in\{1,2,3,4,5\}$. As in the scalar sector, this should be understood as a statement about the roots resolved and retained by the present stability criterion, not as a proof that such modes are absent from the exact spectrum. The near-extremal purely imaginary spectrum exhibits a particularly simple structure. Tables~\ref{scalargoverdampednearextremals1} and \ref{scalargoverdampednearextremalrefineds1} show that, as $a_K\to 1^{-}$, the detected electromagnetic purely imaginary modes approach
\begin{equation}
  \Omega_{N,(s=1)}\simeq -\frac{i}{4}(N+1),\qquad N=0,1,2,\ldots.
\end{equation}
For example, at $a_K=0.99$, Table~\ref{scalargoverdampednearextremals1} gives $\Omega_0=-0.2505i$, $\Omega_1=-0.5001i$, $\Omega_2=-0.7500i$, and $\Omega_3=-1.0000i$, and at $a_K=0.998$, Table~\ref{scalargoverdampednearextremalrefineds1} gives $\Omega_0=-0.2501i$, $\Omega_1=-0.5000i$, $\Omega_2=-0.7500i$, and $\Omega_3=-1.0000i$. At $a_K=0.999$, \cite{KonoplyaPLB2020} reports the low-damping time domain electromagnetic fundamental mode $\Omega_{TD}=0.01245-0.00955i$ in our convention, while our spectral computation detects the purely imaginary sequence $\Omega_0=-0.2500i$, $\Omega_1=-0.5000i$, $\Omega_2=-0.7500i$, and $\Omega_3=-1.0000i$. These frequencies should not be interpreted as the same branch. The time domain value is a low-damped oscillatory fundamental mode, whereas the purely imaginary roots form an overdamped branch. Their coexistence, or the possible loss of one branch under a given numerical extraction criterion, is a feature of the near-extremal regime that deserves further study. Finally, we compare the near-extremal scalar and electromagnetic spectra. There is no evidence for full isospectrality of the oscillatory sector. For example, at $a_K=0.9$, the scalar fundamental mode in Table~\ref{table_s0_nearly_extremal} is $\Omega^{(s=0)}_0=0.0512-0.0792i$, whereas the electromagnetic fundamental mode in Table~\ref{table_s1_nearly_extremal1} is $\Omega^{(s=1)}_0=0.1225-0.0655i$. The two sectors, therefore, remain distinct in their low-lying oscillatory QNMs. However, the purely imaginary near-extremal branches become very close. For instance, at $a_K=0.98$, the scalar table gives the sequence $-0.2538i$,  $-0.5016i$, $-0.7509i$, and $-1.0006i$, whereas the electromagnetic table gives $-0.2515i$, $-0.5003i$, $-0.7501i$, $-1.0001i$. At $a_K=0.998$, the scalar values $-0.2504i$, $-0.5002i$, $-0.7501i$, and $-1.0001i$ are very close to the electromagnetic values $-0.2501i$, $-0.5000i$, $-0.7500i$, and $-1.0000i$. Thus, the data suggest an approximate near-extremal isospectrality of the overdamped purely imaginary ladder, but not of the full scalar and electromagnetic spectra. We regard this as a numerical indication of a common near-extremal spacing structure rather than as a proof of exact isospectrality.

\subsection{The axial gravitational sector}\label{subsec:axial_gravitational_results}

We finally discuss the axial gravitational sector. The word {\it{axial}} should be understood in the restricted sense defined in Sec.~\ref{subsec:axial_grav_potential}. The frequencies in this subsection are those of the metric-led axial effective-source model with $\delta T^{ind}_{ax}=0$. They should not be interpreted as the complete gravitational QNM spectrum of the KSQC black hole for an arbitrary anisotropic source. A nonzero independent axial perturbation of the effective source would lead to a model-dependent coupled system, potentially shifting these frequencies or introducing additional axial modes. As explained in Sec.~\ref{subsec:axial_grav_potential}, this sector should be interpreted as an axial effective source Regge--Wheeler model. The distinction is important. The KSQC spacetime is not Ricci flat, and therefore the vacuum perturbation equation $\delta R_{\mu\nu}=0$ cannot be imposed without further assumptions. This point was emphasised in \cite{KonoplyaPLB2020}, where the earlier gravitational treatment of \cite{SalehASS2016} was criticised. The potential used here coincides algebraically with Eq.~(21) of \cite{SalehASS2016}, but in the present work, it is understood as the axial potential of an effective source model rather than as the result of a vacuum gravitational perturbation calculation. The comparison with the third-order WKB data of \cite{SalehASS2016} is shown in Tables~\ref{table_s2_a0_25_and_a0_5}, \ref{table_s2_a0_75_and_a1_0}, and \ref{table_s2_a1_75_and_a0_99}. Since \cite{SalehASS2016} uses $M=1$, the frequencies reported there can be compared directly with our dimensionless frequency $\Omega=M\omega$. For the fundamental modes, the agreement is very close. For instance, at $\mathfrak{a}=0.25$, $\ell=2$, and $n=0$, \cite{SalehASS2016} gives $\Omega_{WKB}=0.3712-0.0890i$, whereas the spectral method gives $\Omega_{SM}=0.3712-0.0888i$. The relative differences are about $0.13\%$ in the real part and $0.23\%$ in the magnitude of the imaginary part. At $\mathfrak{a}=1.25$, $\ell=2$, and $n=0$, the comparison is $\Omega_{ WKB}=0.3322-0.0857i$, $\Omega_{SM}=0.3325-0.0852i$, corresponding to relative differences of about $0.10\%$ and $0.59\%$, respectively. This confirms that the spectral implementation reproduces the known third-order WKB fundamental frequencies in their range of applicability. The agreement remains good for higher multipoles and low overtones. For example, at $\mathfrak{a}=1.25$, $\ell=5$, and $n=0$, the WKB value $0.9042-0.0909i$ is reproduced by the spectral value $0.9042-0.0909i$. For $\ell=5$, $n=4$, and the same deformation parameter, the values are $\Omega_{WKB}=0.8091-0.8462i$, $\Omega_{SM}=0.8028-0.8575i$, with relative differences of about $0.78\%$ in the real part and $1.34\%$ in the damping rate. Thus, the WKB approximation remains reasonably accurate in this higher-multipole case. As expected, the comparison deteriorates for low multipoles and higher overtones. At $\mathfrak{a}=0.5$, $\ell=2$, and $n=4$, \cite{SalehASS2016} gives $\Omega_{WKB}=0.1663-0.8745i$, whereas the spectral method gives $\Omega_{SM}=0.1951-0.9442i$. The relative differences are about $17.3\%$ in the real part and $8.0\%$ in the damping rate. At $\mathfrak{a}=1.25$, $\ell=2$, and $n=4$, the discrepancy becomes $\Omega_{WKB}=0.1163-0.8533i$, $\Omega_{SM}=0.1392-0.9316i$, corresponding to differences of about $19.7\%$ and $9.2\%$, respectively. This is consistent with the known limitation of low-order WKB methods when the overtone number is not small compared with the multipole number. The dependence on the deformation parameter follows the same qualitative trend reported in \cite{SalehASS2016}. For fixed $\ell$ and $n$, both the oscillation frequency and the damping rate decrease as the deformation parameter increases. The same trend is observed for the higher multipoles displayed in the tables. In the near-extremal entry $\mathfrak{a}=14.04$, corresponding to $a_K=0.99$, the low-lying oscillatory modes are much smaller. The real part increases with $\ell$, while the damping rates of these near-extremal fundamental modes are all of order $3\times10^{-2}$ in the $\Omega=M\omega$ normalisation. The spectral method also returns additional axial overtones not listed in the WKB comparison tables. In Tables~\ref{table_s2_a0_25_and_a0_5}, \ref{table_s2_a0_75_and_a1_0}, and \ref{table_s2_a1_75_and_a0_99}, several entries with $n=5$ and $n=6$ have no WKB counterpart in \cite{SalehASS2016}, while the spectral method still produces stable roots. These entries extend the available axial effective source data beyond the third-order WKB results. We next consider the purely imaginary axial modes. They are reported in Tables~\ref{table_s2_a0_25_and_a0_5_overdamped}, \ref{table_s2_a0_75_and_a1_0_overdamped}, and \ref{table_s2_a1_25_and_a0_99_overdamped}. For moderate deformation, the detected purely imaginary modes are highly damped. At $\mathfrak{a}=0.25$, $\ell=2$, the first seven purely imaginary roots display spacings close to $i/4$. The same trend is observed for $\mathfrak{a}=0.5$, and $\ell=2$. However, the spacing between consecutive roots retained by the stability criterion is not always a single elementary step. For example, at $\mathfrak{a}=0.75$, $\ell=2$, the first two detected purely imaginary roots are $-20.2322i$, and $-20.9792i$, so that $\Delta\Omega=0.7470i$, which is approximately three times $i/4$. Similarly, at $\mathfrak{a}=1$, and $\ell=4$, the first spacing is $\Delta\Omega=0.7441i$. We therefore interpret the data as evidence for an underlying spacing scale close to $i/4$, with possible missing or unresolved intermediate members in the retained numerical sequence, rather than as a strictly uniform ladder at all parameter values. The near-extremal purely imaginary spectrum is much cleaner. At $a_K=0.99$, Table~\ref{table_s2_a1_25_and_a0_99_overdamped} gives, for $\ell=2$, $-0.2492i$, $-0.4994i$, $-0.7496i$, $-0.9997i$, $-1.2498i$, $-1.4998i$, and $-1.7498i$, while for $\ell=3$ it gives $-0.2508i$, $-0.4998i$, $-0.7498i$, $-0.9998i$, $-1.2498i$, $-1.4998i$, $-1.7499i$. The corresponding $\ell=4$ and $\ell=5$ sequences are also close to the same ladder. Thus, in the near-extremal regime, the axial purely imaginary roots approach the pattern
\begin{equation}
    \Omega_n\simeq-\frac{i}{4}(n+1),
    \qquad n=0,1,2,\ldots.
\end{equation}
The small residual dependence on $\ell$ is most visible in the first root and becomes negligible for the higher entries of the detected sequence. It is useful to compare this near-extremal behaviour with the scalar and electromagnetic sectors. At $a_K=0.99$, the minimally coupled scalar sector gives $-0.2520i$, $-0.5008i$, $-0.7505i$, $-1.0003i$, whereas the electromagnetic sector gives $-0.2505i$, $-0.5001i$, $-0.7500i$, $-1.0000i$. The quoted axial gravitational values are very close to both. This indicates an approximate near-extremal isospectrality of the purely imaginary overdamped ladder among the minimally coupled scalar, electromagnetic, and axial effective source sectors. The statement should not be extended to the full spectrum, since the low-lying oscillatory modes of the different sectors remain distinct. Moreover, the current non-minimally coupled scalar table does not provide a corresponding purely imaginary sequence at $a_K=0.99$ because at $a_K=0.9$ the detected non-minimally coupled sequence begins at $-2.7414i$, whereas the minimally coupled scalar and electromagnetic sequences begin near $-2.50i$. Thus, the available data do not support an all-sector isospectrality statement, including the non-minimally coupled scalar field. They support only a common near-extremal spacing structure for those sectors in which the overdamped ladder is resolved.

\section{Conclusions and outlook}\label{sec:conclusions}

In this work, we have carried out a high-precision spectral analysis of QNMs of the KSQC black hole. After factoring out the ingoing behaviour at the event horizon and the outgoing behaviour at spatial infinity, the perturbation equations were reduced to quadratic eigenvalue problems for the dimensionless frequency $\Omega=M\omega$. The resulting matrix pencils were solved with a Chebyshev collocation method, and the physical roots were selected by requiring stability under changes of the spectral resolution. This provides a global numerical approach to the QNM problem and allows one to access not only the low-lying oscillatory modes, but also highly damped and purely imaginary branches that are difficult to isolate with local approximation methods. We have applied this framework to four sectors: minimally coupled scalar perturbations, electromagnetic perturbations, non-minimally coupled scalar perturbations with conformal coupling, and an axial effective source Regge--Wheeler-type gravitational sector. The scalar and electromagnetic sectors are genuine test-field problems on the fixed KSQC geometry. The non-minimally coupled scalar sector incorporates the $R/6$ curvature term associated with the non-Ricci-flat character of the spacetime. The axial gravitational sector requires more care. Since the KSQC geometry is not Ricci flat, one cannot consistently impose the vacuum equation $\delta R_{\mu\nu}=0$. For this reason, we have interpreted the axial potential as an effective source Regge--Wheeler model, obtained by regarding the background as supported by an effective anisotropic stress tensor and by assuming that this source carries no independent axial degree of freedom. Thus, the axial sector studied here should not be identified with the full gravitational perturbation problem of a specified four-dimensional quantum gravity action. A first outcome of the analysis is the validation of the spectral implementation against the available literature. After accounting for the different normalisations used in the previous literature, our results agree with the time-domain data of \cite{KonoplyaPLB2020} in the common range where such data are available. In the scalar and electromagnetic sectors, the agreement with time-domain results is significantly better than with WKB-only data in the strongly deformed, low-multipole regime. This is consistent with the known limitations of WKB methods when the overtone number is not small compared with the multipole number, or when the effective potential is strongly modified. Comparisons with the third-order WKB data of \cite{SalehASS2014, WangJAA2017, SalehASS2016} show very good agreement for fundamental modes and for higher multipoles, while the discrepancies increase for low $\ell$ and higher overtones. The comparison with the Mashhoon and AIM results of \cite{ZhangIJP2023} provides an additional independent benchmark in the scalar sector and confirms that the spectral method reproduces the more accurate numerical estimates in the regimes where the methods overlap. The dependence of the oscillatory modes on the deformation parameter is sector-dependent, but a common qualitative trend is visible in the mass-normalised frequency. In most of the parameter ranges considered, increasing the deformation reduces both the real part of the fundamental mode and the magnitude of its imaginary part. Thus, in this normalisation, the dominant ringdown becomes less oscillatory and less damped as the deformation grows. This statement should not be confused with the behaviour of the horizon normalised frequency $r_h\omega$, because the conversion factor between the latter and  $M\omega$ depends on the deformation parameter. For higher overtones, especially at large $a_K$, the behaviour is more intricate. In several tables, we observe non-monotonic changes in the real parts of the retained roots, indicating a rearrangement of the overtone structure in the strongly deformed regime. We interpret this cautiously, since identifying individual overtone branches can become delicate when several roots approach one another or when the number of stable roots retained by the convergence criterion changes. A central feature revealed by the spectral method is the appearance of purely imaginary, overdamped roots. Such roots occur in the scalar, electromagnetic, non-minimally coupled scalar, and axial effective-source sectors across several regions of parameter space. With the convention $e^{-i\omega t}$, these modes are damped because $\Im{\Omega}<0$. Their spacing is often controlled by the scale $M\kappa=1/4$ where $\kappa$ is the surface gravity of the Schwarzschild black hole. However, the detected sequences are not always strictly uniformly spaced. In some regions, the spacing between two consecutive retained roots is close to two or three times the elementary value $i/4$, and in other regions, purely imaginary roots are not detected even at high numerical precision. Thus, the data are best interpreted as evidence for an underlying surface gravity-controlled overdamped structure, with possible missing or unresolved intermediate members in the retained numerical sequence.  The status of these purely imaginary roots deserves some caution. Purely imaginary QNMs are known to occur in black-hole spectra and are not, by themselves, a numerical pathology. At the same time, the KSQC spacetime is asymptotically flat, and the frequency-domain Green function may contain non-pole contributions associated with late-time tails and branch cuts along the negative imaginary axis. This distinction is familiar from the Schwarzschild problem, where the frequency-domain Green function contains both QNM pole contributions and a branch-cut integral along the negative imaginary axis, the latter being responsible for the late-time tail~\cite{Leaver1986}. For this reason, one must distinguish between isolated poles and roots that merely sample a cut. To test this issue within the present spectral formulation, we performed an additional wide-resolution drift test for the minimally coupled scalar monopole, electromagnetic dipole, and axial effective-source quadrupole sectors at $a_K=0.99$. Instead of using only the triplet $N\in\{190,195,200\}$, we recomputed the imaginary-axis branches using resolution triplets $(N-5,N,N+5)$, with $N\in\{125,140,\ldots,260\}$. For each retained imaginary-axis root, we assigned the nearest ladder integer $m=\operatorname{round}[4(-\operatorname{Im}\Omega)]$ and monitored the residual $4(-\operatorname{Im}\Omega)-m$ as a function of $N$. The resulting drift test shows that fixed ladder levels remain close to resolution-independent plateaus, while increasing $N$ mainly extends the branch to larger damping rather than moving the previously resolved levels. This behaviour supports the interpretation of the displayed imaginary-axis roots as stable spectral features of the Chebyshev boundary-value problem, rather than as roots whose locations drift appreciably with the discretisation. Nevertheless, this test is not a complete analytic proof that every member of the sequence is an isolated pole of the continued Green function. A definitive pole-versus-continuum classification would still benefit from independent diagnostics, such as a Wronskian search, a continued-fraction or shooting computation, a resolvent analysis, or a systematic study of the dependence on the compactification map.

The near-extremal regime $a_K\to 1^{-}$ exhibits a particularly simple and suggestive pattern. In the minimally coupled scalar, electromagnetic, and axial effective-source sectors, the detected purely imaginary modes approach an approximately equally spaced ladder of the form
\begin{equation}
\Omega_n\simeq -\frac{i}{4}(n+1),\qquad n=0,1,2,\ldots .
\end{equation}
Notice that the near-extremal scalar and electromagnetic oscillatory modes are not isospectral, nor are the oscillatory modes of the axial sector identical to them. Nevertheless, the overdamped purely imaginary branches in the sectors where they are resolved approach one another very closely. This suggests an approximate near-extremal degeneracy of the detected overdamped ladders, rather than a full isospectrality of the complete QNM spectra. The non-minimally coupled scalar data are less conclusive in this respect, because the corresponding near-extremal purely imaginary sequence is not resolved in the same way across the full range of parameters considered. The interpretation of this ladder requires care. The limit $a_K\to 1^{-}$ is not a regular extremal black-hole limit analogous to the extremal Reissner--Nordstr\"om case. In the horizon-normalised convention, $a_K=a/r_h$, taking $a_K\to 1^{-}$ at fixed $r_h$ sends $M\to 0$, while the curvature singularity at $r=a$ approaches the event horizon. The surface gravity behaves as $\kappa_p=1/(4M)$, and therefore diverges. Thus a finite value of the mass-normalised frequency $\Omega=M\omega$ corresponds to $r_h\omega\to\infty$, or equivalently to a physical frequency $\omega$ that diverges in proportion to $\kappa_p$. In this sense, the ladder is attached to the black-hole family only for $a_K<1$, where a regular event horizon is still present. The strict limiting object is instead a singular high-temperature degeneration of the geometry. The finite statement suggested by the data is the surface-gravity-scaled relation $\omega_n/\kappa_p=4\Omega_n\simeq -i(n+1)$, not a finite-frequency QNM spectrum of a regular extremal KSQC black hole.

Moreover, although the wide-resolution drift test supports the stability of the plateau-forming imaginary-axis branches, they should also be checked by methods independent of the present Chebyshev discretisation. A direct Wronskian calculation, a continued fraction or shooting approach, or a pseudospectral resolvent analysis would help determine which of the detected roots correspond to isolated QNM poles and which, if any, approximate a branch-cut contribution. Second, the dependence of the overdamped roots on the compactification map and on the choice of collocation grid should be explored systematically. Such tests are particularly important in sectors where double or triple spacing appears in the retained sequence. Third, a dedicated near-extremal analysis is needed to understand the origin of the $-i(n+1)/4$ ladder and its relation to the singular scaling limit $a_K\to1^{-}$. The most important theoretical open problem concerns the full gravitational perturbation theory of the KSQC geometry.The axial effective source model studied here is a closed Regge--Wheeler-type metric sector obtained by imposing the model-defining condition $\delta T^{ind}_{ax}=0$. This condition is not implied by the background equation of state $p_r=-\rho$. Rather, it excludes independent transverse momentum or axial anisotropic-stress perturbations of the effective source. If such perturbations are allowed, the axial metric equation becomes part of a coupled metric-source system, and the corresponding frequencies are model-dependent. However, the polar sector is expected to couple to perturbations of the source or of the effective degrees of freedom that generate the KSQC background. A complete gravitational analysis would therefore require a fully specified effective action or source model, as well as the derivation of the coupled perturbation equations. This lies beyond the scope of the present work, but it is essential for assigning a unique gravitational QNM spectrum to the quantum-corrected geometry. 

\subsection*{Code availability}

\noindent The codes used to assemble the spectral matrices and to post-process the quadratic eigenvalue problems are available at \url{https://github.com/dutykh/KS-quantum/}

\begin{table}[t]
\centering
\caption{\textit{QNMs of massless scalar perturbations ($s=0$) of the KSQC black hole for $\ell=0$ and several values of the deformation parameter $a_K$. We compare our spectral method results with $N=200$ Chebyshev polynomials (final column) against the seventh-order WKB and the time-domain method adopted by \cite{KonoplyaPLB2020}. Entries labelled N/A indicate unavailable data. The notation 'SM' stands for Spectral Method.}}
\label{table_s0_L0}
\setlength\tabcolsep{0.1cm}
\def\arraystretch{1.5}
\begin{tabular}{@{}|c| c| c| c| c| c|c|c|@{}} 
\hline
$a_K$  & $\mathfrak{a}$ & $n$ & $\Omega_{WKB}$ \cite{KonoplyaPLB2020} & $\Omega_{TD}$ \cite{KonoplyaPLB2020} & $\Omega_{SM}$ \\ [0.5ex] 
\hline
$0$    & $0$                  & $0$ & \mbox{N/A}                      & \mbox{N/A}                           & $0.1105-0.1049i$\\ 
       &                     & $1$ & \mbox{N/A}           & \mbox{N/A}         & $0.0861-0.3481i$\\
       &                     & $2$ & \mbox{N/A}           & \mbox{N/A}         & $0.0757-0.6011i$\\ 
       \hline
$0.01$ & $\frac{2\sqrt{1111}}{3333}$ & $0$ & $0.111806-0.105410i$ & $0.11055-0.10528i$ & $0.1105-0.1049i$ \\
       &                     & $1$ & \mbox{N/A}           & \mbox{N/A}                 & $0.0861-0.3481i$ \\
       &                     & $2$ & \mbox{N/A}           & \mbox{N/A}                 & $0.0757-0.6011i$ \\
       \hline
$0.1$  & $\frac{2\sqrt{11}}{33}$     & $0$ & $0.111420-0.105331i$ & $0.11018-0.10519i$ & $0.1101-0.1048i$ \\
       &                     & $1$ & \mbox{N/A}           & \mbox{N/A}                 & $0.0857-0.3479i$ \\
       &                     & $2$ & \mbox{N/A}           & \mbox{N/A}                 & $0.0753-0.6009i$ \\
       \hline
$0.2$  & $\frac{\sqrt{6}}{6}$        & $0$ & $0.110223-0.105085i$ & $0.10894-0.10494i$ & $0.1089-0.1045i$ \\
       &                     & $1$ & \mbox{N/A}           & \mbox{N/A}                 & $0.0843-0.3474i$ \\
       &                     & $2$ & \mbox{N/A}           & \mbox{N/A}                 & $0.0738-0.6002i$ \\
       \hline
$0.3$  & $\frac{6\sqrt{91}}{91}$     & $0$ & $0.108143-0.104638i$ & $0.10702-0.10430i$ & $0.1068-0.1041i$ \\
       &                     & $1$ & \mbox{N/A}           & \mbox{N/A}                 & $0.0819-0.3465i$ \\
       &                     & $2$ & \mbox{N/A}           & \mbox{N/A}                 & $0.0713-0.5990i$ \\
       \hline
$0.4$  & $\frac{4\sqrt{21}}{21}$     & $0$ & $0.105051-0.103937i$ & $0.10398-0.10347i$ & $0.1037-0.1033i$ \\
       &                     & $1$ & \mbox{N/A}           & \mbox{N/A}                 & $0.0785-0.3451i$ \\
       &                     & $2$ & \mbox{N/A}           & \mbox{N/A}                 & $0.0678-0.5968i$ \\    
       \hline
$0.5$  & $\frac{2\sqrt{3}}{3}$       & $0$ & $0.100806-0.102937i$ & $0.09964-0.10228i$ & $0.0994-0.1021i$ \\
       &                     & $1$ & \mbox{N/A}           & \mbox{N/A}                 & $0.0738-0.3427i$ \\
       &                     & $2$ & \mbox{N/A}           & \mbox{N/A}                 & $0.0633-0.5932i$ \\     
       \hline
$0.6$  & $\frac{3}{2}$               & $0$ & $0.095530-0.101832i$ & $0.09349-0.10077i$ & $0.0933-0.1003i$ \\
       &                     & $1$ & \mbox{N/A}           & \mbox{N/A}                 & $0.0676-0.3387i$ \\
       &                     & $2$ & \mbox{N/A}           & \mbox{N/A}                 & $0.0580-0.5870i$ \\
       \hline
$0.7$  & $\frac{14\sqrt{51}}{51}$    & $0$ & $0.091066-0.081670i$ & \mbox{N/A}         & $0.0847-0.0972i$ \\
       &                     & $1$ & \mbox{N/A}           & \mbox{N/A}                 & $0.0596-0.3320i$ \\
       &                     & $2$ & \mbox{N/A}           & \mbox{N/A}                 & $0.0523-0.5764i$ \\
       &                     & $3$ & \mbox{N/A}           & \mbox{N/A}                 & $0.0512-0.8206i$ \\
       &                     & $4$ & \mbox{N/A}           & \mbox{N/A}                 & $0.0537-1.0649i$ \\
       \hline
$0.8$  & $\frac{8}{3}$               & $0$ & $0.106513-0.123592i$ & \mbox{N/A}         & $0.0721-0.0917i$ \\
       &                     & $1$ & \mbox{N/A}           & \mbox{N/A}                 & $0.0497-0.3194i$ \\
       &                     & $2$ & \mbox{N/A}           & \mbox{N/A}                 & $0.0479-0.5575i$ \\
       &                     & $3$ & \mbox{N/A}           & \mbox{N/A}                 & $0.0657-1.2742i$ \\
       &                     & $4$ & \mbox{N/A}           & \mbox{N/A}                 & $0.0740-1.5120i$ \\
       \hline
$0.85$ & $\frac{34\sqrt{111}}{111}$  & $0$ & \mbox{N/A}           & \mbox{N/A}         & $0.0631-0.0869i$ \\
       &                     & $1$ & \mbox{N/A}           & \mbox{N/A}                 & $0.0444-0.3089i$ \\
       &                     & $2$ & \mbox{N/A}           & \mbox{N/A}                 & $0.0469-0.5425i$ \\
       &                     & $3$ & \mbox{N/A}           & \mbox{N/A}                 & $0.0622-1.0113i$ \\
       &                     & $4$ & \mbox{N/A}           & \mbox{N/A}                 & $0.0704-1.2441i$ \\
     [0.5ex] 
 \hline
 \end{tabular}
\end{table}

\begin{table}[t]
\centering
\caption{\textit{QNMs of massless scalar perturbations ($s=0$) of the KSQC black hole for $\ell=0$ and several values of the deformation parameter $a_K$. We compare our spectral method results with $N=200$ Chebyshev polynomials (final column) against the Mashhoon method (MM) and the asymptotic iteration method (AIM) used by \cite{ZhangIJP2023}. The notation 'SM' stands for Spectral Method.}}
\label{table_s0_L0_Zhang}
\setlength\tabcolsep{0.1cm}
\def\arraystretch{1.5}
\begin{tabular}{@{}|c| c| c| c| c| c|c|c|@{}} 
\hline
$a_K$  & $n$ & $\Omega_{MM}$ \cite{ZhangIJP2023} & $\Omega_{AIM}$ \cite{ZhangIJP2023} & $\Omega_{SM}$ \\ [0.5ex] 
\hline
$0.01$ & $0$ & $0.114816-0.114819i$              & $0.110301-0.104634i$               & $0.1105-0.1049i$ \\
       \hline
$0.1$  & $0$ & $0.114424-0.114783i$              & $0.109915-0.104551i$               & $0.1101-0.1048i$ \\
       \hline
$0.2$  & $0$ & $0.113218-0.114665i$              & $0.108723-0.104289i$               & $0.1089-0.1045i$ \\
       \hline
$0.3$  & $0$ & $0.111150-0.114435i$              & $0.106679-0.103820i$               & $0.1068-0.1041i$ \\
       \hline
$0.4$  & $0$ & $0.108120-0.114033i$              & $0.103688-0.103086i$               & $0.1037-0.1033i$ \\
       \hline
$0.5$  & $0$ & $0.103972-0.113346i$              & $0.099589-0.101984i$               & $0.0994-0.1021i$ \\
       \hline
$0.6$  & $0$ & $0.098464-0.112149i$              & $0.094130-0.100321i$               & $0.0933-0.1003i$ \\
       \hline
$0.7$  & $0$ & $0.091229-0.109960i$              & $0.086567-0.095951i$               & $0.0847-0.0972i$ \\
     [0.5ex] 
 \hline
 \end{tabular}
\end{table}

\begin{table}
\centering
\caption{Candidate purely imaginary QNMs for scalar perturbations of a KSQC black hole for $\ell=0$ and several values of the deformation parameter $a_K$. The corresponding results are obtained through our SM, utilising $200$ polynomials with a precision of $200$ digits. In this context, $\Omega$ and $n$ represent the dimensionless frequency and the corresponding overtone, respectively, while $\Delta\Omega=\Omega_n-\Omega_{n+1}$.  Entries labelled N/A indicate unavailable data.}
\label{scalargoverdamped}
\vspace*{1em}
\begin{tabular}{||c|c|c|c|c|c|c|c|c||}
\hline\hline
$a_K$  & $n$ & $\Omega$                 &$\Delta\Omega$  & $a_K$ & $n$ & $\Omega$  &$\Delta\Omega$\\ [0.5ex]
\hline\hline
$0.01$ & $0$ & $0.0000-18.1166i$            & \mbox{N/A}     & $0.5$  & $0$ & \mbox{N/A}             &\mbox{N/A}\\
       & $1$ & $0.0000-18.3667i$            & $0.2501i$      &        & $1$ & \mbox{N/A}             &\mbox{N/A}\\
       & $2$ & $0.0000-18.6164i$            & $0.2497i$      &        & $2$ & \mbox{N/A}             &\mbox{N/A}\\
       & $3$ & $0.0000-18.8667i$            & $0.2503i$      &        & $3$ & \mbox{N/A}             &\mbox{N/A}\\
       & $4$ & $0.0000-19.1166i$            & $0.2499i$      &        & $4$ & \mbox{N/A}             &\mbox{N/A}\\ 
       & $5$ & $0.0000-19.3667i$            & $0.2501i$      &        & $5$ & \mbox{N/A}             &\mbox{N/A}\\
\hline
$0.1$  & $0$ & $0.0000-18.3639i$            & \mbox{N/A}     & $0.6$  & $0$ & \mbox{N/A}             &\mbox{N/A}\\
       & $1$ & $0.0000-18.6139i$            & $0.2500i$      &        & $1$ & \mbox{N/A}             &\mbox{N/A}\\
       & $2$ & $0.0000-18.8642i$            & $0.2503i$      &        & $2$ & \mbox{N/A}             &\mbox{N/A}\\
       & $3$ & $0.0000-19.1140i$            & $0.2498i$      &        & $3$ & \mbox{N/A}             &\mbox{N/A}\\
       & $4$ & $0.0000-19.3642i$            & $0.2502i$      &        & $4$ & \mbox{N/A}             &\mbox{N/A}\\ 
       & $5$ & $0.0000-19.6142i$            & $0.2500i$      &        & $5$ & \mbox{N/A}             &\mbox{N/A}\\
\hline
$0.2$  & $0$ & $0.0000-18.8500i$            & \mbox{N/A}     & $0.7$  & $0$ & \mbox{N/A}             &\mbox{N/A}\\
       & $1$ & $0.0000-19.0999i$            & $0.2499i$      &        & $1$ & \mbox{N/A}             &\mbox{N/A}\\
       & $2$ & $0.0000-19.3499i$            & $0.2500i$      &        & $2$ & \mbox{N/A}             &\mbox{N/A}\\
       & $3$ & $0.0000-19.5996i$            & $0.2497i$      &        & $3$ & \mbox{N/A}             &\mbox{N/A}\\
       & $4$ & $0.0000-19.8495i$            & $0.2499i$      &        & $4$ & \mbox{N/A}             &\mbox{N/A}\\
       & $5$ & $0.0000-20.0994i$            & $0.2499i$      &        & $5$ & \mbox{N/A}             &\mbox{N/A}\\
\hline
$0.3$  & $0$ & $0.0000-19.8065i$            & \mbox{N/A}     & $0.8$  & $0$ & $0.0000-10.0006i$      &\mbox{N/A}\\
       & $1$ & $0.0000-20.0558i$            & $0.2493i$      &        & $1$ & $0.0000-10.2504i$      &$0.2498i$\\
       & $2$ & $0.0000-20.3053i$            & $0.2495i$      &        & $2$ & $0.0000-10.5004i$      &$0.2500i$\\
       & $3$ & $0.0000-20.5544i$            & $0.2491i$      &        & $3$ & $0.0000-10.7504i$      &$0.2500i$\\
       & $4$ & $0.0000-20.8037i$            & $0.2493i$      &        & $4$ & $0.0000-11.0004i$      &$0.2500i$\\ 
       & $5$ & $0.0000-21.0530i$            & $0.2493i$      &        & $5$ & $0.0000-11.2504i$      &$0.2500i$\\
\hline
$0.4$  & $0$ & $0.0000-21.6982i$            & \mbox{N/A}     & $0.85$ & $0$ & $0.0000-5.75055i$      &\mbox{N/A}\\
       & $1$ & $0.0000-21.9459i$            & $0.2477i$      &        & $1$ & $0.0000-6.00053i$      &$0.2500i$\\
       & $2$ & $0.0000-22.1937i$            & $0.2478i$      &        & $2$ & $0.0000-6.25051i$      &$0.2500i$\\
       & $3$ & $0.0000-22.4422i$            & $0.2485i$      &        & $3$ & $0.0000-6.50049i$      &$0.2500i$\\
       & $4$ & $0.0000-22.6901i$            & $0.2479i$      &        & $4$ & $0.0000-6.75047i$      &$0.2500i$\\ 
       & $5$ & $0.0000-22.9382i$            & $0.2481i$      &        & $5$ & $0.0000-7.00046i$      &$0.2500i$\\
       [1ex]
 \hline\hline 
 \end{tabular}
\end{table}

\begin{table}[t]
\centering
\caption{\textit{QNMs of massless scalar perturbations ($s=0$) of the KSQC black hole for several values of $\ell$ and the deformation parameter $\mathfrak{a}=a/M$. We compare our spectral method results with $N=200$ Chebyshev polynomials (final column) against the third-order WKB employed by \cite{SalehASS2014}. Entries labelled N/A indicate unavailable data. The notation 'SM' stands for Spectral Method.}}
\label{table_s0_a0_25_and_a0_5}
\setlength\tabcolsep{0.1cm}
\def\arraystretch{1.5}
\begin{tabular}{@{}|c| c| c| c| c| c|c|c|c|c|c|c|@{}} 
\hline
$a$    & $a_K$   & $\ell$ & $n$ & $\Omega_{WKB}$ \cite{SalehASS2014} & $\Omega_{SM}$ &$a$    & $a_K$   & $\ell$ & $n$ & $\Omega_{WKB}$ \cite{SalehASS2014} & $\Omega_{SM}$ \\ [0.5ex] 
\hline
$0.25$ & $0.12$  & $2$    & $0$ & $0.4807-0.0966i$   & $0.4811-0.0966i$ & $0.5$ & $0.24$ & $2$ & $0$ & $0.4734-0.0962i$& $0.4739-0.0961i$\\ 
       &         &        & $1$ & $0.4602-0.2954i$   & $0.4612-0.2951i$ &       &        &     & $1$ & $0.4530-0.2940i$& $0.4536-0.2938i$\\
       &         &        & $2$ & $0.4288-0.5027i$   & $0.4277-0.5079i$ &       &        &     & $2$ & $0.4207-0.5006i$& $0.4196-0.5060i$\\ 
       &         &        & $3$ & $0.3895-0.7149i$   & $0.3909-0.7374i$ &       &        &     & $3$ & $0.3807-0.7121i$& $0.3823-0.7351i$\\
       &         &        & $4$ & $0.3421-0.9294i$   & $0.3582-0.9791i$ &       &        &     & $4$ & $0.3324-0.9259i$& $0.3492-0.9768i$\\ 
       &         &        & $5$ & \mbox{N/A}         & $0.3317-1.2276i$ &       &        &     & $5$ & \mbox{N/A}      & $0.3226-1.2251i$\\
       &         &        & $6$ & \mbox{N/A}         & $0.3106-1.4791i$ &       &        &     & $6$ & \mbox{N/A}      & $0.3014-1.4765i$\\
       \hline
       &         & $3$    & $0$ & $0.6717-0.0963i$   & $0.6719-0.0963i$ &       &        & $3$ & $0$ & $0.6616-0.0959i$& $0.6618-0.0958i$\\  
       &         &        & $1$ & $0.6568-0.2919i$   & $0.6571-0.2918i$ &       &        &     & $1$ & $0.6465-0.2905i$& $0.6467-0.2904i$\\
       &         &        & $2$ & $0.6311-0.4934i$   & $0.6299-0.4953i$ &       &        &     & $2$ & $0.6203-0.4911i$& $0.6191-0.4931i$\\ 
       &         &        & $3$ & $0.5983-0.7000i$   & $0.5948-0.7103i$ &       &        &     & $3$ & $0.5870-0.6970i$& $0.5835-0.7076i$\\
       &         &        & $4$ & $0.5595-0.9099i$   & $0.5575-0.9376i$ &       &        &     & $4$ & $0.5475-0.9060i$& $0.5456-0.9345i$\\ 
       &         &        & $5$ & \mbox{N/A}         & $0.5223-1.1750i$ &       &        &     & $5$ & \mbox{N/A}      & $0.5100-1.1718i$\\
       &         &        & $6$ & \mbox{N/A}         & $0.4913-1.4193i$ &       &        &     & $6$ & \mbox{N/A}      & $0.4787-1.4160i$\\
       \hline
       &         & $4$    & $0$ & $0.8629-0.0962i$   & $0.8629-0.0962i$ &       &        & $4$ & $0$ & $0.8499-0.0957i$& $0.8500-0.0957i$\\ 
       &         &        & $1$ & $0.8511-0.2904i$   & $0.8513-0.2904i$ &       &        &     & $1$ & $0.8380-0.2890i$& $0.8381-0.2889i$\\
       &         &        & $2$ & $0.8298-0.4887i$   & $0.8290-0.4896i$ &       &        &     & $2$ & $0.8163-0.4864i$& $0.8155-0.4872i$\\ 
       &         &        & $3$ & $0.8016-0.6916i$   & $0.7984-0.6965i$ &       &        &     & $3$ & $0.7876-0.6884i$& $0.7844-0.6935i$\\
       &         &        & $4$ & $0.7680-0.8980i$   & $0.7627-0.9128i$ &       &        &     & $4$ & $0.7535-0.8940i$& $0.7481-0.9093i$\\
       &         &        & $5$ & \mbox{N/A}         & $0.7252-1.1388i$ &       &        &     & $5$ & \mbox{N/A}      & $0.7101-1.1350i$\\
       &         &        & $6$ & \mbox{N/A}         & $0.6888-1.3732i$ &       &        &     & $6$ & \mbox{N/A}      & $0.6732-1.3692i$\\
       \hline
       &         & $5$    & $0$ & $1.0541-0.0960i$   & $1.0541-0.0962i$ &       &        & $5$ & $0$ & $1.0383-0.0959i$& $1.0383-0.0957i$\\ 
       &         &        & $1$ & $1.0444-0.2897i$   & $1.0445-0.2897i$ &       &        &     & $1$ & $1.0287-0.2886i$& $1.0285-0.2882i$\\
       &         &        & $2$ & $1.0264-0.4862i$   & $1.0259-0.4866i$ &       &        &     & $2$ & $1.0107-0.4842i$& $1.0096-0.4842i$\\ 
       &         &        & $3$ & $1.0017-0.6863i$   & $0.9994-0.6889i$ &       &        &     & $3$ & $0.9860-0.6832i$& $0.9827-0.6857i$\\
       &         &        & $4$ & $0.9720-0.8900i$   & $0.9670-0.8982i$ &       &        &     & $4$ & $0.9559-0.8854i$& $0.9497-0.8944i$\\
       &         &        & $5$ & \mbox{N/A}         & $0.9308-1.1154i$ &       &        &     & $5$ & \mbox{N/A}      & $0.9130-1.1111i$\\
       &         &        & $6$ & \mbox{N/A}         & $0.8933-1.3406i$ &       &        &     & $6$ & \mbox{N/A}      & $0.8749-1.3360i$\\
     [0.5ex] 
 \hline
 \end{tabular}
\end{table}

\begin{table}[t]
\centering
\caption{\textit{QNMs of massless scalar perturbations ($s=0$) of the KSQC black hole for several values of $\ell$ and the deformation parameter $\mathfrak{a}=a/M$. We compare our spectral method results with $N=200$ Chebyshev polynomials (final column) against the third-order WKB employed by \cite{SalehASS2014}. Entries labelled N/A indicate unavailable data. The notation 'SM' stands for Spectral Method.}}
\label{table_s0_a0_75_and_a1}
\setlength\tabcolsep{0.1cm}
\def\arraystretch{1.5}
\begin{tabular}{@{}|c| c| c| c| c| c|c|c|c|c|c|c|@{}} 
\hline
$a$    & $a_K$   & $\ell$ & $n$ & $\Omega_{WKB}$ \cite{SalehASS2014} & $\Omega_{SM}$ &$a$    & $a_K$   & $\ell$ & $n$ & $\Omega_{WKB}$ \cite{SalehASS2014} & $\Omega_{SM}$ \\ [0.5ex] 
\hline
$0.75$ & $0.35$  & $2$    & $0$ & $0.4622-0.0954i$   & $0.4626-0.0953i$ & $1$   & $0.45$ & $2$ & $0$ & $0.4479-0.0943i$   & $0.4484-0.0942i$\\ 
       &         &        & $1$ & $0.4412-0.2918i$   & $0.4419-0.2916i$ &       &        &     & $1$ & $0.4263-0.2888i$   & $0.4270-0.2886i$\\
       &         &        & $2$ & $0.4082-0.4972i$   & $0.4070-0.5028i$ &       &        &     & $2$ & $0.3924-0.4924i$   & $0.3911-0.4984i$\\ 
       &         &        & $3$ & $0.3671-0.7074i$   & $0.3689-0.7315i$ &       &        &     & $3$ & $0.3501-0.7009i$   & $0.3522-0.7264i$\\
       &         &        & $4$ & $0.3173-0.9200i$   & $0.3355-0.9729i$ &       &        &     & $4$ & $0.2986-0.9119i$   & $0.3184-0.9674i$\\
       &         &        & $5$ & \mbox{N/A}         & $0.3086-1.2209i$ &       &        &     & $5$ & \mbox{N/A}         & $0.2914-1.2149i$\\
       &         &        & $6$ & \mbox{N/A}         & $0.2873-1.4719i$ &       &        &     & $6$ & \mbox{N/A}         & $0.2702-1.4654i$\\
       \hline
       &         & $3$    & $0$ & $0.6459-0.0951i$   & $0.6461-0.0950i$ &       &        & $3$ & $0$ & $0.6261-0.0940i$   & $0.6262-0.0940i$\\  
       &         &        & $1$ & $0.6304-0.2881i$   & $0.6307-0.2881i$ &       &        &     & $1$ & $0.6101-0.2850i$   & $0.6104-0.2849i$\\
       &         &        & $2$ & $0.6037-0.4874i$   & $0.6023-0.4895i$ &       &        &     & $2$ & $0.5826-0.4824i$   & $0.5811-0.4846i$\\ 
       &         &        & $3$ & $0.5695-0.6919i$   & $0.5659-0.7031i$ &       &        &     & $3$ & $0.5474-0.6851i$   & $0.5437-0.6970i$\\
       &         &        & $4$ & $0.5290-0.8997i$   & $0.5273-0.9295i$ &       &        &     & $4$ & $0.5057-0.8910i$   & $0.5042-0.9227i$\\ 
       &         &        & $5$ & \mbox{N/A}         & $0.4911-1.1665i$ &       &        &     & $5$ & \mbox{N/A}         & $0.4674-1.1591i$\\
       &         &        & $6$ & \mbox{N/A}         & $0.4595-1.4104i$ &       &        &     & $6$ & \mbox{N/A}         & $0.4355-1.4026i$\\
       \hline
       &         & $4$    & $0$ & $0.8298-0.0949i$   & $0.8299-0.0949i$ &       &        & $4$ & $0$ & $0.8043-0.0939i$   & $0.8044-0.0938i$\\ 
       &         &        & $1$ & $0.8176-0.2866i$   & $0.8177-0.2866i$ &       &        &     & $1$ & $0.7917-0.2834i$   & $0.7918-0.2834i$\\
       &         &        & $2$ & $0.7954-0.4826i$   & $0.7945-0.4835i$ &       &        &     & $2$ & $0.7689-0.4774i$   & $0.7680-0.4784i$\\ 
       &         &        & $3$ & $0.7660-0.6832i$   & $0.7626-0.6886i$ &       &        &     & $3$ & $0.7387-0.6761i$   & $0.7351-0.6819i$\\
       &         &        & $4$ & $0.7311-0.8874i$   & $0.7255-0.9036i$ &       &        &     & $4$ & $0.7028-0.8785i$   & $0.6970-0.8957i$\\
       &         &        & $5$ & \mbox{N/A}         & $0.6867-1.1287i$ &       &        &     & $5$ & \mbox{N/A}         & $0.6572-1.1200i$\\
       &         &        & $6$ & \mbox{N/A}         & $0.6492-1.3625i$ &       &        &     & $6$ & \mbox{N/A}         & $0.6191-1.3533i$\\
       \hline
       &         & $5$    & $0$ & $1.0137-0.0949i$   & $1.0138-0.0949i$ &       &        & $5$ & $0$ & $0.9826-0.0938i$   & $0.9827-0.0938i$\\ 
       &         &        & $1$ & $1.0037-0.2858i$   & $1.0037-0.2858i$ &       &        &     & $1$ & $0.9723-0.2826i$   & $0.9723-0.2826i$\\
       &         &        & $2$ & $0.9849-0.4799i$   & $0.9843-0.4803i$ &       &        &     & $2$ & $0.9529-0.4747i$   & $0.9523-0.4751i$\\ 
       &         &        & $3$ & $0.9592-0.6778i$   & $0.9567-0.6806i$ &       &        &     & $3$ & $0.9265-0.6706i$   & $0.9239-0.6735i$\\
       &         &        & $4$ & $0.9282-0.8792i$   & $0.9230-0.8882i$ &       &        &     & $4$ & $0.8947-0.8701i$   & $0.8891-0.8797i$\\
       &         &        & $5$ & \mbox{N/A}         & $0.8854-1.1041i$ &       &        &     & $5$ & \mbox{N/A}         & $0.8506-1.0945i$\\
       &         &        & $6$ & \mbox{N/A}         & $0.8465-1.3283i$ &       &        &     & $6$ & \mbox{N/A}         & $0.8108-1.3179i$\\
     [0.5ex] 
 \hline
 \end{tabular}
\end{table}

\begin{table}[t]
\centering
\caption{\textit{QNMs of massless scalar perturbations ($s=0$) of the KSQC black hole for several values of $\ell$ and the deformation parameter $\mathfrak{a}=a/M$. We compare our spectral method results with $N=200$ Chebyshev polynomials (final column) against the third-order WKB employed by \cite{SalehASS2014}. Entries labelled N/A indicate unavailable data. The notation 'SM' stands for Spectral Method.}}
\label{table_s0_a1_25_and_a1_5}
\setlength\tabcolsep{0.1cm}
\def\arraystretch{1.5}
\begin{tabular}{@{}|c| c| c| c| c| c|c|c|c|c|c|c|@{}} 
\hline
$a$    & $a_K$   & $\ell$ & $n$ & $\Omega_{WKB}$ \cite{SalehASS2014} & $\Omega_{SM}$ &$a$    & $a_K$   & $\ell$ & $n$ & $\Omega_{WKB}$ \cite{SalehASS2014} & $\Omega_SM$ \\ [0.5ex] 
\hline
$1.25$ & $0.53$  & $2$    & $0$ & $0.4317-0.0930i$   & $0.4321-0.0929i$ & $1.5$ & $0.60$ & $2$ & $0$ & $0.4144-0.0915i$   & $0.4148-0.0914i$\\ 
       &         &        & $1$ & $0.4094-0.2852i$   & $0.4100-0.2849i$ &       &        &     & $1$ & $0.3916-0.2809i$   & $0.3920-0.2806i$\\
       &         &        & $2$ & $0.3746-0.4866i$   & $0.3731-0.4931i$ &       &        &     & $2$ & $0.3559-0.4797i$   & $0.3540-0.4869i$\\ 
       &         &        & $3$ & $0.3310-0.6928i$   & $0.3334-0.7202i$ &       &        &     & $3$ & $0.3111-0.6834i$   & $0.3136-0.7128i$\\
       &         &        & $4$ & $0.2778-0.9017i$   & $0.2993-0.9604i$ &       &        &     & $4$ & $0.2562-0.8897i$   & $0.2796-0.9522i$\\ 
       &         &        & $5$ & \mbox{N/A}         & $0.2725-1.2073i$ &       &        &     & $5$ & \mbox{N/A}         & $0.2533-1.1980i$\\
       &         &        & $6$ & \mbox{N/A}         & $0.2517-1.4569i$ &       &        &     & $6$ & \mbox{N/A}         & $0.2333-1.4465i$\\
       \hline
       &         & $3$    & $0$ & $0.6035-0.0927i$   & $0.6036-0.0926i$ &       &        & $3$ & $0$ & $0.5794-0.0911i$   & $0.5796-0.0911i$\\  
       &         &        & $1$ & $0.5870-0.2812i$   & $0.5872-0.2811i$ &       &        &     & $1$ & $0.5625-0.2767i$   & $0.5626-0.2766i$\\
       &         &        & $2$ & $0.5588-0.4762i$   & $0.5571-0.4786i$ &       &        &     & $2$ & $0.5335-0.4690i$   & $0.5316-0.4717i$\\ 
       &         &        & $3$ & $0.5526-0.6766i$   & $0.5185-0.6894i$ &       &        &     & $3$ & $0.4964-0.6666i$   & $0.4920-0.6807i$\\
       &         &        & $4$ & $0.4796-0.8802i$   & $0.4782-0.9141i$ &       &        &     & $4$ & $0.4521-0.8676i$   & $0.4509-0.9041i$\\ 
       &         &        & $5$ & \mbox{N/A}         & $0.4410-1.1498i$ &       &        &     & $5$ & \mbox{N/A}         & $0.4135-1.1389i$\\
       &         &        & $6$ & \mbox{N/A}         & $0.4091-1.3927i$ &       &        &     & $6$ & \mbox{N/A}         & $0.3819-1.3808i$\\
       \hline
       &         & $4$    & $0$ & $0.7753-0.0925i$   & $0.7754-0.0925i$ &       &        & $4$ & $0$ & $0.7444-0.0910i$   & $0.7445-0.0910i$\\ 
       &         &        & $1$ & $0.7623-0.2795i$   & $0.7624-0.2795i$ &       &        &     & $1$ & $0.7311-0.2750i$   & $0.7312-0.2749i$\\
       &         &        & $2$ & $0.7389-0.4711i$   & $0.7378-0.4721i$ &       &        &     & $2$ & $0.7070-0.4637i$   & $0.7057-0.4648i$\\ 
       &         &        & $3$ & $0.7078-0.6674i$   & $0.7040-0.6736i$ &       &        &     & $3$ & $0.6751-0.6572i$   & $0.6709-0.6640i$\\
       &         &        & $4$ & $0.6708-0.8674i$   & $0.6647-0.8860i$ &       &        &     & $4$ & $0.6371-0.8545i$   & $0.6306-0.8746i$\\
       &         &        & $5$ & \mbox{N/A}         & $0.6241-1.1092i$ &       &        &     & $5$ & \mbox{N/A}         & $0.5893-1.0966i$\\
       &         &        & $6$ & \mbox{N/A}         & $0.5854-1.3417i$ &       &        &     & $6$ & \mbox{N/A}         & $0.5503-1.3281i$\\
       \hline
       &         & $5$    & $0$ & $0.9472-0.0925i$   & $0.9473-0.0925i$ &       &        & $5$ & $0$ & $0.9095-0.0909i$   & $0.9096-0.0909i$\\ 
       &         &        & $1$ & $0.9365-0.2787i$   & $0.9366-0.2787i$ &       &        &     & $1$ & $0.8985-0.2741i$   & $0.8986-0.2741i$\\
       &         &        & $2$ & $0.9166-0.4682i$   & $0.9159-0.4687i$ &       &        &     & $2$ & $0.8781-0.4607i$   & $0.8773-0.4612i$\\ 
       &         &        & $3$ & $0.8895-0.6617i$   & $0.8866-0.6649i$ &       &        &     & $3$ & $0.8502-0.6514i$   & $0.8471-0.6549i$\\
       &         &        & $4$ & $0.8567-0.8588i$   & $0.8508-0.8692i$ &       &        &     & $4$ & $0.8166-0.8457i$   & $0.8102-0.8570i$\\
       &         &        & $5$ & \mbox{N/A}         & $0.8112-1.0825i$ &       &        &     & $5$ & \mbox{N/A}         & $0.7696-1.0686i$\\
       &         &        & $6$ & \mbox{N/A}         & $0.7705-1.3048i$ &       &        &     & $6$ & \mbox{N/A}         & $0.7281-1.2896i$\\
     [0.5ex] 
 \hline
 \end{tabular}
\end{table}

\begin{table}[t]
\centering
\caption{\textit{Candidate purely imaginary QNMs for scalar perturbations of a KSQC black hole for several values of $\ell$ and the deformation parameter $\mathfrak{a}=a/M$. The corresponding results are obtained through our SM, utilising $200$ polynomials with a precision of $200$ digits. In this context, $\Omega$ and $n$ represent the dimensionless frequency and the corresponding overtone, respectively, while $\Delta\Omega=\Omega_n-\Omega_{n+1}$. Entries labelled N/A indicate unavailable data.}}
\label{table_s0_a0_25_and_a0_5_overdamped}
\setlength\tabcolsep{0.1cm}
\def\arraystretch{1.5}
\begin{tabular}{@{}|c| c| c| c| c| c|c|c|c|c|c|c|@{}} 
\hline
$a$    & $a_K$   & $\ell$ & $n$ & $\Omega$ & $\Delta\Omega$ & $a$    & $a_K$   & $\ell$ & $n$ & $\Omega$ & $\Delta\Omega$ \\ [0.5ex] 
\hline
$0.25$ & $0.12$  & $2$    & $0$ & $0.0000-18.9823i$   & \mbox{N/A}   & $0.5$ & $0.24$ & $2$ & $0$ & $0.0000-19.7063i$   & \mbox{N/A}\\ 
       &         &        & $1$ & $0.0000-19.2332i$   & $0.2509i$    &       &        &     & $1$ & $0.0000-19.9563i$   & $0.2500i$\\
       &         &        & $2$ & $0.0000-19.4842i$   & $0.2510i$    &       &        &     & $2$ & $0.0000-20.2065i$   & $0.2502i$\\ 
       &         &        & $3$ & $0.0000-19.7345i$   & $0.2453i$    &       &        &     & $3$ & $0.0000-20.4570i$   & $0.2505i$\\
       &         &        & $4$ & $0.0000-19.9853i$   & $0.2508i$    &       &        &     & $4$ & $0.0000-20.7071i$   & $0.2501i$\\ 
       &         &        & $5$ & $0.0000-20.2359i$   & $0.2506i$    &       &        &     & $5$ & $0.0000-20.9574i$   & $0.2503i$\\
       &         &        & $6$ & $0.0000-20.4865i$   & $0.2506i$    &       &        &     & $6$ & $0.0000-21.2077i$   & $0.2503i$\\
       \hline
       &         & $3$    & $0$ & $0.0000-19.6461i$   & \mbox{N/A}   &       &        & $3$ & $0$ & $0.0000-20.3677i$   & \mbox{N/A}\\  
       &         &        & $1$ & $0.0000-19.8974i$   & $0.2513i$    &       &        &     & $1$ & $0.0000-20.6181i$   & $0.2504i$\\
       &         &        & $2$ & $0.0000-20.1484i$   & $0.2510i$    &       &        &     & $2$ & $0.0000-20.8686i$   & $0.2505i$\\ 
       &         &        & $3$ & $0.0000-20.3992i$   & $0.2508i$    &       &        &     & $3$ & $0.0000-21.1192i$   & $0.2506i$\\
       &         &        & $4$ & $0.0000-20.6502i$   & $0.2510i$    &       &        &     & $4$ & $0.0000-21.3697i$   & $0.2505i$\\ 
       &         &        & $5$ & $0.0000-20.9010i$   & $0.2508i$    &       &        &     & $5$ & $0.0000-21.6202i$   & $0.2505i$\\
       &         &        & $6$ & $0.0000-21.1519i$   & $0.2509i$    &       &        &     & $6$ & $0.0000-21.8707i$   & $0.2505i$\\
       \hline
       &         & $4$    & $0$ & $0.0000-20.3070i$   & \mbox{N/A}   &       &        & $4$ & $0$ & $0.0000-20.2740i$   & \mbox{N/A}\\ 
       &         &        & $1$ & $0.0000-20.5585i$   & $0.2515i$    &       &        &     & $1$ & $0.0000-20.7756i$   & $0.5016i$\\
       &         &        & $2$ & $0.0000-20.8097i$   & $0.2512i$    &       &        &     & $2$ & $0.0000-21.0268i$   & $0.2512i$\\ 
       &         &        & $3$ & $0.0000-21.0608i$   & $0.2511i$    &       &        &     & $3$ & $0.0000-21.2776i$   & $0.2508i$\\
       &         &        & $4$ & $0.0000-21.3122i$   & $0.2514i$    &       &        &     & $4$ & $0.0000-21.5284i$   & $0.2508i$\\
       &         &        & $5$ & $0.0000-21.5632i$   & $0.2510i$    &       &        &     & $5$ & $0.0000-21.7792i$   & $0.2508i$\\
       &         &        & $6$ & $0.0000-21.8144i$   & $0.2512i$    &       &        &     & $6$ & $0.0000-22.0300i$   & $0.2508i$\\
       \hline
       &         & $5$    & $0$ & $0.0000-20.2089i$   & \mbox{N/A}   &       &        & $5$ & $0$ & $0.0000-20.9283i$   & \mbox{N/A}\\ 
       &         &        & $1$ & $0.0000-20.9637i$   & $0.7548i$    &       &        &     & $1$ & $0.0000-21.1797i$   & $0.2514i$\\
       &         &        & $2$ & $0.0000-21.2154i$   & $0.2517i$    &       &        &     & $2$ & $0.0000-21.4305i$   & $0.2508i$\\ 
       &         &        & $3$ & $0.0000-21.4670i$   & $0.2516i$    &       &        &     & $3$ & $0.0000-21.6820i$   & $0.2515i$\\
       &         &        & $4$ & $0.0000-21.7184i$   & $0.2514i$    &       &        &     & $4$ & $0.0000-21.9330i$   & $0.2510i$\\
       &         &        & $5$ & $0.0000-21.9699i$   & $0.2515i$    &       &        &     & $5$ & $0.0000-22.1841i$   & $0.2511i$\\
       &         &        & $6$ & $0.0000-22.2214i$   & $0.2515i$    &       &        &     & $6$ & $0.0000-22.4353i$   & $0.2512i$\\
     [0.5ex] 
 \hline
 \end{tabular}
\end{table}

\begin{table}[t]
\centering
\caption{\textit{Candidate purely imaginary QNMs for scalar perturbations of a KSQC black hole for several values of $\ell$ and the deformation parameter $\mathfrak{a}=a/M$. The corresponding results are obtained through our SM, utilising $200$ polynomials with a precision of $200$ digits. In this context, $\Omega$ and $n$ represent the dimensionless frequency and the corresponding overtone, respectively, while $\Delta\Omega=\Omega_n-\Omega_{n+1}$. Entries labelled N/A indicate unavailable data.}}
\label{table_s0_a0_75_and_a1_overdamped}
\setlength\tabcolsep{0.1cm}
\def\arraystretch{1.5}
\begin{tabular}{@{}|c| c| c| c| c| c|c|c|c|c|c|c|@{}} 
\hline
$a$    & $a_K$   & $\ell$ & $n$ & $\Omega$       &$\Delta\Omega$ & $a$   & $a_K$ & $\ell$ & $n$ & $\Omega$ & $\Delta\Omega$ \\ [0.5ex] 
\hline
$0.75$ & $0.35$  & $2$    & $0$ & $0.0000-20.8833i$   & \mbox{N/A}   & $1$   & $0.45$ & $2$ & $0$ & $0.0000-24.2212i$   & \mbox{N/A}\\ 
       &         &        & $1$ & $0.0000-21.1326i$   & $0.2493i$    &       &        &     & $1$ & $0.0000-24.4684i$   & $0.2472i$\\
       &         &        & $2$ & $0.0000-21.3824i$   & $0.2498i$    &       &        &     & $2$ & $0.0000-24.7160i$   & $0.2476i$\\ 
       &         &        & $3$ & $0.0000-21.6315i$   & $0.2491i$    &       &        &     & $3$ & $0.0000-24.9641i$   & $0.2481i$\\
       &         &        & $4$ & $0.0000-21.8809i$   & $0.2494i$    &       &        &     & $4$ & $0.0000-25.2115i$   & $0.2474i$\\ 
       &         &        & $5$ & $0.0000-22.1303i$   & $0.2494i$    &       &        &     & $5$ & $0.0000-25.4591i$   & $0.2476i$\\
       &         &        & $6$ & $0.0000-22.3800i$   & $0.2497i$    &       &        &     & $6$ & $0.0000-25.7070i$   & $0.2479i$\\
       \hline
       &         & $3$    & $0$ & $0.0000-21.2930i$   & \mbox{N/A}   &       &        & $3$ & $0$ & $0.0000-24.1342i$   & \mbox{N/A}\\  
       &         &        & $1$ & $0.0000-21.5424i$   & $0.2494i$    &       &        &     & $1$ & $0.0000-24.6294i$   & $0.4952i$\\
       &         &        & $2$ & $0.0000-21.7918i$   & $0.2494i$    &       &        &     & $2$ & $0.0000-24.8778i$   & $0.2484i$\\ 
       &         &        & $3$ & $0.0000-22.0416i$   & $0.2498i$    &       &        &     & $3$ & $0.0000-25.1256i$   & $0.2478i$\\
       &         &        & $4$ & $0.0000-22.2911i$   & $0.2495i$    &       &        &     & $4$ & $0.0000-25.3732i$   & $0.2476i$\\ 
       &         &        & $5$ & $0.0000-22.5407i$   & $0.2496i$    &       &        &     & $5$ & $0.0000-25.6214i$   & $0.2482i$\\
       &         &        & $6$ & $0.0000-22.7903i$   & $0.2496i$    &       &        &     & $6$ & $0.0000-25.8692i$   & $0.2478i$\\
       \hline
       &         & $4$    & $0$ & $0.0000-21.1999i$   & \mbox{N/A}   &       &        & $4$ & $0$ & $0.0000-24.5408i$   & \mbox{N/A}\\ 
       &         &        & $1$ & $0.0000-21.6993i$   & $0.4994i$    &       &        &     & $1$ & $0.0000-24.7894i$   & $0.2486i$\\
       &         &        & $2$ & $0.0000-21.9498i$   & $0.2505i$    &       &        &     & $2$ & $0.0000-25.0379i$   & $0.2485i$\\ 
       &         &        & $3$ & $0.0000-22.1996i$   & $0.2498i$    &       &        &     & $3$ & $0.0000-25.2855i$   & $0.2476i$\\
       &         &        & $4$ & $0.0000-22.4494i$   & $0.2498i$    &       &        &     & $4$ & $0.0000-25.5338i$   & $0.2483i$\\ 
       &         &        & $5$ & $0.0000-22.6994i$   & $0.2500i$    &       &        &     & $5$ & $0.0000-25.7821i$   & $0.2483i$\\
       &         &        & $6$ & $0.0000-22.9493i$   & $0.2499i$    &       &        &     & $6$ & $0.0000-26.0299i$   & $0.2478i$\\
       \hline
       &         & $5$    & $0$ & $0.0000-21.8520i$   & \mbox{N/A}   &       &        & $5$ & $0$ & $0.0000-24.9451i$   & \mbox{N/A}\\ 
       &         &        & $1$ & $0.0000-22.3522i$   & $0.5002i$    &       &        &     & $1$ & $0.0000-25.1927i$   & $0.2476i$\\
       &         &        & $2$ & $0.0000-22.6028i$   & $0.2506i$    &       &        &     & $2$ & $0.0000-25.4411i$   & $0.2484i$\\ 
       &         &        & $3$ & $0.0000-22.8529i$   & $0.2501i$    &       &        &     & $3$ & $0.0000-25.6900i$   & $0.2489i$\\
       &         &        & $4$ & $0.0000-23.1030i$   & $0.2501i$    &       &        &     & $4$ & $0.0000-25.9380i$   & $0.2480i$\\ 
       &         &        & $5$ & $0.0000-23.3533i$   & $0.2503i$    &       &        &     & $5$ & $0.0000-26.1863i$   & $0.2483i$\\
       &         &        & $6$ & $0.0000-23.6034i$   & $0.2501i$    &       &        &     & $6$ & $0.0000-26.4350i$   & $0.2487i$\\
     [0.5ex] 
 \hline
 \end{tabular}
\end{table}

\begin{table}[t]
\centering
\caption{\textit{QNMs of massless scalar perturbations ($s=0$) of the KSQC black hole for $\ell=0$ and several values of the deformation parameter $a_K$ for the near-extremal case. We compare our spectral method results with $N=200$ Chebyshev polynomials (final column) against the seventh-order WKB  adopted by \cite{KonoplyaPLB2020}. Entries labelled N/A indicate unavailable data. The notation 'SM' stands for Spectral Method.}}
\label{table_s0_nearly_extremal}
\setlength\tabcolsep{0.1cm}
\def\arraystretch{1.5}
\begin{tabular}{@{}|c| c| c| c| c| c|c|c|c|c|@{}} 
\hline
$a_K$  & $\mathfrak{a}$ & $n$ & $\Omega_{WKB}$ \cite{KonoplyaPLB2020} & $\Omega_{SM}$ & $a_K$  & $\mathfrak{a}$ & $n$ & $\Omega_{WKB}$ \cite{KonoplyaPLB2020} & $\Omega_{SM}$\\ [0.5ex] 
\hline
$0.90$  & $\frac{18\sqrt{19}}{19}$ & $0$ & $0.107303-0.022386i$ & $0.0512-0.0792i$ & $0.93$    & $\frac{186\sqrt{1351}}{1351}$       & $0$ & \mbox{N/A}   & $0.0418-0.0717i$\\ 
        &           & $1$ & \mbox{N/A}   & $0.0392-0.2932i$ &        &           & $1$ & \mbox{N/A}   & $0.0361-0.2795i$\\
        &           & $2$ & \mbox{N/A}   & $0.0462-0.5207i$ &        &           & $2$ & \mbox{N/A}   & $0.0436-0.5012i$\\
        &           & $3$ & \mbox{N/A}   & $0.0540-0.7480i$ &        &           & $3$ & \mbox{N/A}   & $0.0477-0.7183i$\\
        &           & $4$ & \mbox{N/A}   & $0.0609-0.9714i$ &        &           & $4$ & \mbox{N/A}   & $0.0513-0.9241i$\\
        &           & $5$ & \mbox{N/A}   & $0.0688-1.1891i$ &        &           & $5$ & \mbox{N/A}   & \mbox{N/A}\\
        &           & $6$ & \mbox{N/A}   & $0.0807-1.4025i$ &        &           & $6$ & \mbox{N/A}   & \mbox{N/A}\\
        &           & $7$ & \mbox{N/A}   & $0.0980-1.6154i$ &        &           & $7$ & \mbox{N/A}   & \mbox{N/A}\\
        &           & $8$ & \mbox{N/A}   & $0.1175-1.8303i$ &        &           & $8$ & \mbox{N/A}   & \mbox{N/A}\\
        \hline
$0.91$  & $\frac{182\sqrt{191}}{573}$ & $0$ & \mbox{N/A} & $0.0483-0.0770i$ & $0.94$    & $\frac{94\sqrt{291}}{291}$       & $0$ & \mbox{N/A}   & $0.0381-0.0683i$\\ 
        &           & $1$ & \mbox{N/A}   & $0.0382-0.2891i$ &        &           & $1$ & \mbox{N/A}   & $0.0349-0.2738i$\\
        &           & $2$ & \mbox{N/A}   & $0.0457-0.5150i$ &        &           & $2$ & \mbox{N/A}   & $0.0414-0.4926i$\\
        &           & $3$ & \mbox{N/A}   & $0.0529-0.7397i$ &        &           & $3$ & \mbox{N/A}   & $0.0422-0.7034i$\\
        &           & $4$ & \mbox{N/A}   & $0.0587-0.9589i$ &        &           & $4$ & \mbox{N/A}   & \mbox{N/A}\\
        &           & $5$ & \mbox{N/A}   & $0.0670-1.1711i$ &        &           & $5$ & \mbox{N/A}   & \mbox{N/A}\\
        &           & $6$ & \mbox{N/A}   & $0.0818-1.3795i$ &        &           & $6$ & \mbox{N/A}   & \mbox{N/A}\\
        \hline
$0.92$  & $\frac{23\sqrt{6}}{12}$  & $0$ & \mbox{N/A} & $0.0452-0.0745i$ & $0.95$    & $\frac{38\sqrt{39}}{39}$ & $0$ & \mbox{N/A}   & $0.0341-0.0643i$\\ 
        &           & $1$ & \mbox{N/A}   & $0.0372-0.2846i$ &        &           & $1$ & \mbox{N/A}   & $0.0335-0.2673i$\\
        &           & $2$ & \mbox{N/A}   & $0.0450-0.5086i$ &        &           & $2$ & \mbox{N/A}   & $0.0375-0.4822i$\\
        &           & $3$ & \mbox{N/A}   & $0.0509-0.7300i$ &        &           & $3$ & \mbox{N/A}   & \mbox{N/A}\\
        &           & $4$ & \mbox{N/A}   & $0.0556-0.9436i$ &        &           & $4$ & \mbox{N/A}   & \mbox{N/A}\\
        &           & $5$ & \mbox{N/A}   & $0.0654-1.1487i$ &        &           & $5$ & \mbox{N/A}   & \mbox{N/A}\\
     [0.5ex] 
 \hline
 \end{tabular}
\end{table}

\begin{table}
\centering
\caption{Candidate purely imaginary QNMs for scalar perturbations of a KSQC black hole for $\ell=0$ and several values of the deformation parameter $a_K$ close to the near-extremal case. The corresponding results are obtained through our SM, utilising $200$ polynomials with a precision of $200$ digits. In this context, $\Omega$ and $n$ represent the dimensionless frequency and the corresponding overtone, respectively, while $\Delta\Omega=\Omega_n-\Omega_{n+1}$. Entries labelled N/A indicate unavailable data.}
\label{scalargoverdampednearextremal}
\vspace*{1em}
\begin{tabular}{||c|c|c|c|c|c|c|c|c||}
\hline\hline
$a_K$  & $n$ & $\Omega$                 &$\Delta\Omega$  & $a_K$ & $n$ & $\Omega$  &$\Delta\Omega$\\ [0.5ex]
\hline\hline
$0.9$  & $0$ & $0.0000-2.5010i$             & \mbox{N/A}     & $0.95$ & $0$ & $0.0000-0.7525i$             &\mbox{N/A}\\
       & $1$ & $0.0000-2.7509i$             & $0.2499i$      &        & $1$ & $0.0000-1.0017i$             &$0.2492i$\\
       & $2$ & $0.0000-3.0008i$             & $0.2499i$      &        & $2$ & $0.0000-1.2512i$             &$0.2495i$\\
       & $3$ & $0.0000-3.2507i$             & $0.2499i$      &        & $3$ & $0.0000-1.5009i$             &$0.2497i$\\
       & $4$ & $0.0000-3.5007i$             & $0.2499i$      &        & $4$ & $0.0000-1.7508i$             &$0.2499i$\\ 
       & $5$ & $0.0000-3.7506i$             & $0.2499i$      &        & $5$ & $0.0000-2.0006i$             &$0.2498i$\\
\hline
$0.91$ & $0$ & $0.0000-2.2510i$             & \mbox{N/A}     & $0.96$ & $0$ & $0.0000-0.5034i$             &\mbox{N/A}\\
       & $1$ & $0.0000-2.5009i$             & $0.2499i$      &        & $1$ & $0.0000-0.7519i$             &$0.2485i$\\
       & $2$ & $0.0000-2.7508i$             & $0.2499i$      &        & $2$ & $0.0000-1.0013i$             &$0.2494i$\\
       & $3$ & $0.0000-3.0007i$             & $0.2499i$      &        & $3$ & $0.0000-1.2510i$             &$0.2497i$\\
       & $4$ & $0.0000-3.2506i$             & $0.2499i$      &        & $4$ & $0.0000-1.5007i$             &$0.2497i$\\ 
       & $5$ & $0.0000-3.5006i$             & $0.2500i$      &        & $5$ & $0.0000-1.7506i$             &$0.2499i$\\
\hline
$0.92$ & $0$ & $0.0000-1.7513i$             & \mbox{N/A}     & $0.97$ & $0$ & $0.0000-0.5025i$             &\mbox{N/A}\\
       & $1$ & $0.0000-2.0011i$             & $0.2498i$      &        & $1$ & $0.0000-0.7514i$             &$0.2489i$\\
       & $2$ & $0.0000-2.2509i$             & $0.2498i$      &        & $2$ & $0.0000-1.0010i$             &$0.2496i$\\
       & $3$ & $0.0000-2.5008i$             & $0.2499i$      &        & $3$ & $0.0000-1.2507i$             &$0.2497i$\\
       & $4$ & $0.0000-2.7507i$             & $0.2499i$      &        & $4$ & $0.0000-1.5006i$             &$0.2499i$\\
       & $5$ & $0.0000-3.0006i$             & $0.2499i$      &        & $5$ & $0.0000-1.7505i$             &$0.2499i$\\
\hline
$0.93$ & $0$ & $0.0000-1.5014i$             & \mbox{N/A}     & $0.98$ & $0$ & $0.0000-0.2538i$      &\mbox{N/A}\\
       & $1$ & $0.0000-1.7511i$             & $0.2497i$      &        & $1$ & $0.0000-0.5016i$      &$0.2478i$\\
       & $2$ & $0.0000-2.0009i$             & $0.2498i$      &        & $2$ & $0.0000-0.7509i$      &$0.2493i$\\
       & $3$ & $0.0000-2.2508i$             & $0.2499i$      &        & $3$ & $0.0000-1.0006i$      &$0.2497i$\\
       & $4$ & $0.0000-2.5007i$             & $0.2499i$      &        & $4$ & $0.0000-1.2505i$      &$0.2499i$\\ 
       & $5$ & $0.0000-2.7506i$             & $0.2499i$      &        & $5$ & $0.0000-1.5004i$      &$0.2499i$\\
\hline
$0.94$ & $0$ & $0.0000-1.0021i$             & \mbox{N/A}     & $0.99$ & $0$ & $0.0000-0.2520i$      &\mbox{N/A}\\
       & $1$ & $0.0000-1.2515i$             & $0.2494i$      &        & $1$ & $0.0000-0.5008i$      &$0.2488i$\\
       & $2$ & $0.0000-1.5012i$             & $0.2497i$      &        & $2$ & $0.0000-0.7505i$      &$0.2497i$\\
       & $3$ & $0.0000-1.7509i$             & $0.2497i$      &        & $3$ & $0.0000-1.0003i$      &$0.2498i$\\
       & $4$ & $0.0000-2.0008i$             & $0.2499i$      &        & $4$ & $0.0000-1.2503i$      &$0.2500i$\\ 
       & $5$ & $0.0000-2.2507i$             & $0.2499i$      &        & $5$ & $0.0000-1.5002i$      &$0.2499i$\\
       [1ex]
 \hline\hline 
 \end{tabular}
\end{table}

\begin{table}
\centering
\caption{Candidate purely imaginary QNMs for scalar perturbations of a KSQC black hole for $\ell=0$ and several values of the deformation parameter $a_K$ close to the near-extremal case. The corresponding results are obtained through our SM, utilising $200$ polynomials with a precision of $200$ digits. In this context, $\Omega$ and $n$ represent the dimensionless frequency and the corresponding overtone, respectively, while $\Delta\Omega=\Omega_n-\Omega_{n+1}$. Entries labelled N/A indicate unavailable data.}
\label{scalargoverdampednearextremalrefined}
\vspace*{1em}
\begin{tabular}{||c|c|c|c|c|c|c|c|c||}
\hline\hline
$a_K$  & $n$ & $\Omega$                 &$\Delta\Omega$  & $a_K$ & $n$ & $\Omega$  &$\Delta\Omega$\\ [0.5ex]
\hline\hline
$0.992$& $0$ & $0.0000-0.2516i$             & \mbox{N/A}     & $0.996$ & $0$ & $0.0000-0.2508i$             &\mbox{N/A}\\
       & $1$ & $0.0000-0.5006i$             & $0.2490i$      &         & $1$ & $0.0000-0.5003i$             &$0.2495i$\\
       & $2$ & $0.0000-0.7504i$             & $0.2498i$      &         & $2$ & $0.0000-0.7502i$             &$0.2499i$\\
       & $3$ & $0.0000-1.0003i$             & $0.2499i$      &         & $3$ & $0.0000-1.0002i$             &$0.2500i$\\
       & $4$ & $0.0000-1.2502i$             & $0.2499i$      &         & $4$ & $0.0000-1.2501i$             &$0.2499i$\\ 
       & $5$ & $0.0000-1.5002i$             & $0.2500i$      &         & $5$ & $0.0000-1.5001i$             &$0.2500i$\\
\hline
$0.994$& $0$ & $0.0000-0.2512i$             & \mbox{N/A}     & $0.998$ & $0$ & $0.0000-0.2504i$             &\mbox{N/A}\\
       & $1$ & $0.0000-0.5005i$             & $0.2493i$      &         & $1$ & $0.0000-0.5002i$             &$0.2498i$\\
       & $2$ & $0.0000-0.7503i$             & $0.2498i$      &         & $2$ & $0.0000-0.7501i$             &$0.2499i$\\
       & $3$ & $0.0000-1.0002i$             & $0.2499i$      &         & $3$ & $0.0000-1.0001i$             &$0.2500i$\\
       & $4$ & $0.0000-1.2502i$             & $0.2500i$      &         & $4$ & $0.0000-1.2501i$             &$0.2500i$\\ 
       & $5$ & $0.0000-1.5001i$             & $0.2499i$      &         & $5$ & $0.0000-1.5001i$             &$0.2500i$\\
       [1ex]
 \hline\hline 
 \end{tabular}
\end{table}

\begin{table}[t]
\centering
\caption{\textit{QNMs of non-minimally coupled massless scalar perturbations ($s=0$) of the KSQC black hole for $\ell=0$ and several values of the deformation parameter $a_K$. We compare our spectral method results with $N=200$ Chebyshev polynomials (final column) against the seventh-order WKB and the time-domain method adopted by \cite{KonoplyaPLB2020}. Entries labelled N/A indicate unavailable data. The notation 'SM' stands for Spectral Method.}}
\label{table_s0_L0_minimally coupled}
\setlength\tabcolsep{0.1cm}
\def\arraystretch{1.5}
\begin{tabular}{@{}|c| c| c| c| c| c|c|c|@{}} 
\hline
$a_K$  & $\mathfrak{a}$ & $n$ & $\Omega_{WKB}$ \cite{KonoplyaPLB2020} & $\Omega_{TD}$ \cite{KonoplyaPLB2020} & $\Omega_{SM}$ \\ [0.5ex] 
\hline
$0.1$  & $\frac{2\sqrt{11}}{33}$     & $0$ & $0.110754-0.104204i$ & $0.10988-0.10529i$ & $0.1101-0.1048i$ \\
       &                             & $1$ & \mbox{N/A}           & \mbox{N/A}         & $0.0857-0.3479i$ \\
       &                             & $2$ & \mbox{N/A}           & \mbox{N/A}         & $0.0753-0.6009i$ \\
       \hline
$0.2$  & $\frac{\sqrt{6}}{6}$        & $0$ & $0.111299-0.105560i$ & $0.11057-0.10667i$ & $0.1089-0.1045i$ \\
       &                             & $1$ & \mbox{N/A}           & \mbox{N/A}         & $0.0843-0.3474i$ \\
       &                             & $2$ & \mbox{N/A}           & \mbox{N/A}         & $0.0738-0.6002i$ \\
       \hline
$0.3$  & $\frac{6\sqrt{91}}{91}$     & $0$ & $0.107627-0.103496i$ & $0.10698-0.10445i$ & $0.1069-0.1041i$ \\
       &                             & $1$ & \mbox{N/A}           & \mbox{N/A}         & $0.0820-0.3465i$ \\
       &                             & $2$ & \mbox{N/A}           & \mbox{N/A}         & $0.0714-0.5990i$ \\
       \hline
$0.4$  & $\frac{4\sqrt{21}}{21}$     & $0$ & $0.104745-0.102783i$ & $0.10401-0.10376i$ & $0.1039-0.1033i$ \\
       &                             & $1$ & \mbox{N/A}           & \mbox{N/A}         & $0.0787-0.3451i$ \\
       &                             & $2$ & \mbox{N/A}           & \mbox{N/A}         & $0.0681-0.5968i$ \\
       \hline
$0.5$  & $\frac{2\sqrt{3}}{3}$       & $0$ & $0.100796-0.101626i$ & $0.09928-0.10284i$ & $0.0998-0.1022i$ \\
       &                             & $1$ & \mbox{N/A}           & \mbox{N/A}         & $0.0744-0.3428i$ \\
       &                             & $2$ & \mbox{N/A}           & \mbox{N/A}         & $0.0642-0.5933i$ \\    
       \hline
$0.6$  & $\frac{3}{2}$               & $0$ & $0.095450-0.099682i$ & $0.09480-0.10068i$ & $0.0944-0.1005i$ \\
       &                             & $1$ & \mbox{N/A}           & \mbox{N/A}         & $0.0691-0.3391i$ \\
       &                             & $2$ & \mbox{N/A}           & \mbox{N/A}         & $0.0601-0.5874i$ \\
       \hline
$0.7$  & $\frac{14\sqrt{51}}{51}$    & $0$ & $0.088422-0.096817i$ & $0.08763-0.09838i$ & $0.0873-0.0978i$ \\
       &                             & $1$ & \mbox{N/A}           & \mbox{N/A}         & $0.0632-0.3329i$ \\
       &                             & $2$ & \mbox{N/A}           & \mbox{N/A}         & $0.0571-0.5775i$ \\
       \hline
$0.8$  & $\frac{8}{3}$               & $0$ & $0.079792-0.092359i$ & $0.07822-0.09322i$ & $0.0777-0.0933i$ \\
       &                             & $1$ & \mbox{N/A}           & \mbox{N/A}         & $0.0574-0.3218i$ \\
       &                             & $2$ & \mbox{N/A}           & \mbox{N/A}         & $0.0579-0.5599i$ \\
       &                             & $3$ & \mbox{N/A}           & \mbox{N/A}         & $0.0649-0.7982i$ \\
       &                             & $4$ & \mbox{N/A}           & \mbox{N/A}         & $0.0743-1.0362i$ \\
       \hline
$0.9$ & $\frac{18\sqrt{19}}{19}$   & $0$ & $0.076856-0.088481i$ & $0.06400-0.08330i$   & $0.0640-0.0841i$ \\
       &                             & $1$ & \mbox{N/A}           & \mbox{N/A}         & $0.0549-0.2984i$ \\
       &                             & $2$ & \mbox{N/A}           & \mbox{N/A}         & $0.0654-0.5224i$ \\
       &                             & $3$ & \mbox{N/A}           & \mbox{N/A}         & $0.0788-0.7445i$ \\
       &                             & $4$ & \mbox{N/A}           & \mbox{N/A}         & $0.0932-0.9630i$ \\
       \hline
$0.99$ & $\frac{198\sqrt{199}}{199}$ & $0$ & \mbox{N/A}           & \mbox{N/A}         & $0.0421-0.0538i$ \\
     [0.5ex] 
 \hline
 \end{tabular}
\end{table}

\begin{table}
\centering
\caption{Candidate purely imaginary QNMs for non-minimally coupled scalar perturbations of a KSQC black hole for $\ell=0$ and several values of the deformation parameter $a_K$. The corresponding results are obtained through our SM, utilising $200$ polynomials with a precision of $200$ digits. In this context, $\Omega$ and $n$ represent the dimensionless frequency and the corresponding overtone, respectively, while $\Delta\Omega=\Omega_n-\Omega_{n+1}$. Entries labelled N/A indicate unavailable data.}
\label{scalargoverdampedmincoup}
\vspace*{1em}
\begin{tabular}{||c|c|c|c|c|c|c|c|c||}
\hline\hline
$a_K$  & $n$ & $\Omega$              &$\Delta\Omega$  & $a_K$ & $n$ & $\Omega$  &$\Delta\Omega$\\ [0.5ex]
\hline\hline
$0.1$  & $0$ & $0.0000-18.3639i$     & \mbox{N/A}     & $0.6$  & $0$ & \mbox{N/A}             &\mbox{N/A}\\
       & $1$ & $0.0000-18.6139i$     & $0.2500i$      &        & $1$ & \mbox{N/A}             &\mbox{N/A}\\
       & $2$ & $0.0000-18.8642i$     & $0.2503i$      &        & $2$ & \mbox{N/A}             &\mbox{N/A}\\
       & $3$ & $0.0000-19.1140i$     & $0.2498i$      &        & $3$ & \mbox{N/A}             &\mbox{N/A}\\
       & $4$ & $0.0000-19.3641i$     & $0.2501i$      &        & $4$ & \mbox{N/A}             &\mbox{N/A}\\ 
       & $5$ & $0.0000-19.6141i$     & $0.2500i$      &        & $5$ & \mbox{N/A}             &\mbox{N/A}\\
\hline
$0.2$  & $0$ & $0.0000-18.8494i$     & \mbox{N/A}     & $0.7$  & $0$ & \mbox{N/A}             &\mbox{N/A}\\
       & $1$ & $0.0000-19.0994i$     & $0.2500i$      &        & $1$ & \mbox{N/A}             &\mbox{N/A}\\
       & $2$ & $0.0000-19.3494i$     & $0.2500i$      &        & $2$ & \mbox{N/A}             &\mbox{N/A}\\
       & $3$ & $0.0000-19.5990i$     & $0.2496i$      &        & $3$ & \mbox{N/A}             &\mbox{N/A}\\
       & $4$ & $0.0000-19.8490i$     & $0.2500i$      &        & $4$ & \mbox{N/A}             &\mbox{N/A}\\ 
       & $5$ & $0.0000-20.0988i$     & $0.2498i$      &        & $5$ & \mbox{N/A}             &\mbox{N/A}\\
\hline
$0.3$  & $0$ & $0.0000-19.8050i$     & \mbox{N/A}     & $0.8$  & $0$ & $0.0000-10.7387i$      &\mbox{N/A}\\
       & $1$ & $0.0000-20.0542i$     & $0.2492i$      &        & $1$ & \mbox{N/A}             &\mbox{N/A}\\
       & $2$ & $0.0000-20.3038i$     & $0.2496i$      &        & $2$ & \mbox{N/A}             &\mbox{N/A}\\
       & $3$ & $0.0000-20.5528i$     & $0.2490i$      &        & $3$ & \mbox{N/A}             &\mbox{N/A}\\
       & $4$ & $0.0000-20.8022i$     & $0.2494i$      &        & $4$ & \mbox{N/A}             &\mbox{N/A}\\
       & $5$ & $0.0000-21.0515i$     & $0.2493i$      &        & $5$ & \mbox{N/A}             &\mbox{N/A}\\
\hline
$0.4$  & $0$ & $0.0000-22.4401i$     & \mbox{N/A}     & $0.9$  & $0$ & $0.0000-2.7414i$      &\mbox{N/A}\\
       & $1$ & $0.0000-22.6876i$     & $0.2475i$      &        & $1$ & $0.0000-2.9914i$      &$0.2500i$\\
       & $2$ & $0.0000-22.9357i$     & $0.2481i$      &        & $2$ & $0.0000-3.2413i$      &$0.2499i$\\
       & $3$ & $0.0000-23.1840i$     & $0.2483i$      &        & $3$ & $0.0000-3.4912i$      &$0.2499i$\\
       & $4$ & $0.0000-23.4318i$     & $0.2478i$      &        & $4$ & $0.0000-3.7411i$      &$0.2499i$\\ 
       & $5$ & $0.0000-23.6800i$     & $0.2482i$      &        & $5$ & $0.0000-3.9910i$      &$0.2499i$\\
\hline
$0.5$  & $0$ & \mbox{N/A}            & \mbox{N/A}     & $0.99$ & $0$ & \mbox{N/A}      &\mbox{N/A}\\
       & $1$ & \mbox{N/A}            & \mbox{N/A}     &        & $1$ & \mbox{N/A}      &\mbox{N/A}\\
       & $2$ & \mbox{N/A}            & \mbox{N/A}     &        & $2$ & \mbox{N/A}      &\mbox{N/A}\\
       & $3$ & \mbox{N/A}            & \mbox{N/A}     &        & $3$ & \mbox{N/A}      &\mbox{N/A}\\
       & $4$ & \mbox{N/A}            & \mbox{N/A}     &        & $4$ & \mbox{N/A}      &\mbox{N/A}\\ 
       & $5$ & \mbox{N/A}            & \mbox{N/A}     &        & $5$ & \mbox{N/A}      &\mbox{N/A}\\
       [1ex]
 \hline\hline 
 \end{tabular}
\end{table}

\begin{table}[t]
\centering
\caption{\textit{QNMs of electromagnetic perturbations ($s=1$) of the KSQC black hole for $\ell=1$ and several values of the deformation parameter $a_K$. We compare our spectral method results with $N=200$ Chebyshev polynomials (final column) against the seventh-order WKB and the time-domain method adopted by \cite{KonoplyaPLB2020}. Entries labelled N/A indicate unavailable data. The notation 'SM' stands for Spectral Method.}}
\label{table_s1_L1}
\setlength\tabcolsep{0.1cm}
\def\arraystretch{1.5}
\begin{tabular}{@{}|c| c| c| c| c| c|c|c|@{}} 
\hline
$a_K$  & $\mathfrak{a}$ & $n$ & $\Omega_{WKB}$ \cite{KonoplyaPLB2020} & $\Omega_{TD}$ \cite{KonoplyaPLB2020} & $\Omega_{SM}$ \\ [0.5ex] 
\hline
$0$    & $0$       & $0$ & \mbox{N/A}                            & \mbox{N/A}                           & $0.2483-0.0925i$\\ 
       &           & $1$ & \mbox{N/A}                            & \mbox{N/A}                           & $0.2145-0.2937i$\\
       &           & $2$ & \mbox{N/A}                            & \mbox{N/A}                           & $0.1748-0.5252i$\\ 
       \hline
$0.01$ & $\frac{2\sqrt{1111}}{3333}$ & $0$ & $0.248253-0.092486i$ & $0.24827-0.09248i$ & $0.2483-0.0925i$ \\
       &                             & $1$ & \mbox{N/A}           & \mbox{N/A}         & $0.2145-0.2937i$ \\
       &                             & $2$ & \mbox{N/A}           & \mbox{N/A}         & $0.1748-0.5252i$ \\
       \hline
$0.1$  & $\frac{2\sqrt{11}}{33}$     & $0$ & $0.247314-0.092362i$ & $0.24739-0.09230i$ & $0.2473-0.0924i$ \\
       &                             & $1$ & \mbox{N/A}           & \mbox{N/A}         & $0.2134-0.2934i$ \\
       &                             & $2$ & \mbox{N/A}           & \mbox{N/A}         & $0.1736-0.5248i$ \\
       \hline
$0.2$  & $\frac{\sqrt{6}}{6}$        & $0$ & $0.244423-0.091976i$ & $0.24449-0.09192i$ & $0.2444-0.0920i$ \\
       &                             & $1$ & \mbox{N/A}           & \mbox{N/A}         & $0.2102-0.2924i$ \\
       &                             & $2$ & \mbox{N/A}           & \mbox{N/A}         & $0.1701-0.5237i$ \\
       \hline
$0.3$  & $\frac{6\sqrt{91}}{91}$     & $0$ & $0.239451-0.091291i$ & $0.23951-0.09125i$ & $0.2395-0.0913i$ \\
       &                             & $1$ & \mbox{N/A}           & \mbox{N/A}         & $0.2046-0.2907i$ \\
       &                             & $2$ & \mbox{N/A}           & \mbox{N/A}         & $0.1640-0.5217i$ \\
       \hline
$0.4$  & $\frac{4\sqrt{21}}{21}$     & $0$ & $0.232132-0.090233i$ & $0.23218-0.09020i$ & $0.2321-0.0902i$ \\
       &                             & $1$ & \mbox{N/A}           & \mbox{N/A}         & $0.1964-0.2881i$ \\
       &                             & $2$ & \mbox{N/A}           & \mbox{N/A}         & $0.1551-0.5185i$ \\   
       \hline
$0.5$  & $\frac{2\sqrt{3}}{3}$       & $0$ & $0.222016-0.088667i$ & $0.22205-0.08866i$ & $0.2220-0.0887i$ \\
       &                             & $1$ & \mbox{N/A}           & \mbox{N/A}         & $0.1851-0.2843i$ \\
       &                             & $2$ & \mbox{N/A}           & \mbox{N/A}         & $0.1430-0.5138i$ \\    
       \hline
$0.6$  & $\frac{3}{2}$               & $0$ & $0.208348-0.086334i$ & $0.20838-0.08638i$ & $0.2084-0.0864i$ \\
       &                             & $1$ & \mbox{N/A}           & \mbox{N/A}         & $0.1698-0.2786i$ \\
       &                             & $2$ & \mbox{N/A}           & \mbox{N/A}         & $0.1271-0.5068i$ \\
       \hline
$0.7$  & $\frac{14\sqrt{51}}{51}$    & $0$ & $0.189700-0.082580i$ & $0.18983-0.08289i$ & $0.1898-0.0829i$ \\
       &                             & $1$ & \mbox{N/A}           & \mbox{N/A}         & $0.1494-0.2700i$ \\
       &                             & $2$ & \mbox{N/A}           & \mbox{N/A}         & $0.1063-0.4964i$ \\
       &                             & $3$ & \mbox{N/A}           & \mbox{N/A}         & $0.0797-0.7387i$ \\
       \hline
$0.8$  & $\frac{8}{3}$               & $0$ & $0.161225-0.077362i$ & $0.16362-0.07709i$ & $0.1636-0.0771i$ \\
       &                             & $1$ & \mbox{N/A}           & \mbox{N/A}         & $0.1209-0.2561i$ \\
       &                             & $2$ & \mbox{N/A}           & \mbox{N/A}         & $0.0784-0.4802i$ \\
       \hline
$0.85$ & $\frac{34\sqrt{111}}{111}$  & $0$ & \mbox{N/A}           & \mbox{N/A}         & $0.1458-0.0725i$ \\
       &                             & $1$ & \mbox{N/A}           & \mbox{N/A}         & $0.1019-0.2456i$ \\
       &                             & $2$ & \mbox{N/A}           & \mbox{N/A}         & $0.0596-0.4686i$ \\
     [0.5ex] 
 \hline
 \end{tabular}
\end{table}

\begin{table}
\centering
\caption{Candidate purely imaginary QNMs for electromagnetic perturbations of a KSQC black hole for $\ell=1$ and several values of the deformation parameter $a_K$. The corresponding results are obtained through our SM, utilising $200$ polynomials with a precision of $200$ digits. In this context, $\Omega$ and $n$ represent the dimensionless frequency and the corresponding overtone, respectively, while $\Delta\Omega=\Omega_n-\Omega_{n+1}$. Entries labelled N/A indicate unavailable data.}
\label{emoverdamped}
\vspace*{1em}
\begin{tabular}{||c|c|c|c|c|c|c|c|c||}
\hline\hline
$a_K$  & $n$ & $\Omega$                 &$\Delta\Omega$  & $a_K$ & $n$ & $\Omega$  &$\Delta\Omega$\\ [0.5ex]
\hline\hline
$0.01$ & $0$ & $0.0000-18.4765i$            & \mbox{N/A}     & $0.5$  & $0$ & $0.0000-24.7869i$      &\mbox{N/A}\\
       & $1$ & $0.0000-18.7270i$            & $0.2505i$      &        & $1$ & $0.0000-25.2793i$      &$0.4924i$\\
       & $2$ & $0.0000-18.9772i$            & $0.2502i$      &        & $2$ & $0.0000-25.5247i$      &$0.2454i$\\
       & $3$ & $0.0000-19.2272i$            & $0.2500i$      &        & $3$ & $0.0000-25.7704i$      &$0.2457i$\\
       & $4$ & $0.0000-19.4774i$            & $0.2502i$      &        & $4$ & $0.0000-26.0168i$      &$0.2464i$\\ 
       & $5$ & $0.0000-19.7275i$            & $0.2501i$      &        & $5$ & $0.0000-26.2624i$      &$0.2456i$\\
\hline
$0.1$  & $0$ & $0.0000-18.7225i$            & \mbox{N/A}     & $0.6$  & $0$ & \mbox{N/A}             &\mbox{N/A}\\
       & $1$ & $0.0000-18.9724i$            & $0.2499i$      &        & $1$ & \mbox{N/A}             &\mbox{N/A}\\
       & $2$ & $0.0000-19.2225i$            & $0.2501i$      &        & $2$ & \mbox{N/A}             &\mbox{N/A}\\
       & $3$ & $0.0000-19.4727i$            & $0.2502i$      &        & $3$ & \mbox{N/A}             &\mbox{N/A}\\
       & $4$ & $0.0000-19.7227i$            & $0.2500i$      &        & $4$ & \mbox{N/A}             &\mbox{N/A}\\ 
       & $5$ & $0.0000-19.9729i$            & $0.2502i$      &        & $5$ & \mbox{N/A}             &\mbox{N/A}\\
\hline
$0.2$  & $0$ & $0.0000-19.2036i$            & \mbox{N/A}     & $0.7$  & $0$ & \mbox{N/A}             &\mbox{N/A}\\
       & $1$ & $0.0000-19.4533i$            & $0.2497i$      &        & $1$ & \mbox{N/A}             &\mbox{N/A}\\
       & $2$ & $0.0000-19.7032i$            & $0.2499i$      &        & $2$ & \mbox{N/A}             &\mbox{N/A}\\
       & $3$ & $0.0000-19.9532i$            & $0.2500i$      &        & $3$ & \mbox{N/A}             &\mbox{N/A}\\
       & $4$ & $0.0000-20.2030i$            & $0.2498i$      &        & $4$ & \mbox{N/A}             &\mbox{N/A}\\
       & $5$ & $0.0000-20.4530i$            & $0.2500i$      &        & $5$ & \mbox{N/A}             &\mbox{N/A}\\
\hline
$0.3$  & $0$ & $0.0000-19.1563i$            & \mbox{N/A}     & $0.8$  & $0$ & $0.0000-9.2501i$       &\mbox{N/A}\\
       & $1$ & $0.0000-19.9043i$            & $0.7480i$      &        & $1$ & $0.0000-9.5000i$       & $0.2499i$ \\
       & $2$ & $0.0000-20.1541i$            & $0.2498i$      &        & $2$ & $0.0000-9.7500i$       & $0.2500i$ \\
       & $3$ & $0.0000-20.4033i$            & $0.2492i$      &        & $3$ & $0.0000-10.000i$       & $0.2500i$ \\
       & $4$ & $0.0000-20.6526i$            & $0.2493i$      &        & $4$ & $0.0000-10.250i$       & $0.2500i$ \\ 
       & $5$ & $0.0000-20.9020i$            & $0.2494i$      &        & $5$ & $0.0000-10.500i$       & $0.2500i$ \\
\hline
$0.4$  & $0$ & $0.0000-21.0467i$            & \mbox{N/A}     & $0.85$ & $0$ & $0.0000-5.2500i$       &\mbox{N/A}\\
       & $1$ & $0.0000-21.2946i$            & $0.2479i$      &        & $1$ & $0.0000-5.5000i$       & $0.2500i$ \\
       & $2$ & $0.0000-21.5427i$            & $0.2481i$      &        & $2$ & $0.0000-5.7500i$       & $0.2500i$ \\
       & $3$ & $0.0000-21.7910i$            & $0.2483i$      &        & $3$ & $0.0000-6.0000i$       & $0.2500i$ \\
       & $4$ & $0.0000-22.0391i$            & $0.2481i$      &        & $4$ & $0.0000-6.2500i$       & $0.2500i$ \\ 
       & $5$ & $0.0000-22.2872i$            & $0.2481i$      &        & $5$ & $0.0000-6.5000i$       & $0.2500i$ \\
       [1ex]
 \hline\hline 
 \end{tabular}
\end{table}

\begin{table}[t]
\centering
\caption{\textit{QNMs of electromagnetic perturbations ($s=1$) of the KSQC black hole for several values of $\ell$ and the deformation parameter $\mathfrak{a}=a/M$. We compare our spectral method results with $N=200$ Chebyshev polynomials (final column) against the third-order WKB employed by \cite{WangJAA2017}. Entries labelled N/A indicate unavailable data. The notation 'SM' stands for Spectral Method.}}
\label{table_s1_a0_2_and_a0_4}
\setlength\tabcolsep{0.1cm}
\def\arraystretch{1.5}
\begin{tabular}{@{}|c| c| c| c| c| c|c|c|c|c|c|c|@{}} 
\hline
$\mathfrak{a}$    & $a_K$   & $\ell$ & $n$ & $\Omega_{WKB}$ \cite{WangJAA2017} & $\Omega_{SM}$ &$a$    & $a_K$   & $\ell$ & $n$ & $\Omega_{WKB}$ \cite{WangJAA2017} & $\Omega_{SM}$ \\ [0.5ex] 
\hline
$0.2$ & $0.099$  & $1$    & $0$ & $0.24492-0.09299i$ & $0.2473-0.0924i$ & $0.4$ & $0.196$& $1$ & $0$ & $0.24214-0.09266i$& $0.2446-0.0920i$\\ 
       &         &        & $1$ & $0.21024-0.29556i$ & $0.2135-0.2934i$ &       &        &     & $1$ & $0.20710-0.29474i$& $0.2104-0.2925i$\\
       &         &        & $2$ & $0.16304-0.50865i$ & $0.1736-0.5248i$ &       &        &     & $2$ & $0.15936-0.50734i$& $0.1702-0.5237i$\\ 
       &         &        & $3$ & $0.10037-0.72506i$ & $0.1450-0.7715i$ &       &        &     & $3$ & $0.09586-0.72340i$& $0.1415-0.7703i$\\
       &         &        & $4$ & \mbox{N/A}         & $0.1253-1.0221i$ &       &        &     & $4$ & \mbox{N/A}        & $0.1218-1.0207i$\\ 
       &         &        & $5$ & \mbox{N/A}         & $0.1110-1.2734i$ &       &        &     & $5$ & \mbox{N/A}        & $0.1075-1.2718i$\\
       \hline
       &         & $2$    & $0$ & $0.45555-0.09496i$ & $0.4560-0.0949i$ &       &        & $2$ & $0$ & $0.45091-0.09463i$& $0.4514-0.0946i$\\  
       &         &        & $1$ & $0.43416-0.29067i$ & $0.4349-0.2904i$ &       &        &     & $1$ & $0.42929-0.28975i$& $0.4300-0.2895i$\\
       &         &        & $2$ & $0.40053-0.49538i$ & $0.3994-0.5011i$ &       &        &     & $2$ & $0.39530-0.49395i$& $0.3942-0.4998i$\\ 
       &         &        & $3$ & $0.35854-0.70499i$ & $0.3607-0.7297i$ &       &        &     & $3$ & $0.35281-0.70305i$& $0.3551-0.7282i$\\
       &         &        & $4$ & \mbox{N/A}         & $0.3267-0.9711i$ &       &        &     & $4$ & \mbox{N/A}        & $0.3209-0.9695i$\\ 
       &         &        & $5$ & \mbox{N/A}         & $0.2995-1.2192i$ &       &        &     & $5$ & \mbox{N/A}        & $0.2935-1.2175i$\\
       &         &        & $6$ & \mbox{N/A}         & $0.2778-1.4702i$ &       &        &     & $6$ & \mbox{N/A}        & $0.2717-1.4684i$\\
       \hline
       &         & $3$    & $0$ & $0.65451-0.09552i$ & $0.6547-0.0955i$ &       &        & $3$ & $0$ & $0.64800-0.09520i$& $0.6482-0.0952i$\\ 
       &         &        & $1$ & $0.63919-0.28948i$ & $0.6395-0.2894i$ &       &        &     & $1$ & $0.63251-0.28855i$& $0.6328-0.2885i$\\
       &         &        & $2$ & $0.61273-0.48956i$ & $0.6114-0.4916i$ &       &        &     & $2$ & $0.60577-0.48806i$& $0.6045-0.4901i$\\ 
       &         &        & $3$ & $0.57891-0.69487i$ & $0.5754-0.7057i$ &       &        &     & $3$ & $0.57158-0.69284i$& $0.5680-0.7039i$\\
       &         &        & $4$ & \mbox{N/A}         & $0.5371-0.9325i$ &       &        &     & $4$ & \mbox{N/A}        & $0.5294-0.9305i$\\
       &         &        & $5$ & \mbox{N/A}         & $0.5012-1.1697i$ &       &        &     & $5$ & \mbox{N/A}        & $0.4932-1.1676i$\\
       &         &        & $6$ & \mbox{N/A}         & $0.4696-1.4138i$ &       &        &     & $6$ & \mbox{N/A}        & $0.4614-1.4116i$\\
       \hline
       &         & $4$    & $0$ & $0.85016-0.09576i$ & $0.8502-0.0958i$ &       &        & $4$ & $0$ & $0.84177-0.09544i$& $0.8418-0.0954i$\\ 
       &         &        & $1$ & $0.83823-0.28902i$ & $0.8384-0.2890i$ &       &        &     & $1$ & $0.82972-0.28808i$& $0.8298-0.2881i$\\
       &         &        & $2$ & $0.81657-0.48649i$ & $0.8157-0.4873i$ &       &        &     & $2$ & $0.80783-0.48497i$& $0.8070-0.4858i$\\ 
       &         &        & $3$ & $0.78785-0.68853i$ & $0.7846-0.6936i$ &       &        &     & $3$ & $0.77880-0.68646i$& $0.7755-0.6916i$\\
       &         &        & $4$ & \mbox{N/A}         & $0.7483-0.9095i$ &       &        &     & $4$ & \mbox{N/A}        & $0.7388-0.9072i$\\
       &         &        & $5$ & \mbox{N/A}         & $0.7103-1.1352i$ &       &        &     & $5$ & \mbox{N/A}        & $0.7004-1.1327i$\\
       &         &        & $6$ & \mbox{N/A}         & $0.6734-1.3694i$ &       &        &     & $6$ & \mbox{N/A}        & $0.6633-1.3668i$\\
       \hline
       &         & $5$    & $0$ & $1.04437-0.09588i$ & $1.0444-0.0959i$ &       &        & $5$ & $0$ & $1.03411-0.09556i$& $1.0342-0.0956i$\\ 
       &         &        & $1$ & $1.03461-0.28880i$ & $1.0347-0.2888i$ &       &        &     & $1$ & $1.02425-0.28785i$& $1.0243-0.2878i$\\
       &         &        & $2$ & $1.01636-0.48473i$ & $1.0158-0.4851i$ &       &        &     & $2$ & $1.00581-0.48319i$& $1.0053-0.4836i$\\ 
       &         &        & $3$ & $0.99144-0.68440i$ & $0.9891-0.6870i$ &       &        &     & $3$ & $0.98062-0.68229i$& $0.9782-0.6849i$\\
       &         &        & $4$ & \mbox{N/A}         & $0.9563-0.8959i$ &       &        &     & $4$ & \mbox{N/A}        & $0.9451-0.8934i$\\
       &         &        & $5$ & \mbox{N/A}         & $0.9198-1.1129i$ &       &        &     & $5$ & \mbox{N/A}        & $0.9082-1.1101i$\\
       &         &        & $6$ & \mbox{N/A}         & $0.8819-1.3378i$ &       &        &     & $6$ & \mbox{N/A}        & $0.8700-1.3348i$\\
     [0.5ex] 
 \hline
 \end{tabular}
\end{table}

\begin{table}[t]
\centering
\caption{\textit{Candidate purely imaginary QNMs for electromagnetic perturbations of a KSQC black hole for several values of $\ell$ and the deformation parameter $\mathfrak{a}=a/M$. The corresponding results are obtained through our SM, utilising $200$ polynomials with a precision of $200$ digits. In this context, $\Omega$ and $n$ represent the dimensionless frequency and the corresponding overtone, respectively, while $\Delta\Omega=\Omega_n-\Omega_{n+1}$. Entries labelled N/A indicate unavailable data.}}
\label{table_s1_a0_2_and_a0_4_overdamped}
\setlength\tabcolsep{0.1cm}
\def\arraystretch{1.5}
\begin{tabular}{@{}|c| c| c| c| c| c|c|c|c|c|c|c|@{}} 
\hline
$\mathfrak{a}$    & $a_K$   & $\ell$ & $n$ & $\Omega$ & $\Delta\Omega$ & $a$    & $a_K$   & $\ell$ & $n$ & $\Omega$ & $\Delta\Omega$ \\ [0.5ex] 
\hline
$0.2$ & $0.099$  & $1$    & $0$ & $0.0000-18.7226i$   & \mbox{N/A}   & $0.4$ & $0.196$& $1$ & $0$ & $0.0000-18.2045i$   & \mbox{N/A}\\ 
       &         &        & $1$ & $0.0000-18.9725i$   & $0.2499i$    &       &        &     & $1$ & $0.0000-19.2047i$   & $1.0002i$\\
       &         &        & $2$ & $0.0000-19.2225i$   & $0.2500i$    &       &        &     & $2$ & $0.0000-19.4545i$   & $0.2498i$\\ 
       &         &        & $3$ & $0.0000-19.4728i$   & $0.2503i$    &       &        &     & $3$ & $0.0000-19.7044i$   & $0.2499i$\\
       &         &        & $4$ & $0.0000-19.7228i$   & $0.2500i$    &       &        &     & $4$ & $0.0000-19.9544i$   & $0.2500i$\\ 
       &         &        & $5$ & $0.0000-19.9729i$   & $0.2501i$    &       &        &     & $5$ & $0.0000-20.2042i$   & $0.2498i$\\
       &         &        & $6$ & $0.0000-20.2230i$   & $0.2501i$    &       &        &     & $6$ & $0.0000-20.4542i$   & $0.2500i$\\
       \hline
       &         & $2$    & $0$ & $0.0000-18.6769i$   & \mbox{N/A}   &       &        & $2$ & $0$ & $0.0000-19.1599i$   & \mbox{N/A}\\  
       &         &        & $1$ & $0.0000-18.9273i$   & $0.2504i$    &       &        &     & $1$ & $0.0000-19.4100i$   & $0.2501i$\\
       &         &        & $2$ & $0.0000-19.1773i$   & $0.2500i$    &       &        &     & $2$ & $0.0000-19.6600i$   & $0.2500i$\\ 
       &         &        & $3$ & $0.0000-19.4281i$   & $0.2508i$    &       &        &     & $3$ & $0.0000-19.9104i$   & $0.2504i$\\
       &         &        & $4$ & $0.0000-19.6783i$   & $0.2502i$    &       &        &     & $4$ & $0.0000-20.1605i$   & $0.2501i$\\ 
       &         &        & $5$ & $0.0000-19.9287i$   & $0.2504i$    &       &        &     & $5$ & $0.0000-20.4107i$   & $0.2502i$\\
       &         &        & $6$ & $0.0000-20.1791i$   & $0.2504i$    &       &        &     & $6$ & $0.0000-20.6609i$   & $0.2502i$\\
       \hline
       &         & $3$    & $0$ & $0.0000-19.3630i$   & \mbox{N/A}   &       &        & $3$ & $0$ & $0.0000-19.8458i$   & \mbox{N/A}\\ 
       &         &        & $1$ & $0.0000-19.6134i$   & $0.2504i$    &       &        &     & $1$ & $0.0000-20.0962i$   & $0.2504i$\\
       &         &        & $2$ & $0.0000-19.8641i$   & $0.2507i$    &       &        &     & $2$ & $0.0000-20.3467i$   & $0.2505i$\\ 
       &         &        & $3$ & $0.0000-20.1150i$   & $0.2509i$    &       &        &     & $3$ & $0.0000-20.5973i$   & $0.2506i$\\
       &         &        & $4$ & $0.0000-20.3656i$   & $0.2506i$    &       &        &     & $4$ & $0.0000-20.8478i$   & $0.2505i$\\ 
       &         &        & $5$ & $0.0000-20.6164i$   & $0.2508i$    &       &        &     & $5$ & $0.0000-21.0983i$   & $0.2505i$\\
       &         &        & $6$ & $0.0000-20.8671i$   & $0.2507i$    &       &        &     & $6$ & $0.0000-21.3488i$   & $0.2505i$\\
       \hline
       &         & $4$    & $0$ & $0.0000-19.2812i$   & \mbox{N/A}   &       &        & $4$ & $0$ & $0.0000-19.7643i$   & \mbox{N/A}\\ 
       &         &        & $1$ & $0.0000-20.0349i$   & $0.7537i$    &       &        &     & $1$ & $0.0000-20.2658i$   & $0.5015i$\\
       &         &        & $2$ & $0.0000-20.2856i$   & $0.2507i$    &       &        &     & $2$ & $0.0000-20.5172i$   & $0.2514i$\\ 
       &         &        & $3$ & $0.0000-20.5368i$   & $0.2512i$    &       &        &     & $3$ & $0.0000-20.7680i$   & $0.2508i$\\
       &         &        & $4$ & $0.0000-20.7880i$   & $0.2512i$    &       &        &     & $4$ & $0.0000-21.0189i$   & $0.2509i$\\ 
       &         &        & $5$ & $0.0000-21.0390i$   & $0.2510i$    &       &        &     & $5$ & $0.0000-21.2699i$   & $0.2510i$\\
       &         &        & $6$ & $0.0000-21.2901i$   & $0.2511i$    &       &        &     & $6$ & $0.0000-21.5207i$   & $0.2508i$\\
       \hline
       &         & $5$    & $0$ & $0.0000-20.6975i$   & \mbox{N/A}   &       &        & $5$ & $0$ & $0.0000-20.4253i$   & \mbox{N/A}\\ 
       &         &        & $1$ & $0.0000-20.9487i$   & $0.2512i$    &       &        &     & $1$ & $0.0000-20.6766i$   & $0.2513i$\\
       &         &        & $2$ & $0.0000-21.2002i$   & $0.2515i$    &       &        &     & $2$ & $0.0000-20.9275i$   & $0.2509i$\\ 
       &         &        & $3$ & $0.0000-21.4517i$   & $0.2515i$    &       &        &     & $3$ & $0.0000-21.1792i$   & $0.2517i$\\
       &         &        & $4$ & $0.0000-21.7031i$   & $0.2514i$    &       &        &     & $4$ & $0.0000-21.4303i$   & $0.2511i$\\ 
       &         &        & $5$ & $0.0000-21.9546i$   & $0.2515i$    &       &        &     & $5$ & $0.0000-21.6815i$   & $0.2512i$\\
       &         &        & $6$ & $0.0000-22.2059i$   & $0.2513i$    &       &        &     & $6$ & $0.0000-21.9328i$   & $0.2513i$\\
     [0.5ex] 
 \hline
 \end{tabular}
\end{table}

\begin{table}[t]
\centering
\caption{\textit{QNMs of electromagnetic perturbations ($s=1$) of the KSQC black hole for several values of $\ell$ and the deformation parameter $\mathfrak{a}=a/M$. We compare our spectral method results with $N=200$ Chebyshev polynomials (final column) against the third-order WKB employed by \cite{WangJAA2017}. Entries labelled N/A indicate unavailable data. The notation 'SM' stands for Spectral Method.}}
\label{table_s1_a0_6_and_a0_8}
\setlength\tabcolsep{0.1cm}
\def\arraystretch{1.5}
\begin{tabular}{@{}|c| c| c| c| c| c|c|c|c|c|c|c|@{}} 
\hline
$\mathfrak{a}$    & $a_K$   & $\ell$ & $n$ & $\Omega_{WKB}$ \cite{WangJAA2017} & $\Omega_{SM}$ &$a$    & $a_K$   & $\ell$ & $n$ & $\Omega_{WKB}$ \cite{WangJAA2017} & $\Omega_{SM}$ \\ [0.5ex] 
\hline
$0.6$ & $0.287$  & $1$    & $0$ & $0.23773-0.09211i$ & $0.2402-0.0914i$ & $0.8$ & $0.371$& $1$ & $0$ & $0.23197-0.09137i$& $0.2345-0.0906i$\\ 
       &         &        & $1$ & $0.20215-0.29340i$ & $0.2054-0.2910i$ &       &        &     & $1$ & $0.19572-0.29156i$& $0.1990-0.2890i$\\
       &         &        & $2$ & $0.15355-0.50527i$ & $0.1649-0.5220i$ &       &        &     & $2$ & $0.14605-0.50235i$& $0.1579-0.5195i$\\ 
       &         &        & $3$ & $0.08875-0.72066i$ & $0.1360-0.7683i$ &       &        &     & $3$ & $0.07959-0.71684i$& $0.1290-0.7655i$\\
       &         &        & $4$ & \mbox{N/A}         & $0.1164-1.0184i$ &       &        &     & $4$ & \mbox{N/A}        & $0.1094-1.0152i$\\ 
       &         &        & $5$ & \mbox{N/A}         & $0.1021-1.2692i$ &       &        &     & $5$ & \mbox{N/A}        & \mbox{N/A} \\
       \hline
       &         & $2$    & $0$ & $0.44353-0.09409i$ & $0.4440-0.0940i$ &       &        & $2$ & $0$ & $0.43389-0.09336i$& $0.4344-0.0933i$\\  
       &         &        & $1$ & $0.42156-0.28825i$ & $0.4223-0.2880i$ &       &        &     & $1$ & $0.41145-0.28621i$& $0.4122-0.2859i$\\
       &         &        & $2$ & $0.38701-0.49159i$ & $0.3858-0.4977i$ &       &        &     & $2$ & $0.37620-0.48834i$& $0.3749-0.4947i$\\ 
       &         &        & $3$ & $0.34374-0.69985i$ & $0.3462-0.7257i$ &       &        &     & $3$ & $0.33194-0.69543i$& $0.3347-0.7223i$\\
       &         &        & $4$ & \mbox{N/A}         & $0.3118-0.9669i$ &       &        &     & $4$ & \mbox{N/A}        & $0.2999-0.9632i$\\ 
       &         &        & $5$ & \mbox{N/A}         & $0.2842-1.2147i$ &       &        &     & $5$ & \mbox{N/A}        & $0.2721-1.2108i$\\
       &         &        & $6$ & \mbox{N/A}         & $0.2623-1.4654i$ &       &        &     & $6$ & \mbox{N/A}        & $0.2501-1.4612i$\\
       \hline
       &         & $3$    & $0$ & $0.63764-0.09467i$ & $0.6378-0.0947i$ &       &        & $3$ & $0$ & $0.62409-0.09395i$& $0.6243-0.0939i$\\ 
       &         &        & $1$ & $0.62189-0.28701i$ & $0.6222-0.2869i$ &       &        &     & $1$ & $0.60801-0.28491i$& $0.6083-0.2848i$\\
       &         &        & $2$ & $0.59472-0.48561i$ & $0.5934-0.4878i$ &       &        &     & $2$ & $0.58029-0.48224i$& $0.5788-0.4845i$\\ 
       &         &        & $3$ & $0.55995-0.68951i$ & $0.5563-0.7010i$ &       &        &     & $3$ & $0.54479-0.68492i$& $0.5411-0.6969i$\\
       &         &        & $4$ & \mbox{N/A}         & $0.5172-0.9272i$ &       &        &     & $4$ & \mbox{N/A}        & $0.5012-0.9227i$\\
       &         &        & $5$ & \mbox{N/A}         & $0.4805-1.1641i$ &       &        &     & $5$ & \mbox{N/A}        & $0.4641-1.1592i$\\
       &         &        & $6$ & \mbox{N/A}         & $0.4485-1.4080i$ &       &        &     & $6$ & \mbox{N/A}        & $0.4318-1.4029i$\\
       \hline
       &         & $4$    & $0$ & $0.82844-0.09491i$ & $0.8285-0.0949i$ &       &        & $4$ & $0$ & $0.81099-0.09419i$& $0.8111-0.0942i$\\ 
       &         &        & $1$ & $0.81618-0.28653i$ & $0.8163-0.2865i$ &       &        &     & $1$ & $0.79847-0.28441i$& $0.7986-0.2844i$\\
       &         &        & $2$ & $0.79393-0.48246i$ & $0.7930-0.4834i$ &       &        &     & $2$ & $0.77576-0.47903i$& $0.7748-0.4800i$\\ 
       &         &        & $3$ & $0.76443-0.68304i$ & $0.7611-0.6884i$ &       &        &     & $3$ & $0.74566-0.67835i$& $0.7422-0.6840i$\\
       &         &        & $4$ & \mbox{N/A}         & $0.7238-0.9034i$ &       &        &     & $4$ & \mbox{N/A}        & $0.7041-0.8982i$\\
       &         &        & $5$ & \mbox{N/A}         & $0.6848-1.1285i$ &       &        &     & $5$ & \mbox{N/A}        & $0.6645-1.1228i$\\
       &         &        & $6$ & \mbox{N/A}         & $0.6472-1.3624i$ &       &        &     & $6$ & \mbox{N/A}        & $0.6263-1.3564i$\\
       \hline
       &         & $5$    & $0$ & $1.01781-0.09503i$ & $1.0178-0.0950i$ &       &        & $5$ & $0$ & $0.99646-0.09431i$& $0.9965-0.0943i$\\ 
       &         &        & $1$ & $1.00778-0.28629i$ & $1.0078-0.2863i$ &       &        &     & $1$ & $0.98621-0.28416i$& $0.9863-0.2841i$\\
       &         &        & $2$ & $0.98903-0.48065i$ & $0.9884-0.4811i$ &       &        &     & $2$ & $0.96707-0.47717i$& $0.9665-0.4776i$\\ 
       &         &        & $3$ & $0.96343-0.67882i$ & $0.9610-0.6815i$ &       &        &     & $3$ & $0.94096-0.67406i$& $0.9384-0.6769i$\\
       &         &        & $4$ & \mbox{N/A}         & $0.9273-0.8894i$ &       &        &     & $4$ & \mbox{N/A}        & $0.9040-0.8838i$\\
       &         &        & $5$ & \mbox{N/A}         & $0.8898-1.1055i$ &       &        &     & $5$ & \mbox{N/A}        & $0.8658-1.0991i$\\
       &         &        & $6$ & \mbox{N/A}         & $0.8510-1.3299i$ &       &        &     & $6$ & \mbox{N/A}        & $0.8263-1.3230i$\\
     [0.5ex] 
 \hline
 \end{tabular}
\end{table}

\begin{table}[t]
\centering
\caption{\textit{Candidate purely imaginary QNMs for electromagnetic perturbations of a KSQC black hole for several values of $\ell$ and the deformation parameter $\mathfrak{a}=a/M$. The corresponding results are obtained through our SM, utilising $200$ polynomials with a precision of $200$ digits. In this context, $\Omega$ and $n$ represent the dimensionless frequency and the corresponding overtone, respectively, while $\Delta\Omega=\Omega_n-\Omega_{n+1}$. }}
\label{table_s1_a0_6_and_a0_8_overdamped}
\setlength\tabcolsep{0.1cm}
\def\arraystretch{1.5}
\begin{tabular}{@{}|c| c| c| c| c| c|c|c|c|c|c|c|@{}} 
\hline
$\mathfrak{a}$    & $a_K$   & $\ell$ & $n$ & $\Omega$ & $\Delta\Omega$ & $a$    & $a_K$   & $\ell$ & $n$ & $\Omega$ & $\Delta\Omega$ \\ [0.5ex] 
\hline
$0.6$  & $0.287$ & $1$    & $0$ & $0.0000-19.9128i$   & \mbox{N/A}   & $0.8$ & $0.371$& $1$ & $0$ & $0.0000-15.9915i$   & \mbox{N/A}\\ 
       &         &        & $1$ & $0.0000-20.1625i$   & $0.2497i$    &       &        &     & $1$ & $0.0000-20.5869i$   & $4.5954i$\\
       &         &        & $2$ & $0.0000-20.4120i$   & $0.2495i$    &       &        &     & $2$ & $0.0000-20.8350i$   & $0.2481i$\\ 
       &         &        & $3$ & $0.0000-20.6613i$   & $0.2493i$    &       &        &     & $3$ & $0.0000-21.0837i$   & $0.2487i$\\
       &         &        & $4$ & $0.0000-20.9108i$   & $0.2495i$    &       &        &     & $4$ & $0.0000-21.3323i$   & $0.2486i$\\ 
       &         &        & $5$ & $0.0000-21.1602i$   & $0.2494i$    &       &        &     & $5$ & $0.0000-21.5808i$   & $0.2485i$\\
       &         &        & $6$ & $0.0000-21.4097i$   & $0.2495i$    &       &        &     & $6$ & $0.0000-21.8294i$   & $0.2486i$\\
       \hline
       &         & $2$    & $0$ & $0.0000-19.8694i$   & \mbox{N/A}     &        &     & $2$ & $0$ & $0.0000-20.7934i$   & \mbox{N/A}\\  
       &         &        & $1$ & $0.0000-20.1192i$   & $0.2498i$    &       &        &     & $1$ & $0.0000-21.0422i$   & $0.2488i$\\
       &         &        & $2$ & $0.0000-20.3690i$   & $0.2498i$    &       &        &     & $2$ & $0.0000-21.2912i$   & $0.2490i$\\ 
       &         &        & $3$ & $0.0000-20.6186i$   & $0.2496i$    &       &        &     & $3$ & $0.0000-21.5398i$   & $0.2486i$\\
       &         &        & $4$ & $0.0000-20.8683i$   & $0.2497i$    &       &        &     & $4$ & $0.0000-21.7887i$   & $0.2489i$\\ 
       &         &        & $5$ & $0.0000-21.1180i$   & $0.2497i$    &       &        &     & $5$ & $0.0000-22.0375i$   & $0.2488i$\\
       &         &        & $6$ & $0.0000-21.3677i$   & $0.2497i$    &       &        &     & $6$ & $0.0000-22.2863i$   & $0.2488i$\\
       \hline
       &         & $3$    & $0$ & $0.0000-19.8052i$   & \mbox{N/A}   &       &        & $3$ & $0$ & $0.0000-21.2298i$   & \mbox{N/A}\\ 
       &         &        & $1$ & $0.0000-20.3055i$   & $0.4998i$    &       &        &     & $1$ & $0.0000-21.4785i$   & $0.2487i$\\
       &         &        & $2$ & $0.0000-20.5551i$   & $0.2496i$    &       &        &     & $2$ & $0.0000-21.7277i$   & $0.2492i$\\ 
       &         &        & $3$ & $0.0000-20.8053i$   & $0.2502i$    &       &        &     & $3$ & $0.0000-21.9770i$   & $0.2493i$\\
       &         &        & $4$ & $0.0000-21.0554i$   & $0.2501i$    &       &        &     & $4$ & $0.0000-22.2260i$   & $0.2490i$\\ 
       &         &        & $5$ & $0.0000-21.3054i$   & $0.2500i$    &       &        &     & $5$ & $0.0000-22.4751i$   & $0.2491i$\\
       &         &        & $6$ & $0.0000-21.5554i$   & $0.2500i$    &       &        &     & $6$ & $0.0000-22.7242i$   & $0.2491i$\\
       \hline
       &         & $4$    & $0$ & $0.0000-20.4746i$   & \mbox{N/A}   &       &        & $4$ & $0$ & $0.0000-21.8993i$   & \mbox{N/A}\\ 
       &         &        & $1$ & $0.0000-20.7250i$   & $0.2504i$    &       &        &     & $1$ & $0.0000-22.1485i$   & $0.2492i$\\
       &         &        & $2$ & $0.0000-20.9759i$   & $0.2509i$    &       &        &     & $2$ & $0.0000-22.3980i$   & $0.2495i$\\ 
       &         &        & $3$ & $0.0000-21.2260i$   & $0.2501i$    &       &        &     & $3$ & $0.0000-22.6476i$   & $0.2496i$\\
       &         &        & $4$ & $0.0000-21.4765i$   & $0.2505i$    &       &        &     & $4$ & $0.0000-22.8969i$   & $0.2493i$\\ 
       &         &        & $5$ & $0.0000-21.7269i$   & $0.2504i$    &       &        &     & $5$ & $0.0000-23.1464i$   & $0.2495i$\\
       &         &        & $6$ & $0.0000-21.9773i$   & $0.2504i$    &       &        &     & $6$ & $0.0000-23.3958i$   & $0.2494i$\\
       \hline
       &         & $5$    & $0$ & $0.0000-21.1345i$   & \mbox{N/A}   &       &        & $5$ & $0$ & $0.0000-21.8091i$   & \mbox{N/A}\\ 
       &         &        & $1$ & $0.0000-21.3853i$   & $0.2508i$    &       &        &     & $1$ & $0.0000-22.0587i$   & $0.2496i$\\
       &         &        & $2$ & $0.0000-21.6364i$   & $0.2511i$    &       &        &     & $2$ & $0.0000-22.3081i$   & $0.2494i$\\ 
       &         &        & $3$ & $0.0000-21.8869i$   & $0.2505i$    &       &        &     & $3$ & $0.0000-22.5585i$   & $0.2504i$\\
       &         &        & $4$ & $0.0000-22.1378i$   & $0.2509i$    &       &        &     & $4$ & $0.0000-22.8081i$   & $0.2496i$\\ 
       &         &        & $5$ & $0.0000-22.3885i$   & $0.2507i$    &       &        &     & $5$ & $0.0000-23.0579i$   & $0.2498i$\\
       &         &        & $6$ & $0.0000-22.6392i$   & $0.2507i$    &       &        &     & $6$ & $0.0000-23.3078i$   & $0.2499i$\\
     [0.5ex] 
 \hline
 \end{tabular}
\end{table}

\begin{table}[t]
\centering
\caption{\textit{QNMs of electromagnetic perturbations ($s=1$) of the KSQC black hole for several values of $\ell$ and the deformation parameter $\mathfrak{a}=a/M$. We compare our spectral method results with $N=200$ Chebyshev polynomials (final column) against the third-order WKB employed by \cite{WangJAA2017}. Entries labelled N/A indicate unavailable data. The notation 'SM' stands for Spectral Method.}}
\label{table_s1_a1_0_and_a1_2}
\setlength\tabcolsep{0.1cm}
\def\arraystretch{1.5}
\begin{tabular}{@{}|c| c| c| c| c| c|c|c|c|c|c|c|@{}} 
\hline
$\mathfrak{a}$    & $a_K$   & $\ell$ & $n$ & $\Omega_{WKB}$ \cite{WangJAA2017} & $\Omega_{SM}$  &$a$    & $a_K$   & $\ell$ & $n$ & $\Omega_{WKB}$ \cite{WangJAA2017} & $\Omega_{SM}$  \\ [0.5ex] 
\hline
$1$    & $0.447$ & $1$    & $0$ & $0.22519-0.09048i$ & $0.2277-0.0896i$ & $1.2$ & $0.514$& $1$ & $0$ & $0.21773-0.08944i$& $0.2203-0.0884i$\\ 
       &         &        & $1$ & $0.18822-0.28928i$ & $0.1915-0.2865i$ &       &        &     & $1$ & $0.18003-0.28658i$& $0.1831-0.2836i$\\
       &         &        & $2$ & $0.13734-0.49868i$ & $0.1498-0.5165i$ &       &        &     & $2$ & $0.12791-0.49429i$& $0.1410-0.5129i$\\ 
       &         &        & $3$ & $0.06902-0.71199i$ & $0.1209-0.7620i$ &       &        &     & $3$ & $0.05766-0.70615i$& $0.1123-0.7578i$\\
       &         &        & $4$ & \mbox{N/A}         & $0.1015-1.0111i$ &       &        &     & $4$ & \mbox{N/A}        & $0.0933-1.0063i$\\ 
       \hline
       &         & $2$    & $0$ & $0.42250-0.09246i$ & $0.4230-0.0924i$ &       &        & $2$ & $0$ & $0.40991-0.09140i$& $0.4104-0.0913i$\\  
       &         &        & $1$ & $0.39955-0.28366i$ & $0.4002-0.2833i$ &       &        &     & $1$ & $0.38642-0.28067i$& $0.3870-0.2803i$\\
       &         &        & $2$ & $0.36352-0.48428i$ & $0.3621-0.4910i$ &       &        &     & $2$ & $0.34958-0.47948i$& $0.3480-0.4866i$\\ 
       &         &        & $3$ & $0.31814-0.68988i$ & $0.3212-0.7180i$ &       &        &     & $3$ & $0.30304-0.68329i$& $0.3063-0.7130i$\\
       &         &        & $4$ & \mbox{N/A}         & $0.2860-0.9585i$ &       &        &     & $4$ & \mbox{N/A}        & $0.2708-0.9530i$\\ 
       &         &        & $5$ & \mbox{N/A}         & $0.2581-1.2057i$ &       &        &     & $5$ & \mbox{N/A}        & $0.2430-1.1995i$\\
       &         &        & $6$ & \mbox{N/A}         & $0.2362-1.4556i$ &       &        &     & $6$ & \mbox{N/A}        & $0.2213-1.4488i$\\
       \hline
       &         & $3$    & $0$ & $0.60808-0.09304i$ & $0.6083-0.0930i$ &       &        & $3$ & $0$ & $0.59036-0.09199i$& $0.5905-0.0920i$\\ 
       &         &        & $1$ & $0.59162-0.28229i$ & $0.5919-0.2822i$ &       &        &     & $1$ & $0.57350-0.27921i$& $0.5737-0.2791i$\\
       &         &        & $2$ & $0.56328-0.47803i$ & $0.5617-0.4804i$ &       &        &     & $2$ & $0.54451-0.47306i$& $0.5428-0.4756i$\\ 
       &         &        & $3$ & $0.52696-0.67916i$ & $0.5231-0.6918i$ &       &        &     & $3$ & $0.50733-0.67235i$& $0.5032-0.6858i$\\
       &         &        & $4$ & \mbox{N/A}         & $0.4825-0.9169i$ &       &        &     & $4$ & \mbox{N/A}        & $0.4619-0.9101i$\\
       &         &        & $5$ & \mbox{N/A}         & $0.4449-1.1531i$ &       &        &     & $5$ & \mbox{N/A}        & $0.4239-1.1458i$\\
       &         &        & $6$ & \mbox{N/A}         & $0.4124-1.3964i$ &       &        &     & $6$ & \mbox{N/A}        & $0.3913-1.3886i$\\
       \hline
       &         & $4$    & $0$ & $0.79037-0.09329i$ & $0.7904-0.0933i$ &       &        & $4$ & $0$ & $0.76753-0.09223i$& $0.7676-0.0922i$\\ 
       &         &        & $1$ & $0.77755-0.28176i$ & $0.7777-0.2817i$ &       &        &     & $1$ & $0.75440-0.27865i$& $0.7545-0.2786i$\\
       &         &        & $2$ & $0.75432-0.47473i$ & $0.7533-0.4757i$ &       &        &     & $2$ & $0.73063-0.46967i$& $0.7295-0.4707i$\\ 
       &         &        & $3$ & $0.72354-0.67248i$ & $0.7198-0.6784i$ &       &        &     & $3$ & $0.69914-0.66553i$& $0.6952-0.6718i$\\
       &         &        & $4$ & \mbox{N/A}         & $0.6810-0.8917i$ &       &        &     & $4$ & \mbox{N/A}        & $0.6554-0.8840i$\\
       &         &        & $5$ & \mbox{N/A}         & $0.6406-1.1157i$ &       &        &     & $5$ & \mbox{N/A}        & $0.6143-1.1071i$\\
       &         &        & $6$ & \mbox{N/A}         & $0.6019-1.3488i$ &       &        &     & $6$ & \mbox{N/A}        & $0.5751-1.3396i$\\
       \hline
       &         & $5$    & $0$ & $0.97123-0.09341i$ & $0.9713-0.0934i$ &       &        & $5$ & $0$ & $0.94328-0.09236i$& $0.9433-0.0924i$\\ 
       &         &        & $1$ & $0.96074-0.28150i$ & $0.9608-0.2815i$ &       &        &     & $1$ & $0.93253-0.27837i$& $0.9326-0.2784i$\\
       &         &        & $2$ & $0.94116-0.47283i$ & $0.9405-0.4733i$ &       &        &     & $2$ & $0.91249-0.46770i$& $0.9118-0.4682i$\\ 
       &         &        & $3$ & $0.91445-0.66809i$ & $0.9117-0.6711i$ &       &        &     & $3$ & $0.88516-0.66105i$& $0.8823-0.6643i$\\
       &         &        & $4$ & \mbox{N/A}         & $0.8765-0.8768i$ &       &        &     & $4$ & \mbox{N/A}        & $0.8462-0.8685i$\\
       &         &        & $5$ & \mbox{N/A}         & $0.8375-1.0912i$ &       &        &     & $5$ & \mbox{N/A}        & $0.8063-1.0818i$\\
       &         &        & $6$ & \mbox{N/A}         & $0.7972-1.3144i$ &       &        &     & $6$ & \mbox{N/A}        & $0.7653-1.3041i$\\
     [0.5ex] 
 \hline
 \end{tabular}
\end{table}

\begin{table}[t]
\centering
\caption{\textit{Candidate purely imaginary QNMs for electromagnetic perturbations of a KSQC black hole for several values of $\ell$ and the deformation parameter $\mathfrak{a}=a/M$. The corresponding results are obtained through our SM, utilising $200$ polynomials with a precision of $200$ digits. In this context, $\Omega$ and $n$ represent the dimensionless frequency and the corresponding overtone, respectively, while $\Delta\Omega=\Omega_n-\Omega_{n+1}$. Entries labelled N/A indicate unavailable data.}}
\label{table_s1_a1_0_and_a1_2_overdamped}
\setlength\tabcolsep{0.1cm}
\def\arraystretch{1.5}
\begin{tabular}{@{}|c| c| c| c| c| c|c|c|c|c|c|c|@{}} 
\hline
$\mathfrak{a}$    & $a_K$   & $\ell$ & $n$ & $\Omega$ & $\Delta\Omega$ & $a$    & $a_K$   & $\ell$ & $n$ & $\Omega$ & $\Delta\Omega$ \\ 
\hline
$1$    & $0.447$ & $1$    & $0$ & $0.0000-21.7104i$   & \mbox{N/A}   &$1.2$  & $0.514$& $1$ & $0$ & $0.0000-26.2133i$   & \mbox{N/A}\\ 
       &         &        & $1$ & $0.0000-21.9578i$   & $0.2474i$    &       &        &     & $1$ & $0.0000-26.7056i$   & $0.4923i$\\
       &         &        & $2$ & $0.0000-22.2055i$   & $0.2477i$    &       &        &     & $2$ & $0.0000-26.9500i$   & $0.2444i$\\ 
       &         &        & $3$ & $0.0000-22.4523i$   & $0.2468i$    &       &        &     & $3$ & $0.0000-27.1950i$   & $0.2450i$\\
       &         &        & $4$ & $0.0000-22.6998i$   & $0.2475i$    &       &        &     & $4$ & \mbox{N/A}          & \mbox{N/A}\\ 
       &         &        & $5$ & $0.0000-22.9470i$   & $0.2472i$    &       &        &     & $5$ & \mbox{N/A}          & \mbox{N/A}\\
       &         &        & $6$ & $0.0000-23.1942i$   & $0.2472i$    &       &        &     & $6$ & \mbox{N/A}          & \mbox{N/A}\\
       \hline
       &         & $2$    & $0$ & $0.0000-22.4129i$   & \mbox{N/A}   &       &        & $2$ & $0$ & $0.0000-27.4082i$   & \mbox{N/A}\\  
       &         &        & $1$ & $0.0000-22.6608i$   & $0.2479i$    &       &        &     & $1$ & \mbox{N/A}   & \mbox{N/A}\\
       &         &        & $2$ & $0.0000-22.9087i$   & $0.2479i$    &       &        &     & $2$ & \mbox{N/A}   & \mbox{N/A}\\ 
       &         &        & $3$ & $0.0000-23.1556i$   & $0.2469i$    &       &        &     & $3$ & \mbox{N/A}   & \mbox{N/A}\\
       &         &        & $4$ & $0.0000-23.4033i$   & $0.2477i$    &       &        &     & $4$ & \mbox{N/A}   & \mbox{N/A}\\ 
       &         &        & $5$ & $0.0000-23.6508i$   & $0.2475i$    &       &        &     & $5$ & \mbox{N/A}   & \mbox{N/A}\\
       &         &        & $6$ & $0.0000-23.8980i$   & $0.2472i$    &       &        &     & $6$ & \mbox{N/A}   & \mbox{N/A}\\
       \hline
       &         & $3$    & $0$ & $0.0000-23.0974i$   & \mbox{N/A}   &       &        & $3$ & $0$ & \mbox{N/A}   & \mbox{N/A}\\ 
       &         &        & $1$ & $0.0000-23.3455i$   & $0.2481i$    &       &        &     & $1$ & \mbox{N/A}   & \mbox{N/A}\\
       &         &        & $2$ & $0.0000-23.5937i$   & $0.2482i$    &       &        &     & $2$ & \mbox{N/A}   & \mbox{N/A}\\ 
       &         &        & $3$ & $0.0000-23.8409i$   & $0.2472i$    &       &        &     & $3$ & \mbox{N/A}   & \mbox{N/A}\\
       &         &        & $4$ & $0.0000-24.0888i$   & $0.2479i$    &       &        &     & $4$ & \mbox{N/A}   & \mbox{N/A}\\ 
       &         &        & $5$ & $0.0000-24.3366i$   & $0.2478i$    &       &        &     & $5$ & \mbox{N/A}   & \mbox{N/A}\\
       &         &        & $6$ & $0.0000-24.5841i$   & $0.2475i$    &       &        &     & $6$ & \mbox{N/A}   & \mbox{N/A}\\
       \hline
       &         & $4$    & $0$ & $0.0000-23.7667i$   & \mbox{N/A}   &       &        & $4$ & $0$ & \mbox{N/A}   & \mbox{N/A}\\ 
       &         &        & $1$ & $0.0000-24.0148i$   & $0.2481i$    &       &        &     & $1$ & \mbox{N/A}   & \mbox{N/A}\\
       &         &        & $2$ & $0.0000-24.2633i$   & $0.2485i$    &       &        &     & $2$ & \mbox{N/A}   & \mbox{N/A}\\ 
       &         &        & $3$ & $0.0000-24.5109i$   & $0.2476i$    &       &        &     & $3$ & \mbox{N/A}   & \mbox{N/A}\\
       &         &        & $4$ & $0.0000-24.7590i$   & $0.2481i$    &       &        &     & $4$ & \mbox{N/A}   & \mbox{N/A}\\ 
       &         &        & $5$ & $0.0000-25.0072i$   & $0.2482i$    &       &        &     & $5$ & \mbox{N/A}   & \mbox{N/A}\\
       &         &        & $6$ & $0.0000-25.2550i$   & $0.2478i$    &       &        &     & $6$ & \mbox{N/A}   & \mbox{N/A}\\
       \hline
       &         & $5$    & $0$ & $0.0000-24.1774i$   & \mbox{N/A}   &       &        & $5$ & $0$ & \mbox{N/A}   & \mbox{N/A}\\ 
       &         &        & $1$ & $0.0000-24.4253i$   & $0.2479i$    &       &        &     & $1$ & \mbox{N/A}   & \mbox{N/A}\\
       &         &        & $2$ & $0.0000-24.6733i$   & $0.2480i$    &       &        &     & $2$ & \mbox{N/A}   & \mbox{N/A}\\ 
       &         &        & $3$ & $0.0000-24.9222i$   & $0.2489i$    &       &        &     & $3$ & \mbox{N/A}   & \mbox{N/A}\\
       &         &        & $4$ & $0.0000-25.1703i$   & $0.2481i$    &       &        &     & $4$ & \mbox{N/A}   & \mbox{N/A}\\ 
       &         &        & $5$ & $0.0000-25.4185i$   & $0.2482i$    &       &        &     & $5$ & \mbox{N/A}   & \mbox{N/A}\\
       &         &        & $6$ & $0.0000-25.6671i$   & $0.2486i$    &       &        &     & $6$ & \mbox{N/A}   & \mbox{N/A}\\
     [0.5ex] 
 \hline
 \end{tabular}
\end{table}

\begin{table}[t]
\centering
\caption{\textit{QNMs of electromagnetic perturbations ($s=1$) of the KSQC black hole for several values of $\ell$ and $\mathfrak{a}=a/M=1.5$. We compare our spectral method results with $N=200$ Chebyshev polynomials against the third-order WKB employed by \cite{WangJAA2017}. Entries labelled N/A indicate unavailable data. The notation 'SM' stands for Spectral Method.}}
\label{table_s1_a1_5_QNM_and_overdamped}
\setlength\tabcolsep{0.1cm}
\def\arraystretch{1.5}
\begin{tabular}{@{}|c| c| c| c| c| c| c| c|@{}} 
\hline
$\ell$ & $n$ & $\Omega_{WKB}$ \cite{WangJAA2017} & $\Omega_{SM}$ \\ [0.5ex] 
\hline
$1$    & $0$ & $0.20586-0.08771i$       & $0.2084-0.0864i$ \\ 
       & $1$ & $0.16724-0.28189i$       & $0.1698-0.2786i$ \\
       & $2$ & $0.11336-0.48651i$       & $0.1271-0.5068i$ \\ 
       & $3$ & $0.04029-0.69570i$       & $0.0990-0.7507i$ \\
       & $4$ & \mbox{N/A}               & $0.0809-0.9982i$ \\ 
       \hline
$2$    & $0$ & $0.38976-0.08958i$       & $0.3902-0.0894i$ \\ 
       & $1$ & $0.36551-0.27548i$       & $0.3659-0.2751i$ \\
       & $2$ & $0.32754-0.47108i$       & $0.3254-0.4790i$ \\ 
       & $3$ & $0.27932-0.67170i$       & $0.2828-0.7041i$ \\
       & $4$ & \mbox{N/A}               & $0.2472-0.9430i$ \\ 
       & $5$ & \mbox{N/A}               & $0.2198-1.1885i$ \\
       & $6$ & \mbox{N/A}               & $0.1988-1.4364i$ \\
$3$    & $0$ & $0.56198-0.09016i$       & $0.5621-0.0901i$ \\ 
       & $1$ & $0.54454-0.27387i$       & $0.5447-0.2738i$ \\
       & $2$ & $0.51462-0.46440i$       & $0.5125-0.4672i$ \\ 
       & $3$ & $0.47622-0.66042i$       & $0.4716-0.6752i$ \\
       & $4$ & \mbox{N/A}               & $0.4293-0.8982i$ \\ 
       & $5$ & \mbox{N/A}               & $0.3910-1.1327i$ \\
       & $6$ & \mbox{N/A}               & $0.3586-1.3745i$ \\
       \hline
$4$    & $0$ & $0.73094-0.09040i$       & $0.7310-0.0904i$ \\ 
       & $1$ & $0.71734-0.27323i$       & $0.7174-0.2732i$ \\
       & $2$ & $0.69278-0.46084i$       & $0.6915-0.4620i$ \\ 
       & $3$ & $0.66027-0.65339i$       & $0.6559-0.6604i$ \\
       & $4$ & \mbox{N/A}               & $0.6148-0.8704i$ \\ 
       & $5$ & \mbox{N/A}               & $0.5727-1.0921i$ \\
       & $6$ & \mbox{N/A}               & $0.5330-1.3235i$ \\
       \hline
$5$    & $0$ & $0.89850-0.09052i$       & $0.8985-0.0905i$ \\ 
       & $1$ & $0.88737-0.27292i$       & $0.8874-0.2729i$ \\
       & $2$ & $0.86663-0.45877i$       & $0.8658-0.4593i$ \\ 
       & $3$ & $0.83841-0.64875i$       & $0.8352-0.6524i$ \\
       & $4$ & \mbox{N/A}               & $0.7978-0.8540i$ \\ 
       & $5$ & \mbox{N/A}               & $0.7567-1.0652i$ \\
       & $6$ & \mbox{N/A}               & $0.7147-1.2860i$ \\
     [0.5ex] 
 \hline
 \end{tabular}
\end{table}

\begin{table}[t]
\centering
\caption{\textit{QNMs of electromagnetic perturbations ($s=1$) of the KSQC black hole for $\ell=1$ and several values of the deformation parameter $a_K$ for the near-extremal case. We compare our spectral method results with $N=200$ Chebyshev polynomials (final column) against the time-domain method adopted by \cite{KonoplyaPLB2020}. Entries labelled N/A indicate unavailable data. The notation 'SM' stands for Spectral Method.}}
\label{table_s1_nearly_extremal1}
\setlength\tabcolsep{0.1cm}
\def\arraystretch{1.5}
\begin{tabular}{@{}|c| c| c| c| c| c|c|c|c|c|@{}} 
\hline
$a_K$  & $\mathfrak{a}$ & $n$ & $\Omega_{TD}$ \cite{KonoplyaPLB2020} & $\Omega_{SM}$ & $a_K$  & $\mathfrak{a}$ & $n$ & $\Omega_{TD}$ \cite{KonoplyaPLB2020} & $\Omega_{SM}$\\ [0.5ex] 
\hline
$0.90$  & $\frac{18\sqrt{19}}{19}$ & $0$ & $0.12252-0.06553i$ & $0.1225-0.0655i$ & $0.93$ & $\frac{186\sqrt{1351}}{1351}$ & $0$ & $0.10419-0.05913i$ & $0.1042-0.0591i$\\ 
        &           & $1$ & \mbox{N/A}   & $0.0771-0.2309i$ &        &           & $1$ & \mbox{N/A}   & $0.0565-0.2189i$\\
        \hline
$0.91$  & $\frac{182\sqrt{191}}{573}$ & $0$ & $0.11688-0.06365i$ & $0.1169-0.0637i$ & $0.94$    & $\frac{94\sqrt{291}}{291}$ & $0$ &$0.09693-0.05634i$ & $0.0969-0.0563i$\\ 
        &           & $1$ & \mbox{N/A}   & $0.0709-0.2273i$ &        &           & $1$ & \mbox{N/A}   & $0.0476-0.2142i$\\
        \hline 
$0.92$  & $\frac{23\sqrt{6}}{12}$  & $0$ & $0.11080-0.06154i$ & $0.1108-0.0615i$ & $0.95$    & $\frac{38\sqrt{39}}{39}$ & $0$ & $0.08887-0.05307i$   & $0.0889-0.0531i$\\ 
        &           & $1$ & \mbox{N/A}   & $0.0641-0.2233i$ &        &           & $1$ & \mbox{N/A}   & $0.0368-0.2088i$\\    
     [0.5ex] 
 \hline
 \end{tabular}
\end{table}

\begin{table}[t]
\centering
\caption{\textit{QNMs of electromagnetic perturbations ($s=1$) of the KSQC black hole for $\ell=1$ and several values of the deformation parameter $a_K$ for the near-extremal case. We compare our spectral method results with $N=200$ Chebyshev polynomials (final column) against the time-domain method adopted by \cite{KonoplyaPLB2020}. Entries labelled N/A indicate unavailable data. The notation 'SM' stands for Spectral Method.}}
\label{table_s1_nearly_extremal2}
\setlength\tabcolsep{0.1cm}
\def\arraystretch{1.5}
\begin{tabular}{@{}|c| c| c| c| c| c|c|c|c|c|@{}} 
\hline
$a_K$  & $\mathfrak{a}$ & $n$ & $\Omega_{TD}$ \cite{KonoplyaPLB2020} & $\Omega_{SM}$ & $a_K$  & $\mathfrak{a}$ & $n$ & $\Omega_{TD}$ \cite{KonoplyaPLB2020} & $\Omega_{SM}$\\ [0.5ex] 
\hline
$0.96$  & $\frac{48}{7}$ & $0$ & $0.07978-0.04906i$ & $0.0798-0.0491i$ & $0.992$    & $\frac{248\sqrt{249}}{249}$ & $0$ & $0.03563-0.02538i$ & $0.0356-0.0254i$\\ 
        &                & $1$ & \mbox{N/A}         & $0.0224-0.2026i$ &            &                             & $1$ & \mbox{N/A}         & \mbox{N/A}\\
        \hline
$0.97$  & $\frac{194\sqrt{591}}{591}$ & $0$ & $0.06926-0.04414i$ & $0.0693-0.0441i$ & $0.994$    & $\frac{994\sqrt{2991}}{2991}$ & $0$ &$0.03080-0.02229i$ & $0.0308-0.0223i$\\ 
        \hline
$0.98$  & $\frac{98\sqrt{11}}{33}$  & $0$ & $0.05659-0.03762i$ & $0.0566-0.0376i$ & $0.996$    & $\frac{498\sqrt{499}}{499}$ & $0$ & $0.02508-0.01848i$   & $0.0251-0.0185i$\\ 
        \hline
$0.99$  & $\frac{198\sqrt{199}}{199}$  & $0$ & $0.03990-0.02803i$ & $0.03990-0.0280i$ & $0.998$    & $\frac{998\sqrt{111}}{333}$ & $0$ & $0.01767-0.01333i$   & $0.0177-0.0133i$\\ 
     [0.5ex] 
 \hline
 \end{tabular}
\end{table}

\begin{table}
\centering
\caption{Candidate purely imaginary QNMs for electromagnetic perturbations of a KSQC black hole for $\ell=1$ and several values of the deformation parameter $a_K$ close to the near-extremal case. The corresponding results are obtained through our SM, utilising $200$ polynomials with a precision of $200$ digits. In this context, $\Omega$ and $n$ represent the dimensionless frequency and the corresponding overtone, respectively, while $\Delta\Omega=\Omega_n-\Omega_{n+1}$. Entries labelled N/A indicate unavailable data.}
\label{scalargoverdampednearextremals1}
\vspace*{1em}
\begin{tabular}{||c|c|c|c|c|c|c|c|c||}
\hline\hline
$a_K$  & $n$ & $\Omega$           &$\Delta\Omega$  & $a_K$ & $n$ & $\Omega$  &$\Delta\Omega$\\ [0.5ex]
\hline\hline
$0.9$  & $0$ & $0.0000-2.5001i$   & \mbox{N/A}     & $0.95$  & $0$ & $0.0000-0.7506i$             &\mbox{N/A}\\
       & $1$ & $0.0000-2.7501i$   & $0.2500i$      &         & $1$ & $0.0000-1.0003i$             &$0.2497i$\\
       & $2$ & $0.0000-3.0001i$   & $0.2500i$      &         & $2$ & $0.0000-1.2502i$             &$0.2499i$\\
       & $3$ & $0.0000-3.2501i$   & $0.2500i$      &         & $3$ & $0.0000-1.5001i$             &$0.2499i$\\
       & $4$ & $0.0000-3.5000i$   & $0.2500i$      &         & $4$ & $0.0000-1.7501i$             &$0.2500i$\\ 
       & $5$ & $0.0000-3.7500i$   & $0.2500i$      &         & $5$ & $0.0000-2.0001i$             &$0.2500i$\\
\hline
$0.91$ & $0$ & $0.0000-2.0002i$   & \mbox{N/A}     & $0.96$  & $0$ & $0.0000-0.5010i$             &\mbox{N/A}\\
       & $1$ & $0.0000-2.2501i$   & $0.2499i$      &         & $1$ & $0.0000-0.7504i$             &$0.2494i$\\
       & $2$ & $0.0000-2.5001i$   & $0.2500i$      &         & $2$ & $0.0000-1.0002i$             &$0.2498i$\\
       & $3$ & $0.0000-2.7501i$   & $0.2500i$      &         & $3$ & $0.0000-1.2501i$             &$0.2499i$\\
       & $4$ & $0.0000-3.0001i$   & $0.2500i$      &         & $4$ & $0.0000-1.5001i$             &$0.2500i$\\ 
       & $5$ & $0.0000-3.2500i$   & $0.2499i$      &         & $5$ & $0.0000-1.7501i$             &$0.2500i$\\
\hline
$0.92$ & $0$ & $0.0000-1.7502i$   & \mbox{N/A}     & $0.97$  & $0$ & $0.0000-0.5006i$             &\mbox{N/A}\\
       & $1$ & $0.0000-2.0001i$   & $0.2499i$      &         & $1$ & $0.0000-0.7502i$             &$0.2496i$\\
       & $2$ & $0.0000-2.2501i$   & $0.2500i$      &         & $2$ & $0.0000-1.0001i$             &$0.2499i$\\
       & $3$ & $0.0000-2.5001i$   & $0.2500i$      &         & $3$ & $0.0000-1.2501i$             &$0.2500i$\\
       & $4$ & $0.0000-2.7501i$   & $0.2500i$      &         & $4$ & $0.0000-1.5000i$             &$0.2499i$\\ 
       & $5$ & $0.0000-3.0000i$   & $0.2499i$      &         & $5$ & $0.0000-1.7500i$             &$0.2500i$\\
\hline
$0.93$ & $0$ & $0.0000-1.2503i$   & \mbox{N/A}     & $0.98$ & $0$  & $0.0000-0.2515i$      &\mbox{N/A}\\
       & $1$ & $0.0000-1.5002i$   & $0.2499i$      &         & $1$ & $0.0000-0.5003i$             &$0.2488i$\\
       & $2$ & $0.0000-1.7501i$   & $0.2499i$      &         & $2$ & $0.0000-0.7501i$             &$0.2498i$\\
       & $3$ & $0.0000-2.0001i$   & $0.2500i$      &         & $3$ & $0.0000-1.0001i$             &$0.2500i$\\
       & $4$ & $0.0000-2.2501i$   & $0.2500i$      &         & $4$ & $0.0000-1.2500i$             &$0.2499i$\\ 
       & $5$ & $0.0000-2.5001i$   & $0.2500i$      &         & $5$ & $0.0000-1.5000i$             &$0.2500i$\\
\hline
$0.94$ & $0$ & $0.0000-1.0004i$   & \mbox{N/A}     & $0.99$  & $0$ & $0.0000-0.2505i$      &\mbox{N/A}\\
       & $1$ & $0.0000-1.2502i$   & $0.2498i$      &         & $1$ & $0.0000-0.5001i$             &$0.2496i$\\
       & $2$ & $0.0000-1.5001i$   & $0.2499i$      &         & $2$ & $0.0000-0.7500i$             &$0.2499i$\\
       & $3$ & $0.0000-1.7501i$   & $0.2500i$      &         & $3$ & $0.0000-1.0000i$             &$0.2500i$\\
       & $4$ & $0.0000-2.0001i$   & $0.2500i$      &         & $4$ & $0.0000-1.2500i$             &$0.2500i$\\ 
       & $5$ & $0.0000-2.2501i$   & $0.2500i$      &         & $5$ & $0.0000-1.5000i$             &$0.2500i$\\
       [1ex]
 \hline\hline 
 \end{tabular}
\end{table}

\begin{table}
\centering
\caption{Candidate purely imaginary QNMs for electromagnetic perturbations of a KSQC black hole for $\ell=1$ and several values of the deformation parameter $a_K$ close to the near-extremal case. The corresponding results are obtained through our SM, utilising $200$ polynomials with a precision of $200$ digits. In this context, $\Omega$ and $n$ represent the dimensionless frequency and the corresponding overtone, respectively, while $\Delta\Omega=\Omega_n-\Omega_{n+1}$. Entries labelled N/A indicate unavailable data.}
\label{scalargoverdampednearextremalrefineds1}
\vspace*{1em}
\begin{tabular}{||c|c|c|c|c|c|c|c|c||}
\hline\hline
$a_K$  & $n$ & $\Omega$               &$\Delta\Omega$  & $a_K$ & $n$ & $\Omega$  &$\Delta\Omega$\\ [0.5ex]
\hline\hline
$0.992$& $0$ & $0.0000-0.2504i$             & \mbox{N/A}     & $0.996$ & $0$ & $0.0000-0.2501i$             &\mbox{N/A}\\
       & $1$ & $0.0000-0.5001i$             & $0.2497i$      &         & $1$ & $0.0000-0.5000i$             &$0.2499i$\\
       & $2$ & $0.0000-0.7500i$             & $0.2499i$      &         & $2$ & $0.0000-0.7500i$             &$0.2500i$\\
       & $3$ & $0.0000-1.0000i$             & $0.2500i$      &         & $3$ & $0.0000-1.0000i$             &$0.2500i$\\
       & $4$ & $0.0000-1.2500i$             & $0.2500i$      &         & $4$ & $0.0000-1.2500i$             &$0.2500i$\\ 
       & $5$ & $0.0000-1.5000i$             & $0.2500i$      &         & $5$ & $0.0000-1.5000i$             &$0.2500i$\\
\hline
$0.994$& $0$ & $0.0000-0.2503i$             & \mbox{N/A}     & $0.998$ & $0$ & $0.0000-0.2501i$             &\mbox{N/A}\\
       & $1$ & $0.0000-0.5001i$             & $0.2498i$      &         & $1$ & $0.0000-0.5000i$             &$0.2499i$\\
       & $2$ & $0.0000-0.7500i$             & $0.2499i$      &         & $2$ & $0.0000-0.7500i$             &$0.2500i$\\
       & $3$ & $0.0000-1.0000i$             & $0.2500i$      &         & $3$ & $0.0000-1.0000i$             &$0.2500i$\\
       & $4$ & $0.0000-1.2500i$             & $0.2500i$      &         & $4$ & $0.0000-1.2500i$             &$0.2500i$\\ 
       & $5$ & $0.0000-1.5000i$             & $0.2500i$      &         & $5$ & $0.0000-1.5000i$             &$0.2500i$\\
       [1ex]
 \hline\hline 
 \end{tabular}
\end{table}

\begin{table}[t]
\centering
\caption{\textit{QNMs of axial gravitational perturbations ($s=2$) of the KSQC black hole for several values of $\ell$ and the deformation parameter $\mathfrak{a}=a/M$. We compare our spectral method results with $N=200$ Chebyshev polynomials (final column) against the third-order WKB employed by \cite{SalehASS2016}. Entries labelled N/A indicate unavailable data. The notation 'SM' stands for Spectral Method.}}
\label{table_s2_a0_25_and_a0_5}
\setlength\tabcolsep{0.1cm}
\def\arraystretch{1.5}
\begin{tabular}{@{}|c| c| c| c| c| c|c|c|c|c|c|c|@{}} 
\hline
$\mathfrak{a}$    & $a_K$   & $\ell$ & $n$ & $\Omega_{WKB}$ \cite{SalehASS2016} & $\Omega_{SM}$  &$a$    & $a_K$   & $\ell$ & $n$ & $\Omega_{WKB}$ \cite{SalehASS2016} & $\Omega_{SM}$ \\ [0.5ex] 
\hline
$0.25$ & $0.12$  & $2$    & $0$ & $0.3712-0.0890i$   & $0.3717-0.08880i$ & $0.5$ & $0.24$ & $2$ & $0$ & $0.3654-0.0886i$  & $0.3659-0.0883i$\\ 
       &         &        & $1$ & $0.3439-0.2745i$   & $0.3445-0.2735i$ &       &        &     & $1$ & $0.3377-0.2732i$   & $0.3383-0.2721i$\\
       &         &        & $2$ & $0.3005-0.4704i$   & $0.2986-0.4776i$ &       &        &     & $2$ & $0.2936-0.4684i$   & $0.2914-0.4757i$\\ 
       &         &        & $3$ & $0.2447-0.6720i$   & $0.2487-0.7045i$ &       &        &     & $3$ & $0.2368-0.6694i$   & $0.2405-0.7025i$\\
       &         &        & $4$ & $0.1755-0.8776i$   & $0.2043-0.9462i$ &       &        &     & $4$ & $0.1663-0.8745i$   & $0.1951-0.9442i$\\ 
       &         &        & $5$ & \mbox{N/A}         & $0.1656-1.1950i$ &       &        &     & $5$ & \mbox{N/A}         & $0.1548-1.1933i$\\
       &         &        & $6$ & \mbox{N/A}         & $0.1287-1.4475i$ &       &        &     & $6$ & \mbox{N/A}         & $0.1148-1.4464i$\\
       \hline
       &         & $3$    & $0$ & $0.5961-0.0926i$   & $0.5963-0.0925i$ &       &        & $3$ & $0$ & $0.5870-0.0921i$   & $0.5872-0.0920i$\\  
       &         &        & $1$ & $0.5791-0.2809i$   & $0.5794-0.2808i$ &       &        &     & $1$ & $0.5697-0.2795i$   & $0.5700-0.2793i$\\
       &         &        & $2$ & $0.5498-0.4759i$   & $0.5483-0.4783i$ &       &        &     & $2$ & $0.5399-0.4736i$   & $0.5383-0.4761i$\\ 
       &         &        & $3$ & $0.5121-0.6764i$   & $0.5083-0.6894i$ &       &        &     & $3$ & $0.5016-0.6732i$   & $0.4977-0.6866i$\\
       &         &        & $4$ & $0.4671-0.8802i$   & $0.4663-0.9146i$ &       &        &     & $4$ & $0.4558-0.8762i$   & $0.4551-0.9116i$\\ 
       &         &        & $5$ & \mbox{N/A}         & $0.4273-1.1511i$ &       &        &     & $5$ & \mbox{N/A}         & $0.4157-1.1479i$\\
       &         &        & $6$ & \mbox{N/A}         & $0.3935-1.3948i$ &       &        &     & $6$ & \mbox{N/A}         & $0.3815-1.3916i$\\
       \hline
       &         & $4$    & $0$ & $0.8049-0.0940i$   & $0.8050-0.0940i$ &       &        & $4$ & $0$ & $0.7927-0.0935i$   & $0.7927-0.0935i$\\ 
       &         &        & $1$ & $0.7922-0.2839i$   & $0.7923-0.2838i$ &       &        &     & $1$ & $0.7798-0.2824i$   & $0.7799-0.2824i$\\
       &         &        & $2$ & $0.7692-0.4782i$   & $0.7683-0.4791i$ &       &        &     & $2$ & $0.7564-0.4758i$   & $0.7555-0.4768i$\\ 
       &         &        & $3$ & $0.7387-0.6772i$   & $0.7352-0.6829i$ &       &        &     & $3$ & $0.7254-0.6739i$   & $0.7218-0.6798i$\\
       &         &        & $4$ & $0.7024-0.8799i$   & $0.6967-0.8970i$ &       &        &     & $4$ & $0.6885-0.8758i$   & $0.6827-0.8935i$\\
       &         &        & $5$ & \mbox{N/A}         & $0.6565-1.1217i$ &       &        &     & $5$ & \mbox{N/A}         & $0.6420-1.1178i$\\
       &         &        & $6$ & \mbox{N/A}         & $0.6179-1.3553i$ &       &        &     & $6$ & \mbox{N/A}         & $0.6029-1.3513i$\\
       \hline
       &         & $5$    & $0$ & $1.0070-0.0947i$   & $1.0070-0.0947i$ &       &        & $5$ & $0$ & $0.9918-0.0942i$   & $0.9918-0.0942i$\\ 
       &         &        & $1$ & $0.9968-0.2853i$   & $0.9969-0.2853i$ &       &        &     & $1$ & $0.9814-0.2838i$   & $0.9815-0.2838i$\\
       &         &        & $2$ & $0.9778-0.4791i$   & $0.9773-0.4795i$ &       &        &     & $2$ & $0.9621-0.4767i$   & $0.9615-0.4771i$\\ 
       &         &        & $3$ & $0.9519-0.6767i$   & $0.9494-0.6795i$ &       &        &     & $3$ & $0.9358-0.6734i$   & $0.9332-0.6763i$\\
       &         &        & $4$ & $0.9206-0.8778i$   & $0.9153-0.8869i$ &       &        &     & $4$ & $0.9040-0.8736i$   & $0.8986-0.8830i$\\
       &         &        & $5$ & \mbox{N/A}         & $0.8773-1.1027i$ &       &        &     & $5$ & \mbox{N/A}         & $0.8600-1.0984i$\\
       &         &        & $6$ & \mbox{N/A}         & $0.8380-1.3269i$ &       &        &     & $6$ & \mbox{N/A}         & $0.8202-1.3222i$\\
     [0.5ex] 
 \hline
 \end{tabular}
\end{table}

\begin{table}[t]
\centering
\caption{\textit{Candidate purely imaginary QNMs for axial gravitational perturbations of a KSQC black hole for several values of $\ell$ and the deformation parameter $\mathfrak{a}=a/M$. The corresponding results are obtained through our SM, utilising $200$ polynomials with a precision of $200$ digits. In this context, $\Omega$ and $n$ represent the dimensionless frequency and the corresponding overtone, respectively, while $\Delta\Omega=\Omega_n-\Omega_{n+1}$. Entries labelled N/A indicate unavailable data.}}
\label{table_s2_a0_25_and_a0_5_overdamped}
\setlength\tabcolsep{0.1cm}
\def\arraystretch{1.5}
\begin{tabular}{@{}|c| c| c| c| c| c|c|c|c|c|c|c|@{}} 
\hline
$\mathfrak{a}$    & $a_K$   & $\ell$ & $n$ & $\Omega$ & $\Delta\Omega$ & $a$    & $a_K$   & $\ell$ & $n$ & $\Omega$ & $\Delta\Omega$ \\ [0.5ex] 
\hline
$0.25$ & $0.12$  & $2$    & $0$ & $0.0000-18.3282i$   & \mbox{N/A}   & $0.5$ & $0.24$ & $2$ & $0$ & $0.0000-19.0529i$   & \mbox{N/A}\\ 
       &         &        & $1$ & $0.0000-18.5786i$   & $0.2504i$    &       &        &     & $1$ & $0.0000-19.3030i$   & $0.2501i$\\
       &         &        & $2$ & $0.0000-18.8283i$   & $0.2497i$    &       &        &     & $2$ & $0.0000-19.5531i$   & $0.2501i$\\ 
       &         &        & $3$ & $0.0000-19.0789i$   & $0.2506i$    &       &        &     & $3$ & $0.0000-19.8027i$   & $0.2496i$\\
       &         &        & $4$ & $0.0000-19.3289i$   & $0.2500i$    &       &        &     & $4$ & $0.0000-20.0527i$   & $0.2500i$\\ 
       &         &        & $5$ & $0.0000-19.5792i$   & $0.2503i$    &       &        &     & $5$ & $0.0000-20.3026i$   & $0.2499i$\\
       &         &        & $6$ & $0.0000-19.8294i$   & $0.2502i$    &       &        &     & $6$ & $0.0000-20.5524i$   & $0.2498i$\\
       \hline
       &         & $3$    & $0$ & $0.0000-19.0282i$   & \mbox{N/A}   &       &        & $3$ & $0$ & $0.0000-19.0005i$   & \mbox{N/A}\\  
       &         &        & $1$ & $0.0000-19.2785i$   & $0.2503i$    &       &        &     & $1$ & $0.0000-19.5011i$   & $0.5006i$\\
       &         &        & $2$ & $0.0000-19.5290i$   & $0.2505i$    &       &        &     & $2$ & $0.0000-19.7509i$   & $0.2498i$\\ 
       &         &        & $3$ & $0.0000-19.7796i$   & $0.2506i$    &       &        &     & $3$ & $0.0000-20.0013i$   & $0.2504i$\\
       &         &        & $4$ & $0.0000-20.0301i$   & $0.2505i$    &       &        &     & $4$ & $0.0000-20.2514i$   & $0.2501i$\\ 
       &         &        & $5$ & $0.0000-20.2806i$   & $0.2505i$    &       &        &     & $5$ & $0.0000-20.5016i$   & $0.2502i$\\
       &         &        & $6$ & $0.0000-20.5311i$   & $0.2505i$    &       &        &     & $6$ & $0.0000-20.7518i$   & $0.2502i$\\
       \hline
       &         & $4$    & $0$ & $0.0000-18.9611i$   & \mbox{N/A}   &       &        & $4$ & $0$ & $0.0000-18.6804i$   & \mbox{N/A}\\ 
       &         &        & $1$ & $0.0000-19.2121i$   & $0.2510i$    &       &        &     & $1$ & $0.0000-19.1819i$   & $0.5015i$\\
       &         &        & $2$ & $0.0000-19.4632i$   & $0.2511i$    &       &        &     & $2$ & $0.0000-19.4320i$   & $0.2501i$\\ 
       &         &        & $3$ & $0.0000-19.7140i$   & $0.2508i$    &       &        &     & $3$ & $0.0000-19.6828i$   & $0.2508i$\\
       &         &        & $4$ & $0.0000-19.9650i$   & $0.2510i$    &       &        &     & $4$ & $0.0000-19.9334i$   & $0.2506i$\\ 
       &         &        & $5$ & $0.0000-20.2158i$   & $0.2508i$    &       &        &     & $5$ & $0.0000-20.1839i$   & $0.2505i$\\
       &         &        & $6$ & $0.0000-20.4667i$   & $0.2509i$    &       &        &     & $6$ & $0.0000-20.4345i$   & $0.2506i$\\
       \hline
       &         & $5$    & $0$ & $0.0000-19.8863i$   & \mbox{N/A}   &       &        & $5$ & $0$ & $0.0000-20.3547i$   & \mbox{N/A}\\ 
       &         &        & $1$ & $0.0000-20.1377i$   & $0.2514i$    &       &        &     & $1$ & $0.0000-20.6054i$   & $0.2507i$\\
       &         &        & $2$ & $0.0000-20.3887i$   & $0.2510i$    &       &        &     & $2$ & $0.0000-20.8563i$   & $0.2509i$\\ 
       &         &        & $3$ & $0.0000-20.6402i$   & $0.2515i$    &       &        &     & $3$ & $0.0000-21.1073i$   & $0.2510i$\\
       &         &        & $4$ & $0.0000-20.8914i$   & $0.2512i$    &       &        &     & $4$ & $0.0000-21.3581i$   & $0.2508i$\\ 
       &         &        & $5$ & $0.0000-21.1426i$   & $0.2512i$    &       &        &     & $5$ & $0.0000-21.6090i$   & $0.2509i$\\
       &         &        & $6$ & $0.0000-21.3938i$   & $0.2512i$    &       &        &     & $6$ & $0.0000-21.8598i$   & $0.2508i$\\
     [0.5ex] 
 \hline
 \end{tabular}
\end{table}

\begin{table}[t]
\centering
\caption{\textit{QNMs of axial gravitational perturbations ($s=2$) of the KSQC black hole for several values of $\ell$ and the deformation parameter $\mathfrak{a}=a/M$. We compare our spectral method results with $N=200$ Chebyshev polynomials (final column) against the third-order WKB employed by \cite{SalehASS2016}. Entries labelled N/A indicate unavailable data. The notation 'SM' stands for Spectral Method.}}
\label{table_s2_a0_75_and_a1_0}
\setlength\tabcolsep{0.1cm}
\def\arraystretch{1.5}
\begin{tabular}{@{}|c| c| c| c| c| c|c|c|c|c|c|c|@{}} 
\hline
$\mathfrak{a}$    & $a_K$   & $\ell$ & $n$ & $\Omega_{WKB}$ \cite{SalehASS2016} & $\Omega$ (SM) &$a$    & $a_K$   & $\ell$ & $n$ & $\Omega_{WKB}$ \cite{SalehASS2016} & $\Omega$ (SM) \\ [0.5ex] 
\hline
$0.75$ & $0.35$  & $2$    & $0$ & $0.3564-0.0879i$   & $0.3569-0.0875i$ & $1$   & $0.45$ & $2$ & $0$ & $0.3451-0.0869i$   & $0.3455-0.0865i$\\ 
       &         &        & $1$ & $0.3281-0.2712i$   & $0.3285-0.2699i$ &       &        &     & $1$ & $0.3161-0.2684i$   & $0.3162-0.2669i$\\
       &         &        & $2$ & $0.2831-0.4652i$   & $0.2803-0.4727i$ &       &        &     & $2$ & $0.2699-0.4608i$   & $0.2662-0.4685i$\\ 
       &         &        & $3$ & $0.2247-0.6651i$   & $0.2278-0.6992i$ &       &        &     & $3$ & $0.2098-0.6592i$   & $0.2116-0.6947i$\\
       &         &        & $4$ & $0.1522-0.8694i$   & $0.1806-0.9410i$ &       &        &     & $4$ & $0.1350-0.8623i$   & $0.1617-0.9367i$\\ 
       &         &        & $5$ & \mbox{N/A}         & $0.1374-1.1905i$ &       &        &     & $5$ & \mbox{N/A}         & $0.1130-1.1871i$\\
       \hline
       &         & $3$    & $0$ & $0.5728-0.0912i$   & $0.5730-0.0912i$ &       &        & $3$ & $0$ & $0.5549-0.0901i$   & $0.5551-0.0901i$\\  
       &         &        & $1$ & $0.5551-0.2771i$   & $0.5554-0.2770i$ &       &        &     & $1$ & $0.5367-0.2739i$   & $0.5370-0.2738i$\\
       &         &        & $2$ & $0.5247-0.4698i$   & $0.5230-0.4724i$ &       &        &     & $2$ & $0.5054-0.4647i$   & $0.5035-0.4674i$\\ 
       &         &        & $3$ & $0.4854-0.6680i$   & $0.4814-0.6821i$ &       &        &     & $3$ & $0.4650-0.6611i$   & $0.4607-0.6759i$\\
       &         &        & $4$ & $0.4384-0.8698i$   & $0.4378-0.9066i$ &       &        &     & $4$ & $0.4164-0.8610i$   & $0.4161-0.8997i$\\ 
       &         &        & $5$ & \mbox{N/A}         & $0.3977-1.1427i$ &       &        &     & $5$ & \mbox{N/A}         & $0.3752-1.1354i$\\
       &         &        & $6$ & \mbox{N/A}         & $0.3630-1.3861i$ &       &        &     & $6$ & \mbox{N/A}         & $0.3400-1.3785i$\\
       \hline
       &         & $4$    & $0$ & $0.7737-0.0927i$   & $0.7738-0.0927i$ &       &        & $4$ & $0$ & $0.7497-0.0916i$   & $0.7498-0.0916i$\\ 
       &         &        & $1$ & $0.7605-0.2800i$   & $0.7606-0.2799i$ &       &        &     & $1$ & $0.7361-0.2767i$   & $0.7362-0.2767i$\\
       &         &        & $2$ & $0.7366-0.4719i$   & $0.7356-0.4729i$ &       &        &     & $2$ & $0.7116-0.4666i$   & $0.7104-0.4677i$\\ 
       &         &        & $3$ & $0.7049-0.6686i$   & $0.7011-0.6748i$ &       &        &     & $3$ & $0.6790-0.6614i$   & $0.6750-0.6680i$\\
       &         &        & $4$ & $0.6671-0.8691i$   & $0.6611-0.8876i$ &       &        &     & $4$ & $0.6400-0.8600i$   & $0.6338-0.8797i$\\
       &         &        & $5$ & \mbox{N/A}         & $0.6195-1.1114i$ &       &        &     & $5$ & \mbox{N/A}         & $0.5912-1.1027i$\\
       &         &        & $6$ & \mbox{N/A}         & $0.5797-1.3445i$ &       &        &     & $6$ & \mbox{N/A}         & $0.5507-1.3353i$\\
       \hline
       &         & $5$    & $0$ & $0.9681-0.0934i$   & $0.9682-0.0934i$ &       &        & $5$ & $0$ & $0.9382-0.0923i$   & $0.9383-0.0923i$\\ 
       &         &        & $1$ & $0.9576-0.2814i$   & $0.9576-0.2814i$ &       &        &     & $1$ & $0.9273-0.2782i$   & $0.9274-0.2782i$\\
       &         &        & $2$ & $0.9378-0.4727i$   & $0.9372-0.4732i$ &       &        &     & $2$ & $0.9070-0.4674i$   & $0.9063-0.4679i$\\ 
       &         &        & $3$ & $0.9109-0.6680i$   & $0.9082-0.6710i$ &       &        &     & $3$ & $0.8793-0.6607i$   & $0.8764-0.6639i$\\
       &         &        & $4$ & $0.8783-0.8668i$   & $0.8726-0.8767i$ &       &        &     & $4$ & $0.8458-0.8576i$   & $0.8399-0.8681i$\\
       &         &        & $5$ & \mbox{N/A}         & $0.8332-1.0912i$ &       &        &     & $5$ & \mbox{N/A}         & $0.7994-1.0815i$\\
       &         &        & $6$ & \mbox{N/A}         & $0.7925-1.3145i$ &       &        &     & $6$ & \mbox{N/A}         & $0.7577-1.3040i$\\
     [0.5ex] 
 \hline
 \end{tabular}
\end{table}

\begin{table}[t]
\centering
\caption{\textit{Candidate purely imaginary QNMs for axial gravitational perturbations of a KSQC black hole for several values of $\ell$ and the deformation parameter $\mathfrak{a}=a/M$. The corresponding results are obtained through our SM, utilising $200$ polynomials with a precision of $200$ digits. In this context, $\Omega$ and $n$ represent the dimensionless frequency and the corresponding overtone, respectively, while $\Delta\Omega=\Omega_n-\Omega_{n+1}$. Entries labelled N/A indicate unavailable data.}}
\label{table_s2_a0_75_and_a1_0_overdamped}
\setlength\tabcolsep{0.1cm}
\def\arraystretch{1.5}
\begin{tabular}{@{}|c| c| c| c| c| c|c|c|c|c|c|c|@{}} 
\hline
$\mathfrak{a}$    & $a_K$   & $\ell$ & $n$ & $\Omega$ & $\Delta\Omega$ & $a$    & $a_K$   & $\ell$ & $n$ & $\Omega$ & $\Delta\Omega$ \\ [0.5ex] 
\hline
$0.75$ & $0.35$  & $2$    & $0$ & $0.0000-20.2322i$   & \mbox{N/A}   & $1$   & $0.45$ & $2$ & $0$ & $0.0000-23.8162i$   & \mbox{N/A}\\ 
       &         &        & $1$ & $0.0000-20.9792i$   & $0.7470i$    &       &        &     & $1$ & $0.0000-24.0632i$   & $0.2470i$\\
       &         &        & $2$ & $0.0000-21.2279i$   & $0.2487i$    &       &        &     & $2$ & $0.0000-24.3102i$   & $0.2470i$\\ 
       &         &        & $3$ & $0.0000-21.4768i$   & $0.2489i$    &       &        &     & $3$ & $0.0000-24.5580i$   & $0.2478i$\\
       &         &        & $4$ & $0.0000-21.7259i$   & $0.2491i$    &       &        &     & $4$ & $0.0000-24.8051i$   & $0.2471i$\\ 
       &         &        & $5$ & $0.0000-21.9747i$   & $0.2488i$    &       &        &     & $5$ & $0.0000-25.0523i$   & $0.2472i$\\
       &         &        & $6$ & $0.0000-22.2237i$   & $0.2490i$    &       &        &     & $6$ & $0.0000-25.2998i$   & $0.2475i$\\
       \hline
       &         & $3$    & $0$ & $0.0000-20.9278i$   & \mbox{N/A}   &       &        & $3$ & $0$ & $0.0000-23.7680i$   & \mbox{N/A}\\  
       &         &        & $1$ & $0.0000-21.1769i$   & $0.2491i$    &       &        &     & $1$ & $0.0000-24.0156i$   & $0.2476i$\\
       &         &        & $2$ & $0.0000-21.4261i$   & $0.2492i$    &       &        &     & $2$ & $0.0000-24.2626i$   & $0.2470i$\\ 
       &         &        & $3$ & $0.0000-21.6755i$   & $0.2494i$    &       &        &     & $3$ & $0.0000-24.5107i$   & $0.2481i$\\
       &         &        & $4$ & $0.0000-21.9246i$   & $0.2491i$    &       &        &     & $4$ & $0.0000-24.7582i$   & $0.2475i$\\ 
       &         &        & $5$ & $0.0000-22.1739i$   & $0.2493i$    &       &        &     & $5$ & $0.0000-25.0055i$   & $0.2473i$\\
       &         &        & $6$ & $0.0000-22.4231i$   & $0.2492i$    &       &        &     & $6$ & $0.0000-25.2533i$   & $0.2478i$\\
       \hline
       &         & $4$    & $0$ & $0.0000-20.6095i$   & \mbox{N/A}   &       &        & $4$ & $0$ & $0.0000-23.4542i$   & \mbox{N/A}\\ 
       &         &        & $1$ & $0.0000-20.8594i$   & $0.2499i$    &       &        &     & $1$ & $0.0000-24.1983i$   & $0.7441i$\\
       &         &        & $2$ & $0.0000-21.1091i$   & $0.2497i$    &       &        &     & $2$ & $0.0000-24.4465i$   & $0.2482i$\\ 
       &         &        & $3$ & $0.0000-21.3586i$   & $0.2495i$    &       &        &     & $3$ & $0.0000-24.6945i$   & $0.2480i$\\
       &         &        & $4$ & $0.0000-21.6083i$   & $0.2497i$    &       &        &     & $4$ & $0.0000-24.9420i$   & $0.2475i$\\ 
       &         &        & $5$ & $0.0000-21.8578i$   & $0.2495i$    &       &        &     & $5$ & $0.0000-25.1901i$   & $0.2481i$\\
       &         &        & $6$ & $0.0000-22.1074i$   & $0.2496i$    &       &        &     & $6$ & $0.0000-25.4379i$   & $0.2478i$\\
       \hline
       &         & $5$    & $0$ & $0.0000-20.5287i$   & \mbox{N/A}   &       &        & $5$ & $0$ & $0.0000-23.3777i$   & \mbox{N/A}\\ 
       &         &        & $1$ & $0.0000-21.0285i$   & $0.4998i$    &       &        &     & $1$ & $0.0000-23.8747i$   & $0.4970i$\\
       &         &        & $2$ & $0.0000-21.2789i$   & $0.2504i$    &       &        &     & $2$ & $0.0000-24.1224i$   & $0.2477i$\\ 
       &         &        & $3$ & $0.0000-21.5288i$   & $0.2499i$    &       &        &     & $3$ & $0.0000-24.3706i$   & $0.2482i$\\
       &         &        & $4$ & $0.0000-21.7787i$   & $0.2499i$    &       &        &     & $4$ & $0.0000-24.6191i$   & $0.2485i$\\ 
       &         &        & $5$ & $0.0000-22.0287i$   & $0.2500i$    &       &        &     & $5$ & $0.0000-24.8670i$   & $0.2479i$\\
       &         &        & $6$ & $0.0000-22.2786i$   & $0.2499i$    &       &        &     & $6$ & $0.0000-25.1152i$   & $0.2482i$\\
     [0.5ex] 
 \hline
 \end{tabular}
\end{table}

\begin{table}[t]
\centering
\caption{\textit{QNMs of axial gravitational perturbations ($s=2$) of the KSQC black hole for several values of $\ell$ and the deformation parameter $\mathfrak{a}=a/M$. We compare our spectral method results with $N=200$ Chebyshev polynomials (final column) against the third-order WKB employed by \cite{SalehASS2016}. Entries labelled N/A indicate unavailable data. The notation 'SM' stands for Spectral Method.}}
\label{table_s2_a1_75_and_a0_99}
\setlength\tabcolsep{0.1cm}
\def\arraystretch{1.5}
\begin{tabular}{@{}|c| c| c| c| c| c|c|c|c|c|c|c|@{}} 
\hline
$\mathfrak{a}$    & $a_K$   & $\ell$ & $n$ & $\Omega_{WKB}$ \cite{SalehASS2016} & $\Omega_{SM}$ &$a$    & $a_K$   & $\ell$ & $n$ & $\Omega_{WKB}$ \cite{SalehASS2016} & $\Omega_{SM}$ \\ [0.5ex] 
\hline
$1.25$ & $0.53$  & $2$    & $0$ & $0.3322-0.0857i$   & $0.3325-0.0852i$ &$14.04$& $0.99$ & $2$ & $0$ & \mbox{N/A}   & $0.0711-0.0319i$\\ 
       &         &        & $1$ & $0.3025-0.2651i$   & $0.3021-0.2633i$ &       &        &     & $1$ & \mbox{N/A}   & \mbox{N/A}\\
       &         &        & $2$ & $0.2552-0.4554i$   & $0.2501-0.4635i$ &       &        &     & $2$ & \mbox{N/A}   & \mbox{N/A}\\ 
       &         &        & $3$ & $0.1933-0.6519i$   & $0.1930-0.6893i$ &       &        &     & $3$ & \mbox{N/A}   & \mbox{N/A}\\
       &         &        & $4$ & $0.1163-0.8533i$   & $0.1392-0.9316i$ &       &        &     & $4$ & \mbox{N/A}   & \mbox{N/A}\\ 
       \hline
       &         & $3$    & $0$ & $0.5346-0.0888i$   & $0.5347-0.0888i$ &       &        & $3$ & $0$ & \mbox{N/A}   & $0.1210-0.0336i$\\  
       &         &        & $1$ & $0.5158-0.2700i$   & $0.5160-0.2699i$ &       &        &     & $1$ & \mbox{N/A}   & $0.0995-0.1055i$\\
       &         &        & $2$ & $0.4836-0.4584i$   & $0.4814-0.4614i$ &       &        &     & $2$ & \mbox{N/A}   & \mbox{N/A}\\ 
       &         &        & $3$ & $0.4420-0.6525i$   & $0.4373-0.6684i$ &       &        &     & $3$ & \mbox{N/A}   & \mbox{N/A}\\
       &         &        & $4$ & $0.3918-0.8501i$   & $0.3915-0.8913i$ &       &        &     & $4$ & \mbox{N/A}   & \mbox{N/A}\\ 
       &         &        & $5$ & \mbox{N/A}         & $0.3500-1.1264i$ &       &        &     & $5$ & \mbox{N/A}   & \mbox{N/A}\\
       &         &        & $6$ & \mbox{N/A}         & $0.3144-1.3689i$ &       &        &     & $6$ & \mbox{N/A}   & \mbox{N/A}\\
       \hline
       &         & $4$    & $0$ & $0.7224-0.0902i$   & $0.7225-0.0902i$ &       &        & $4$ & $0$ & \mbox{N/A}   & $0.1661-0.0336i$\\ 
       &         &        & $1$ & $0.7084-0.2728i$   & $0.7085-0.2727i$ &       &        &     & $1$ & \mbox{N/A}   & $0.1537-0.1048i$\\
       &         &        & $2$ & $0.6832-0.4602i$   & $0.6819-0.4614i$ &       &        &     & $2$ & \mbox{N/A}   & \mbox{N/A}\\ 
       &         &        & $3$ & $0.6496-0.6526i$   & $0.6453-0.6597i$ &       &        &     & $3$ & \mbox{N/A}   & \mbox{N/A}\\
       &         &        & $4$ & $0.6095-0.8487i$   & $0.6029-0.8698i$ &       &        &     & $4$ & \mbox{N/A}   & \mbox{N/A}\\
       &         &        & $5$ & \mbox{N/A}         & $0.5593-1.0919i$ &       &        &     & $5$ & \mbox{N/A}   & \mbox{N/A}\\
       &         &        & $6$ & \mbox{N/A}         & $0.5182-1.3238i$ &       &        &     & $6$ & \mbox{N/A}   & \mbox{N/A}\\
       \hline
       &         & $5$    & $0$ & $0.9042-0.0909i$   & $0.9042-0.0909i$ &       &        & $5$ & $0$ & \mbox{N/A}   & $0.2088-0.0332i$\\ 
       &         &        & $1$ & $0.8930-0.2742i$   & $0.8930-0.2742i$ &       &        &     & $1$ & \mbox{N/A}   & $0.2009-0.1030i$\\
       &         &        & $2$ & $0.8721-0.4609i$   & $0.8713-0.4614i$ &       &        &     & $2$ & \mbox{N/A}   & \mbox{N/A}\\ 
       &         &        & $3$ & $0.8436-0.6517i$   & $0.8404-0.6552i$ &       &        &     & $3$ & \mbox{N/A}   & \mbox{N/A}\\
       &         &        & $4$ & $0.8091-0.8462i$   & $0.8028-0.8575i$ &       &        &     & $4$ & \mbox{N/A}   & \mbox{N/A}\\
       &         &        & $5$ & \mbox{N/A}         & $0.7611-1.0695i$ &       &        &     & $5$ & \mbox{N/A}   & \mbox{N/A}\\
       &         &        & $6$ & \mbox{N/A}         & $0.7185-1.2910i$ &       &        &     & $6$ & \mbox{N/A}   & \mbox{N/A}\\
     [0.5ex] 
 \hline
 \end{tabular}
\end{table}

\begin{table}[t]
\centering
\caption{\textit{Candidate purely imaginary QNMs for axial gravitational perturbations of a KSQC black hole for several values of $\ell$ and the deformation parameter $\mathfrak{a}=a/M$. The corresponding results are obtained through our SM, utilising $200$ polynomials with a precision of $200$ digits. In this context, $\Omega$ and $n$ represent the dimensionless frequency and the corresponding overtone, respectively, while $\Delta\Omega=\Omega_n-\Omega_{n+1}$. Entries labelled N/A indicate unavailable data.}}
\label{table_s2_a1_25_and_a0_99_overdamped}
\setlength\tabcolsep{0.1cm}
\def\arraystretch{1.5}
\begin{tabular}{@{}|c| c| c| c| c| c|c|c|c|c|c|c|@{}} 
\hline
$\mathfrak{a}$    & $a_K$   & $\ell$ & $n$ & $\Omega$ & $\Delta\Omega$ & $a$    & $a_K$   & $\ell$ & $n$ & $\Omega$ & $\Delta\Omega$ \\ [0.5ex] 
\hline
$1.25$ & $0.53$  & $2$    & $0$ & \mbox{N/A}   & \mbox{N/A}&    $14.04$& $0.99$ & $2$ & $0$ & $0.0000-0.2492i$   & \mbox{N/A}\\ 
       &         &        & $1$ & \mbox{N/A}   & \mbox{N/A}    &       &        &     & $1$ & $0.0000-0.4994i$   & $0.2502i$\\
       &         &        & $2$ & \mbox{N/A}   & \mbox{N/A}    &       &        &     & $2$ & $0.0000-0.7496i$   & $0.2502i$\\ 
       &         &        & $3$ & \mbox{N/A}   & \mbox{N/A}    &       &        &     & $3$ & $0.0000-0.9997i$   & $0.2501i$\\
       &         &        & $4$ & \mbox{N/A}   & \mbox{N/A}    &       &        &     & $4$ & $0.0000-1.2498i$   & $0.2501i$\\ 
       &         &        & $5$ & \mbox{N/A}   & \mbox{N/A}    &       &        &     & $5$ & $0.0000-1.4998i$   & $0.2500i$\\
       &         &        & $6$ & \mbox{N/A}   & \mbox{N/A}    &       &        &     & $6$ & $0.0000-1.7498i$   & $0.2500i$\\
       \hline
       &         & $3$    & $0$ & \mbox{N/A}   & \mbox{N/A}    &       &        & $3$ & $0$ & $0.0000-0.2508i$   & \mbox{N/A}\\  
       &         &        & $1$ & \mbox{N/A}   & \mbox{N/A}    &       &        &     & $1$ & $0.0000-0.4998i$   & $0.2490i$\\
       &         &        & $2$ & \mbox{N/A}   & \mbox{N/A}    &       &        &     & $2$ & $0.0000-0.7498i$   & $0.2500i$\\ 
       &         &        & $3$ & \mbox{N/A}   & \mbox{N/A}    &       &        &     & $3$ & $0.0000-0.9998i$   & $0.2500i$\\
       &         &        & $4$ & \mbox{N/A}   & \mbox{N/A}    &       &        &     & $4$ & $0.0000-1.2498i$   & $0.2500i$\\ 
       &         &        & $5$ & \mbox{N/A}   & \mbox{N/A}    &       &        &     & $5$ & $0.0000-1.4998i$   & $0.2500i$\\
       &         &        & $6$ & \mbox{N/A}   & \mbox{N/A}    &       &        &     & $6$ & $0.0000-1.7499i$   & $0.2501i$\\
       \hline
       &         & $4$    & $0$ & \mbox{N/A}   & \mbox{N/A}    &       &        & $4$ & $0$ & $0.0000-0.2527i$   & \mbox{N/A}\\ 
       &         &        & $1$ & \mbox{N/A}   & \mbox{N/A}    &       &        &     & $1$ & $0.0000-0.5002i$   & $0.2475i$\\
       &         &        & $2$ & \mbox{N/A}   & \mbox{N/A}    &       &        &     & $2$ & $0.0000-0.7499i$   & $0.2497i$\\ 
       &         &        & $3$ & \mbox{N/A}   & \mbox{N/A}    &       &        &     & $3$ & $0.0000-0.9999i$   & $0.2500i$\\
       &         &        & $4$ & \mbox{N/A}   & \mbox{N/A}    &       &        &     & $4$ & $0.0000-1.2499i$   & $0.2500i$\\ 
       &         &        & $5$ & \mbox{N/A}   & \mbox{N/A}    &       &        &     & $5$ & $0.0000-1.4999i$   & $0.2500i$\\
       &         &        & $6$ & \mbox{N/A}   & \mbox{N/A}    &       &        &     & $6$ & $0.0000-1.7499i$   & $0.2500i$\\
       \hline
       &         & $5$    & $0$ & $0.0000-13.2837$   & \mbox{N/A}      &  &     & $5$ & $0$ & $0.0000-0.2549i$   & \mbox{N/A}\\ 
       &         &        & $1$ & \mbox{N/A}   & \mbox{N/A}    &       &        &     & $1$ & $0.0000-0.5008i$   & $0.2459i$\\
       &         &        & $2$ & \mbox{N/A}   & \mbox{N/A}    &       &        &     & $2$ & $0.0000-0.7502i$   & $0.2494i$\\ 
       &         &        & $3$ & \mbox{N/A}   & \mbox{N/A}    &       &        &     & $3$ & $0.0000-1.0000i$   & $0.2498i$\\
       &         &        & $4$ & \mbox{N/A}   & \mbox{N/A}    &       &        &     & $4$ & $0.0000-1.2499i$   & $0.2499i$\\ 
       &         &        & $5$ & \mbox{N/A}   & \mbox{N/A}    &       &        &     & $5$ & $0.0000-1.4999i$   & $0.2500i$\\
       &         &        & $6$ & \mbox{N/A}   & \mbox{N/A}    &       &        &     & $6$ & $0.0000-1.7499i$   & $0.2500i$\\
     [0.5ex] 
 \hline
 \end{tabular}
\end{table}

\bibliography{bib}

\end{document}